\documentclass{aastex63}

\received{March 15, 2021}
\accepted{\today}
\submitjournal{ApJ}

\shorttitle{Time-Symmetric GRB Pulses and Light Curve Morphology}
\shortauthors{Hakkila}
\graphicspath{{./}{figures/}}

\begin{document}

\title{How Temporal Symmetry Defines Morphology in BATSE Gamma-Ray Burst Pulse Light Curves}

\correspondingauthor{Jon Hakkila}
\email{jon.hakkila@uah.edu}

\author{Jon Hakkila}
\affiliation{Department of Physics and Astronomy, The University of Alabama in Huntsville, 301 Sparkman Drive, Huntsville, AL 35899, USA}
\affiliation{The Graduate School, University of Charleston, SC at the College of Charleston, 66 George St., Charleston, SC 29424-0001, USA}
\affiliation{Department of Physics and Astronomy, College of Charleston, 66 George St. Charleston, SC 29424-0001, USA}

\begin{abstract}

We present compelling evidence that most gamma-ray burst (GRB) pulse light curves can be characterized by a smooth single-peaked component coupled with a more complex emission structure that is temporally-symmetric around the time of the pulse peak. The model successfully fits 86\% of BATSE GRB pulses bright enough for structure properties to be measured. Surprisingly, a GRB pulse's light curve morphology can be accurately predicted by the pulse asymmetry and the stretching/compression needed to align the structural components preceding the temporal mirror with the time-reversed components following it. Such a prediction is only possible because GRB pulses exhibit temporal symmetry. Time-asymmetric pulses include {\em FREDs}, {\em rollercoaster pulses}, and {\em asymmetric u-pulses}, while time-symmetric pulses include {\em u-pulses} and {\em crowns}. Each morphological type is characterized by specific asymmetries, stretching parameters, durations, and alignments between the smooth and structured components, and a delineation in the asymmetry/stretching distribution suggests that symmetric pulses and asymmetric pulses may belong to separate populations. Furthermore, pulses belonging to the short GRB class exhibit similar morphologies to the long GRB class, but appear to simply occur on shorter timescales. 

\end{abstract}

\keywords{Gamma-ray bursts (629), Light curve classification (1954), Astrostatistics techniques (1886)}


\section{Introduction} \label{sec:intro}
Although pulses have been long recognized as the basic units of gamma-ray burst (GRB) prompt emission \citep{nor96}, there is surprisingly little agreement about the observational definition of a GRB pulse. With limited available insights into the microphysical mechanisms that produce GRB prompt emission ({\em e.g.,} \cite{dai17}), researchers have resorted to fitting GRB light curves by applying empirical mathematical functions to photon counts data. For the most part, smooth, single-peaked, temporally-asymmetric, monotonically-increasing and decreasing functions have been used, which we hereafter define to be {\em monotonic pulse models}. Most monotonic pulse models have allowed for longer decay times than rise times ({\em e.g.}, \cite{nor96,ste96,lee00a,lee00b,koc03,nor05,nem12}), although a few are time-symmetric ({\em e.g.}, \cite{bha12}). Historically, monotonic pulse models have been used to fit GRB emission episodes spanning wide ranges of duration and intensity without regard to whether or not the data seem to warrant the use of these models. 

Empirical monotonic pulse models can be used to fit the general characteristics of FRED (Fast Rise Exponential Decay) pulses from long/intermediate GRBs ({\em e.g.}, \cite{nor86,nor96,pen97,nor05,hak18b}), short GRBs ({\em e.g.}, \cite{nor11,hak14,hak18a}), and the x-ray flares found in GRB afterglows ({\em e.g.}, \cite{mar10,hak16}), even though these event durations range from milliseconds to hundreds of seconds and even though their mean energies range from keV to MeV.

Spectral evolution plays an important secondary role in characterizing the smoothly-varying light curves of FRED pulses, whose spectral properties are clearly convolved with their intensity variations.
FREDs evolve from hard-to-soft, with asymmetric FREDs showing more pronounced spectral evolution than symmetric FREDs ({\em e.g.} \cite{cri99,ryd99,hak11,hak15}). When early pulse emission can be observed, pulses are shown to start near-simultaneously at all energies \citep{hak09}. The bandwidth of the observing instrument thus contributes to the part of the spectrum observed, and thus to the pulse shape \citep{hak15,pre16}.


Not all GRB pulses are FREDs, and not all FREDs can be characterized by monotonic pulse models. FREDs make up only a fraction of all GRB pulses. Furthermore, the assumption that all GRB pulses can be characterized by smooth, single-peaked functions is grossly inadequate.  The vast majority of pulses exhibit intensity variations far in excess of what can be attributed to Poisson background fluctuations \citep{hak14}, and some of these rapid fluctuations occur on timescales as short as a few milliseconds \citep{mac12}. If every non-Poisson variation in a GRB light curve constitutes a pulse, then GRBs should be composed of many (sometimes hundreds of) pulses, all of which have the peculiar property of clumping together and of being directly associated with longer, smoother underlying emission episodes. Are GRB pulses the emission episodes, or are GRB pulses the fluctuations? 

The process of characterizing pulses is complicated by the fact that GRBs are observed in low signal-to-noise-ratio regimes. Thus, separating emission from noise is instrument-dependent, and additional clues about the characteristics of GRB pulses can only be obtained after accounting for signal-to-noise ratio and other instrumental properties. This approach finds that GRB pulses are not solely explained by monotonic models of the emission episodes, but are also characterized by {\em residual structures} that overlay the monotonic pulses \citep{bor01,hak14}. The structures typically exhibit considerable complexity for pulses observed at high signal-to-noise ratios. For moderate signal-to-noise ratios found in isolated FRED pulses, the residuals take the form of a smooth triple-peaked wave that have been seen in GRB pulses observed by BATSE, Swift, Fermi's GBM, and Suzaku \citep{hak14,hak15,hak18a,hak18b}, as well as in x-ray flares \citep{hak16}. This triple-peaked structure deviates most from monotonicity during the pulse rise (the {\em precursor peak}), near the pulse maximum (the {\em central peak}), and during the pulse decay (the {\em decay peak}).  At low signal-to-noise ratios only the monotonic pulse component can be reliably recovered.

In fact, the residual structures found in isolated FRED pulses do not constitute unrelated events, but are instead causally linked to each other as well as to the underlying monotonic component. The causal linkage is even stranger than can be attributed to simple clustering because structures overlaying a pulse are {\em temporally symmetric} (also referred to as {\em time-reversed}): they can be folded at a {\em time of reflection}, then {\em temporally stretched} until the time-reversed residuals at the beginning of the pulse match those at the end of the pulse \citep{hak18b}. The asymmetry of the residual structure correlates with the asymmetry of the monotonic component. The time of reflection at which the residuals can be folded generally coincides with the time of monotonic component's peak intensity or is slightly offset from it \citep{hak18a,hak18b}. The residuals are aligned in sequences similar to palindromes (anagrams that produce words when the letters are read in either forward or reverse order), where the {\em time-forward} structural component preceding the time of reflection connects to the folded {\em time-reversed} structural component following the time of reflection. The stretching parameter $s_{\rm mirror}$ indicates the amount by which the residuals need to be stretched so that the time-forward and time-reversed components match.

The residual structures appear related to hardness resets during the pulse spectral evolution process. GRB pulses have been previously described as having either ``hard-to-soft'' or ``intensity tracking'' behaviors ({\em e.g.,} \cite{whe73, gol83, nor86, pac92}), but this oversimplified bimodal classification scheme has instead represents a continuum of behaviors ({\em e.g.,} \cite{kar94, bha94, for95,bor01}). This quandry can be reconciled by treating each GRB pulse as a {\em structured} episode rather than one in which pulses are statistically-significant intensity peaks: the general hard-to-soft evolution fluctuates rather than being smooth and gradual because GRB pulse spectra re-harden around the times that they re-brighten \citep{lia96, hak15, hak18a, hak18b}. Pulses appear to evolve hard-to-soft if their central peaks are softer than their initial emission, whereas they follow intensity tracking behaviors if their central peaks are harder than their initial emission. Depending on the relative hardnesses of the peaks, there exists a wide range of intermediary behaviors \citep{hak15} bounding these extremes.

FRED-like pulses (see Figure 15 of \cite{hak15}) are not the only ones exhibiting temporally-reversed structures. These features are also found in pulses with more complex light curves (see Figure 5 of \cite{hak19}), suggesting that measurement of temporal symmetry might be a defining characteristic of GRB pulses rather than just a quirk. Despite the strong correlation found between pulse asymmetry and and residual stretching, a large scatter is present in the relation (see Figures 18b and 19 of \cite{hak19} and Figure 17 of \cite{hak15}) that cannot be explained by random Poisson variations. The large intrinsic scatter between two strongly-correlated variables suggests that some additional important information is present and needs to be accounted for.


The goals of this study are to 1) explore the ubiquity of temporally-reversible pulse structure in GRB pulses using a large sequential GRB database, and 2) explain the large dispersion in the pulse asymmetry vs.~residual stretching relationship. To perform this study, we use GRB observations from BATSE (the Burst And Transient Source Experiment on NASA's Compton Gamma Ray Observatory; \cite{fish13}). Recognizing the importance of underlying pulse symmetry, we augment the pulse-fitting approach used in \cite{hak19} by also accounting for symmetric GRB pulse shapes. 

\section{Modeling Monotonic GRB Pulse Shapes} \label{sec:style}

We choose to fit GRB pulse light curves using simple generic mathematical models containing a minimum number of free parameters spanning a wide range of asymmetries. 

\subsection{Time Asymmetric Pulse Shapes}\label{sec:norrismodel}

The basis of our asymmetric pulse model is the pulse intensity function of \cite{nor05}:

\begin{equation}\label{eqn:function} 
I(t) = A \lambda e^{[-\tau_1/(t - t_s) - (t - t_s)/\tau_2]},
\end{equation}
where $t$ is time since trigger, $A$ is the pulse amplitude, $t_s$ is the pulse start time, $\tau_1$ is the pulse rise parameter, $\tau_2$ is the pulse decay parameter, and the normalization constant $\lambda$ is $\lambda = \exp{[2 (\tau_1/\tau_2)^{1/2}]}$. 
The intensity is the count rate obtained from the total counts observed in BATSE energy channels 1 ($20 - 50$ keV),
2 ($50 - 100$ keV), 3 ($100 - 300$ keV), and 4 (300 keV $- 1$ MeV).
Poisson statistics and a two-parameter background counts model of the form 
\begin{equation}\label{eqn:background} 
B=B_0+BS \times t,
\end{equation}
where $B$ is the background counts in each bin and $B_0$ and $BS$ are constants denoting the mean background (counts) and the rate of change of the mean background (counts/s)). This model can be used to produce observable pulse parameters such as pulse peak time ($\tau_{\rm peak} = t_s + \sqrt{\tau_1 \tau_2}$), pulse duration ($w$) and pulse asymmetry ($\kappa$). As a result of the smooth rapid rise and more gradual fall of the pulse model, $w$ and $\kappa$ are measured relative to some fraction of the peak intensity. Following \cite{hak11}, we measure $w$ and $\kappa$ at $I_{\rm meas}/I_{\rm peak}=e^{-3}$ (corresponding to $0.05 I_{\rm peak}$), or

\begin{equation}\label{eqn:duration}
w = \tau_2 [9 + 12\mu]^{1/2},
\end{equation}
where $\mu = \sqrt{\tau_1/\tau_2}$, and

\begin{equation}\label{eqn:asymmetry}
\kappa \equiv [1 + 4\mu/3]^{-1/2}.
\end{equation}
Asymmetries range from symmetric ($\kappa=0$ when $\tau_1 \gg \tau_2$ and $\mu \rightarrow \infty$)
to asymmetric ($\kappa=1$ when $\tau_1 \ll \tau_2$ and $\mu \rightarrow 0$) with longer decay than rise times.

\subsection{Time-Symmetric Pulse Shapes}\label{sec:normalmodel}

Many GRB pulses have symmetric shapes that are not easily fitted by the \cite{nor05} pulse model. The \cite{nor05} model requires that symmetric pulses have $\tau_1 >> \tau_2$ as well as large $t_s$ values. The interrelationship between $t_s$, $\tau_1$, and $\tau_2$, combined with the difficulty in separating $t_s$ from the background, makes it difficult to optimize a fit to Equation \ref{eqn:function} for symmetric GRB pulses.

In these circumstances, it is simpler to assume that the pulse shape is itself time-symmetric, such as a Gaussian distribution function of the form
\begin{equation}\label{eqn:normfunction} 
I(t) = \frac{C}{\sigma\sqrt{2\pi}} \exp{\left[-\left(\frac{t-t_0}{\sqrt{2}\sigma}\right)^2\right]},
\end{equation}
where $C$ is the pulse amplitude, $t_0$ is the time at which $C$ occurs, and $\sigma^2$ is the variance. The pulse background is modeled as before using $B_0$ and $BS$.

We test the compatibility of these two mathematical models by comparing the behavior of the symmetric \cite{nor05} function to that of Equation \ref{eqn:function} in the vicinity of the pulse peak using a Taylor series:

\begin{equation}\label{eqn:Taylor}
I(x)=\sum_{n=0}^{\infty} \frac{I^n(a)}{n!}(x-a)^n,
\end{equation}
where $I^n(a)$ represents the $n$th derivative of $I(x)$ with respect to $x=t-t_s$ and $a$ signifies that the function is to be evaluated at $x=a=\sqrt{\tau_1\tau_2}$.

The first four terms of the expansion produce



\begin{equation}\label{eqn:d1}
I(x)= A\left[1-\frac{(x-\sqrt{\tau_1 \tau_2})^2}{\sqrt{\tau_1 \tau_2^3}} +\left(\frac{1}{\tau_1\tau_2^2} - \frac{2}{3\sqrt{\tau_1\tau_2^{3}}}\right)(x-\sqrt{\tau_1 \tau_2})^3\right]+...
\end{equation}

The Taylor series expansion for a Gaussian distribution function is


\begin{equation}\label{eqn:Taylor1}
I(t)=\frac{A'}{\sigma\sqrt{2\pi}}\exp\left[-\frac{(t-t_0)^2}{2\sigma^2}\right]=\frac{A'}{\sigma\sqrt(2\pi)}\sum_{n=0}^{\infty} \left(-\frac{1}{2\sigma^2}\right)^n\frac{(t-t_0)^{2n}}{n!},
\end{equation}

Matching like terms between Equation \ref{eqn:d1} and Equation \ref{eqn:Taylor1}, we find that to order $t^2$,

\begin{equation}\label{eqn:sigma}
\sigma=\frac{\sqrt{2}}{2}\tau_1^{1/4}\tau_2^{3/4},
\end{equation}

\begin{equation}\label{eqn:B}
A'=A\sqrt{(\pi/2)} \tau_1^{1/4}\tau_2^{3/4},
\end{equation}
and
\begin{equation}\label{eqn:t0}
t_0=t_s+\sqrt{\tau_1\tau_2}.
\end{equation}

The asymmetry ($\kappa$) and duration ($w$) of a Gaussian pulse can be obtained by applying the condition $\tau_1 >> \tau_2$ to Equation \ref{eqn:asymmetry} and Equation \ref{eqn:duration} to produce 
\begin{equation}\label{eqn:symmetry} 
\kappa \approx 0
\end{equation} 
and
\begin{equation}\label{eqn:normdur} 
w \approx \sqrt{12\tau_1^{1/2}\tau_2^{3/2}}=\sqrt{24}\sigma\approx4.9\sigma.
\end{equation} 

The monotonic pulse duration defined here is generally large enough to encompass most structure that is directly associated with a particular pulse, but may not be large enough to account for faint structures that occasionally start well in advance of some pulses and continue long after they have ended. Since we expect that such structures will be consistent with the temporal mirroring found in previous studies, we allow ourselves the option of increasing the pulse duration definition when such extended pulse features are present.

\subsection{BATSE GRB Light Curves}\label{sec:database}

The bursts used in this study are primarily BATSE GRBs for which 64 ms data are available\footnote{{\tt https://heasarc.gsfc.nasa.gov/FTP/compton/data/batse/ascii\_data/64ms/}}. Since 64 ms resolution is inadequate for studying structure in short duration GRB pulses, we also fit some short GRB pulses for which 4 ms resolution data are available \citep{hak18a}. During its operation, BATSE observed 2702 GRBs in its role as a full-sky, high surface-area instrument spanning the 20 keV to 1 MeV energy range. Many of these GRBs also have 64 ms and/or sufficient TTE data available.

Several systematic biases are present in BATSE data that potentially affect the analysis of pulse structure: 1) BATSE's 64 ms-resolution data begins two seconds before a burst trigger: prior to the trigger, 4-channel counts exist as lower-resolution, 1024 ms binned data. 2) Compton Gamma-Ray Observatory (CGRO) data suffer from transmission gaps that occasionally show up in 64 ms data. 3) The gamma-ray background rate varies slowly with time during CGRO's eccentric orbit, and other emission sources are often present in BATSE's eight large area detectors. Since the gamma-ray sky contains many faint variable sources, the gamma-ray background detected by BATSE is not always best characterized by Poisson statistics. 4) The counts data used for triggering and for subsequent GRB characterization are not produced by a one-to-one matching between incoming and detected photons: high-energy photons (often very many of them) have downscattered into lower-energy channels. Our analysis attempts to account for as many of these limitations as is possible.

\section{Analysis}\label{sec:analysis}

\subsection{Procedure} \label{sec:procedure}

GRB pulses are extracted and residuals are fitted from BATSE data using the following systematic process. 

\paragraph{Identification of potential pulses.} Bursts with poor quality 64 ms data are eliminated from the analysis via a visual examination of summed four-channel BATSE light curves. The visual inspection is also used is used to estimate the number of emission episodes in each burst. The counts data are rebinned on several timescales longer than 64 ms to better ascertain the number of distinct emission episodes, and the BATSE 4B Catalog Comments file (\texttt{https://gammaray.nsstc.nasa.gov/batse/grb/catalog/4b/tables/4br\_grossc.comments}; \citep{pac99}) is reviewed to determine if there are any peculiarities associated with the burst that can help with this determination. Identifying the number of structured pulses in a burst is the most qualitative step in our analysis, given the aforementioned issues with non-constant background rates, low signal-to-noise ratios, complexity of potential GRB structures, and existence of faint time-reversed structures that have been found to extend outside of formal pulse durations. When it is unclear how many pulses might exist in a complex and/or noisy light curves, we apply Occam's Razor to assume the smallest number of pulses. If a pulse background contains evidence of an overlapping event ({\em e.g.,} solar flare, CYG X-1 event, particle event), then we exclude the burst from our sample. If the burst duration is so short that the number pulses and/or evidence of structure cannot be ascertained, then we exclude the burst from our sample. If the number of pulses in a burst cannot clearly be determined using the aforementioned criteria, then we set the burst aside as unfittable. However, since we wish to learn from the process, we allow ourselves the option of reevaluating our initial determination at a later time, if evidence supports it. 

\paragraph{Potential pulses that are too faint or too short to fit.} Once we have identified the number of pulses in a specific GRB, the background in the vicinity of each pulse is fitted. Goodness-of-fit is determined by applying $\chi^2$ tests to the 64ms data and 4ms binned TTE data. The number of bins in BATSE 64 ms data are found from time periods containing two different temporal resolutions: BATSE bursts have 1024 ms resolution ending two seconds prior to the trigger ({\em the resolution change time}), and 64 ms resolution following that time. At some later time (typically around 300-400 s after the trigger), the resolution reverts to 1024 ms. When fitted to a background model alone, an episode can be characterized by a $\chi^2$ test where the number of degrees of freedom equals the number of data bins in the pulse duration window minus two (the number of background fitting parameters). For 64ms data, the number of data bins is the number of 64 ms bins found after the resolution change time and before the end of the pulse duration plus the number of 1024 ms bins (full or partial) found between the beginning of the pulse and the resolution change time. For 4 ms TTE data, the number of bins is the number of bins in the duration window. A few GRBs are so faint that a background-only fit produces a reasonable $\chi^2$ value, as characterized by a p-value of $p > 0.05$. These episodes are excluded from our sample on the basis that they are too faint to characterize as pulsed emission.

\paragraph{Monotonic pulse fits.} In examining the remaining pulses, we assume that each pulse can be fitted with the time-asymmetric \cite{nor05} model. The Gaussian model is applied if the pulse is too time-symmetric to be easily fitted with the \cite{nor05} model, or if a pulse is so faint that its shape cannot be easily ascertained (the Gaussian model has one fewer degree of freedom than the \cite{nor05} model). 
The number of degrees of freedom for these tests is the number of data bins in the pulse duration window minus six for the \cite{nor05} model (four parameters plus two background parameters) or minus five for the Gaussian model (three parameters plus two background parameters), and where the number of bins are calculated as in the preceding paragraph. A fit that includes the pulse structure is not needed if the p-value obtained from this test is acceptable ($p \ge 0.05$). A small number of pulses are time-asymmetric in the wrong sense (slow rise, fast decay): we do not attempt to fit these if neither a \cite{nor05} model nor a Gaussian model suffices. We set these pulses aside as unfittable. 

\begin{figure}
\epsscale{0.50}
\plotone{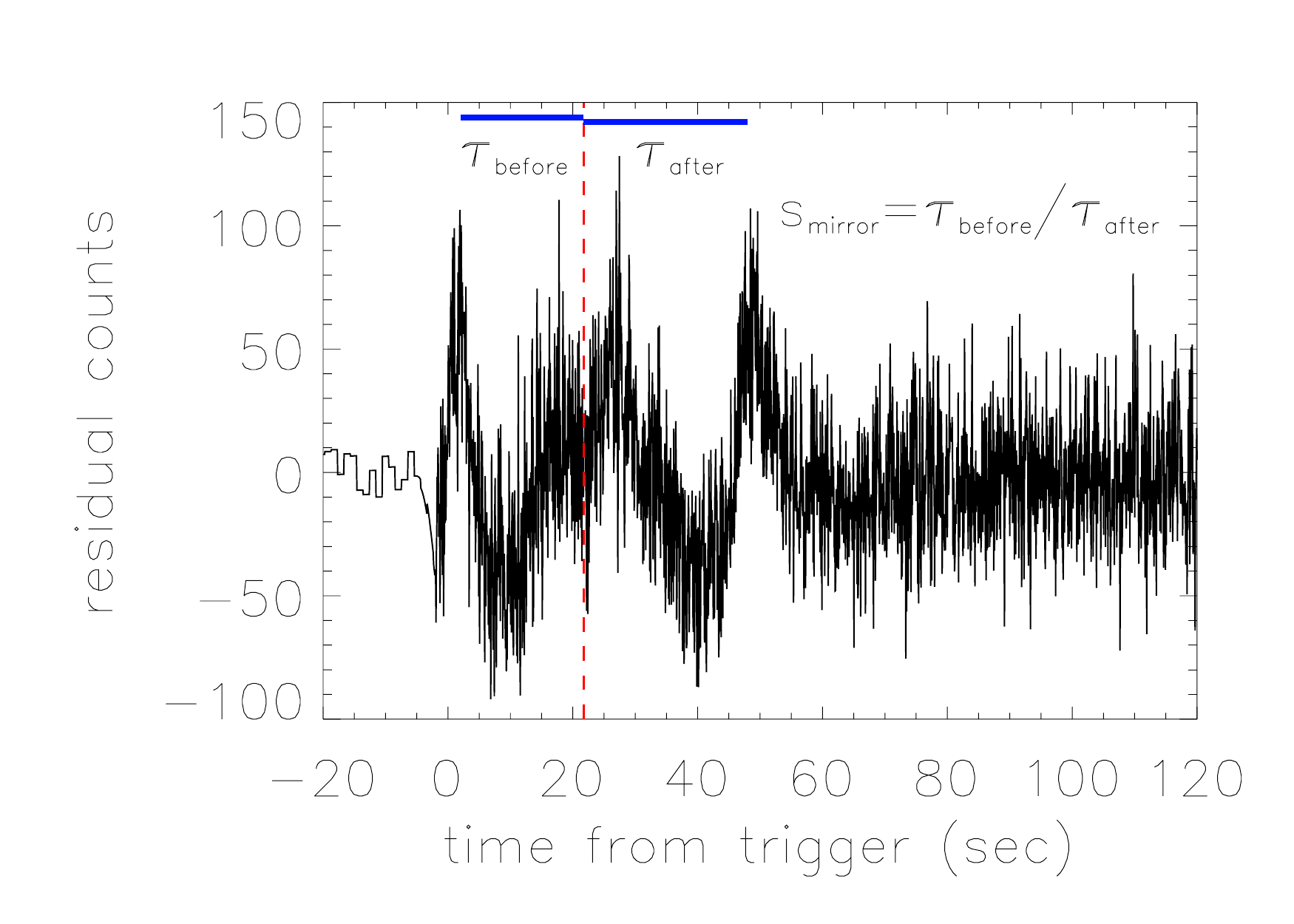}
\caption{Residuals of BATSE pulse 659. The residuals are time-reversed around the time of reflection $t_{\rm 0; mirror}$, which is indicated by a vertical dashed red line. The amount of stretching $s_{\rm mirror}$ is obtained from the ratio of the temporal scale of events prior to $t_{\rm 0; mirror}$ relative to the temporal scale of events following $t_{\rm 0; mirror}$. For BATSE pulse 659, $t_{\rm 0; mirror}=21.75 s$ and $s_{\rm mirror}=0.71$. \label{fig:fig1}}\end{figure}

\begin{figure}[htb]
\centering
  \begin{tabular}{@{}cc@{}}
    \includegraphics[width=.25\textwidth]{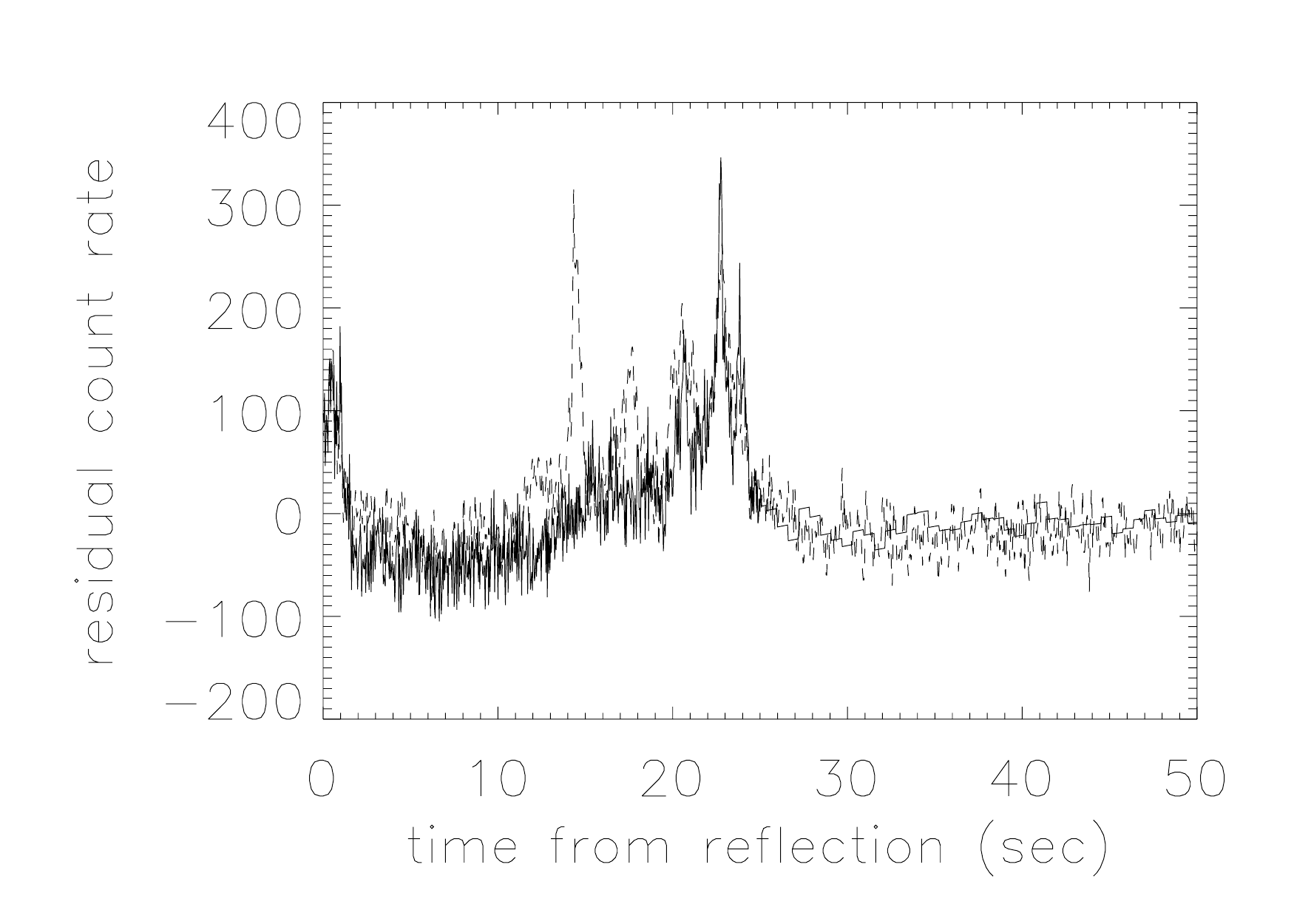} & 
    \includegraphics[width=.25\textwidth]{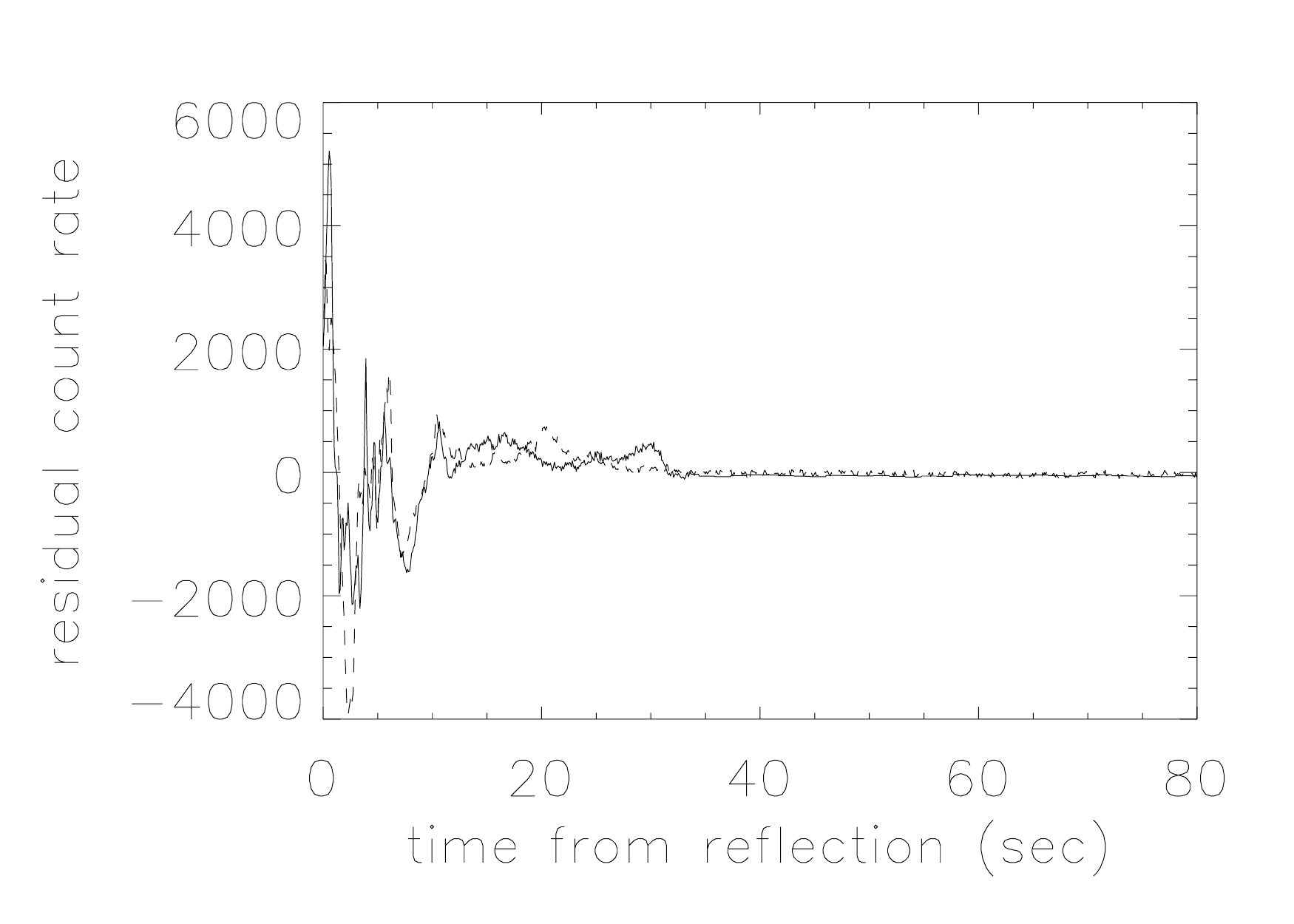} \\
    \includegraphics[width=.25\textwidth]{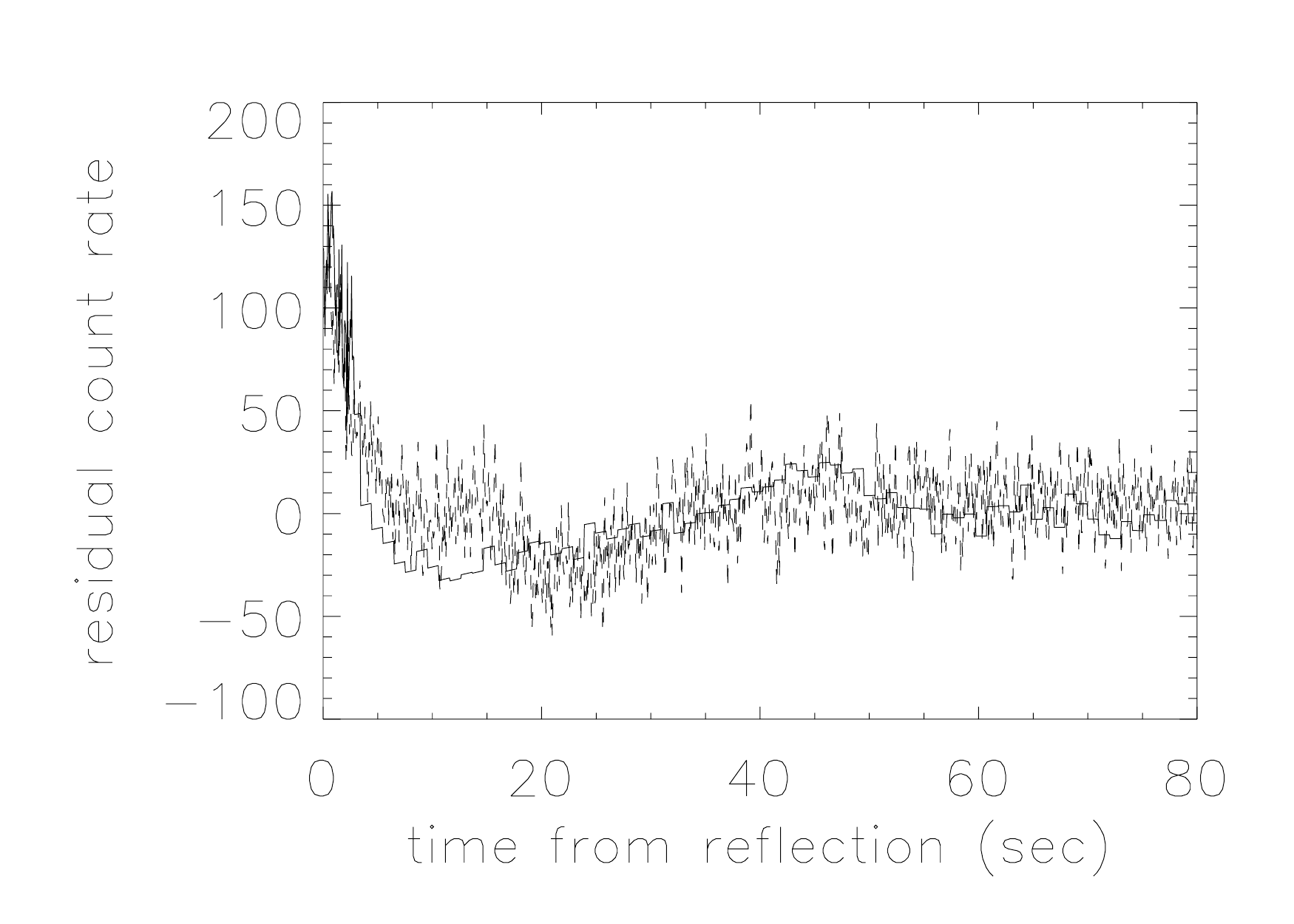} & 
    \includegraphics[width=.25\textwidth]{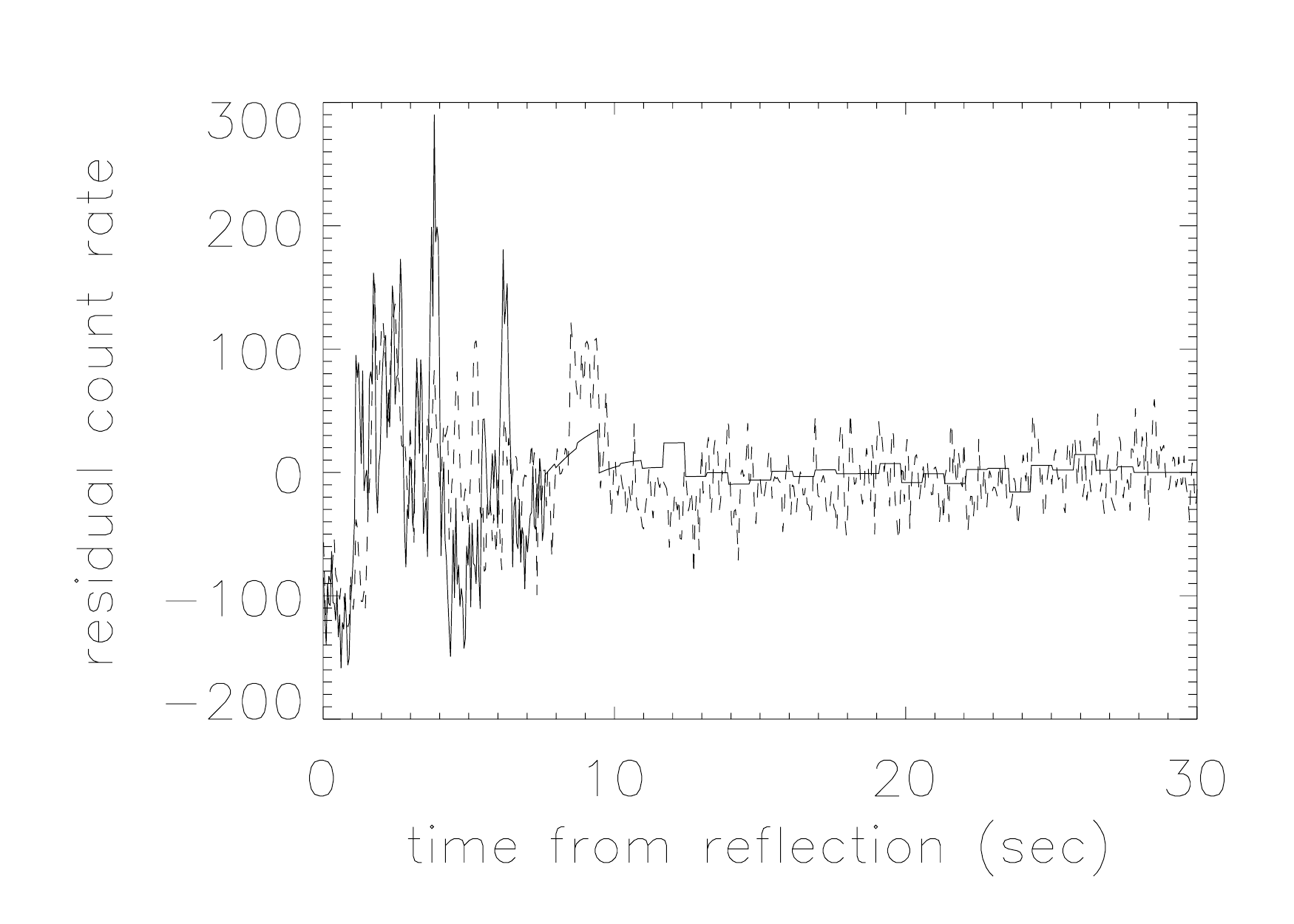} \\
    \includegraphics[width=.25\textwidth]{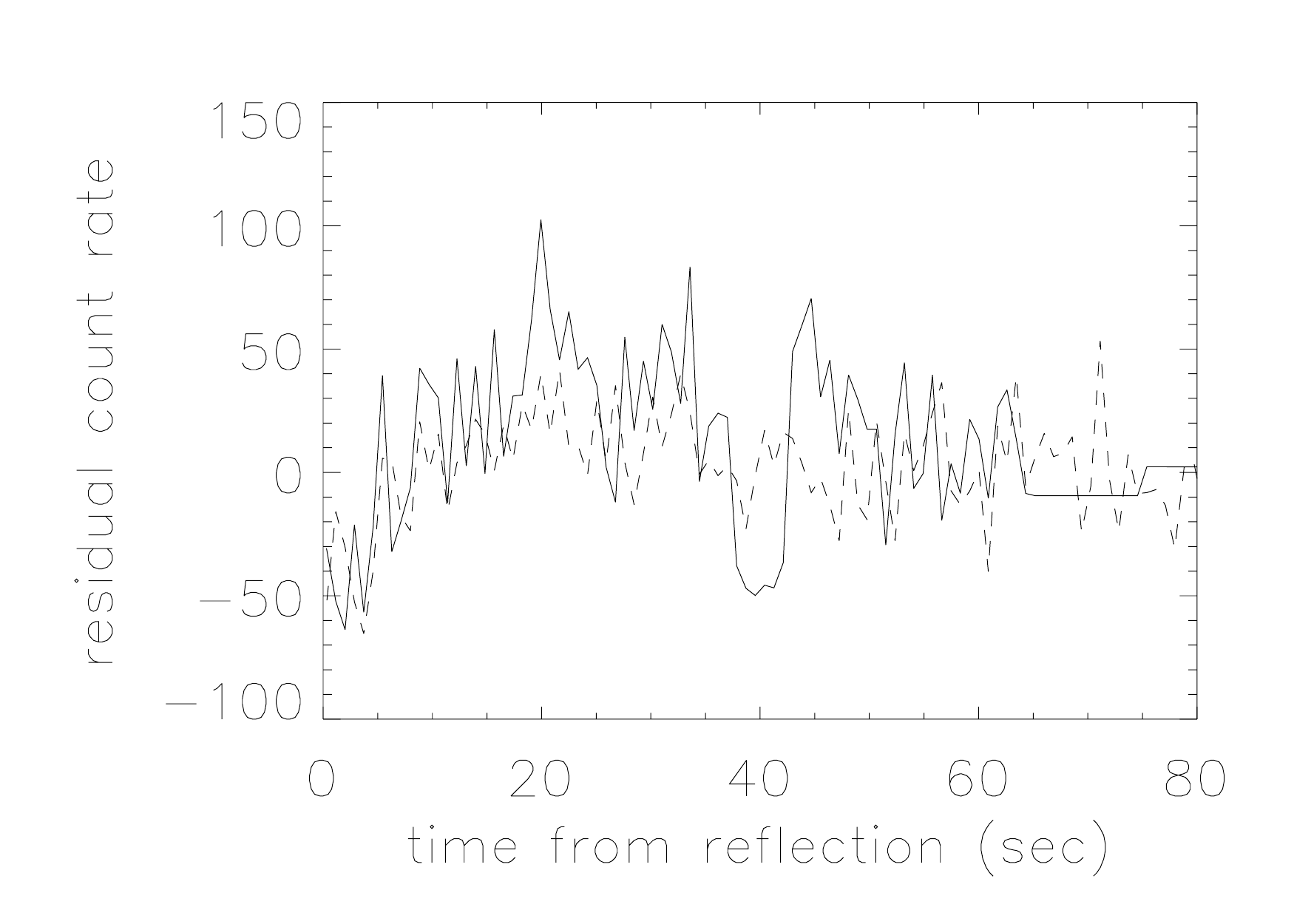}  & 
    \includegraphics[width=.25\textwidth]{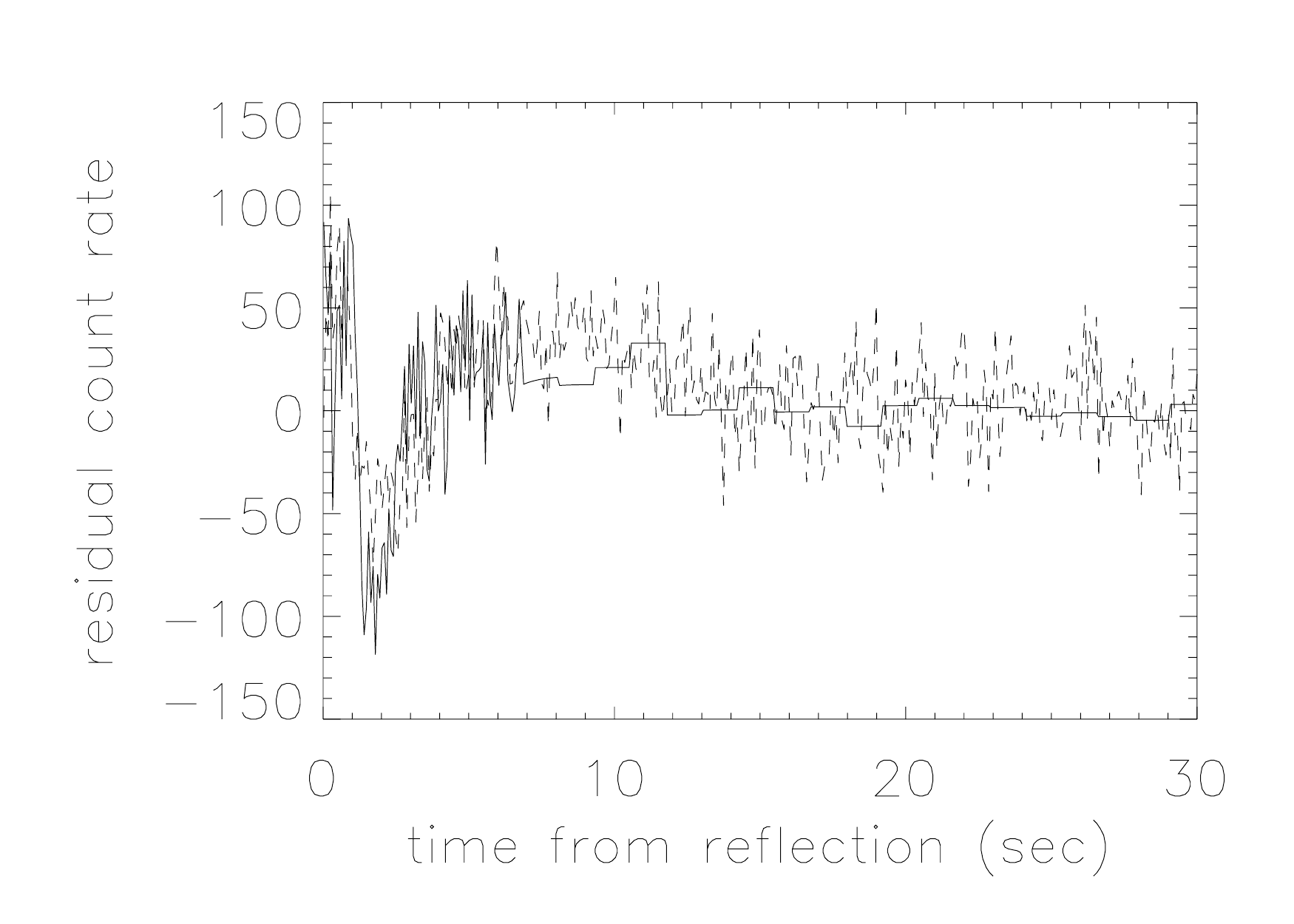}   \\
  \end{tabular}
  \caption{Folded and matched residuals of BATSE pulses 121 (upper left), 249 (upper right), 1046 (center left), 1303 (center right), 1580 (upper left), and 1717 (upper right),. The solid line represents time-reversed rates from prior to the time of reflection, while the dashed line represents rates following the time of reflection. Temporal units refer to time prior to the time of reflection. (lower right). \label{fig:fig2}}
\end{figure}

\paragraph{Characterizing temporally symmetric pulse structure.}\label{sect:CCF}
Once a pulse's monotonic component has been fitted, the residuals are obtained by subtracting this component from the data. Only two parameters are needed to characterize the pulse residuals' temporal symmetry \citep{hak18b}: the time of reflection $t_{\rm 0; mirror}$ at which the residuals are folded and the stretching parameter ($s_{\rm mirror}$) used to align the time-forward and time-reversed pieces. The amount by which the time-reversed residuals are stretched is given by
\begin{equation}
s_{\rm mirror}=\tau_{\rm before}/\tau_{\rm after},
\end{equation}
where $\tau_{\rm before}$ is the temporal scale of structures prior to $t_{\rm 0; mirror}$ and $\tau_{\rm after}$ is the temporal scale of time-reversed structures following $t_{\rm 0; mirror}$, as determined from the maximum value of the CCF (cross-correlation function) obtained. Note that $s_{\rm mirror} \approx 0$ when $\tau_{\rm before} \ll \tau_{\rm after}$ (asymmetric residuals) and that $s_{\rm mirror} = 1$ when $\tau_{\rm before} = \tau_{\rm after}$ (symmetric residuals). 
Figure \ref{fig:fig1} demonstrates how $t_{\rm 0; mirror}$ and $s_{\rm mirror}$ are calculated for the residuals of BATSE pulse 659.
 Initial estimates of $t_{\rm 0; mirror}$ and $s_{\rm mirror}$ are obtained from $t_0$ and $s$ which are found from fitting the residuals with the \cite{hak14} model (for a more detailed description of this process, see \cite{hak14} and \cite{hak15}). The cross-correlation function (CCF) is then applied iteratively by folding the time- forward part of the residual wave at various values of $t_{\rm 0; mirror}$ and stretching it by $s_{\rm mirror}$ until it most closely aligns with the time-reversed part of the wave (see \cite{hak14,hak18b,hak19}). Examples of time-reversed and stretched residual fits are shown in Figure \ref{fig:fig2}.

\paragraph{Temporally-symmetric pulse fits.} \label{sect:conmodel} A temporally-symmetric pulse model is constructed in the following manner. The residuals from $t < t_{\rm 0; mirror}$ are temporally reversed, stretched by $1/s_{\rm mirror}$, and used to model the data after $t = t_{\rm 0; mirror}$. The residuals from $t > t_{\rm 0; mirror}$ are temporally reversed, compressed by $s_{\rm mirror}$, and used to model the data before $t = t_{\rm 0; mirror}$. Linear interpolation is used on the folded part of the light curve during this process, since the temporal binning prior to the time at which the residuals are being reflected and stretched differs from the binning after that time. This process assumes that count rates prior to $t_{\rm 0; mirror}$ can be matched to time-inverted rates following $t_{\rm 0; mirror}$ without adjustment; {\em there is no attempt to otherwise normalize pre-mirrored counts with post-mirrored counts}. The temporally-symmetric model is constructed by adding the background model, the pulse model, and the time-reversed residual model. 

\begin{figure}[htb]
\centering
  \begin{tabular}{@{}ccc@{}}
    \includegraphics[width=.25\textwidth]{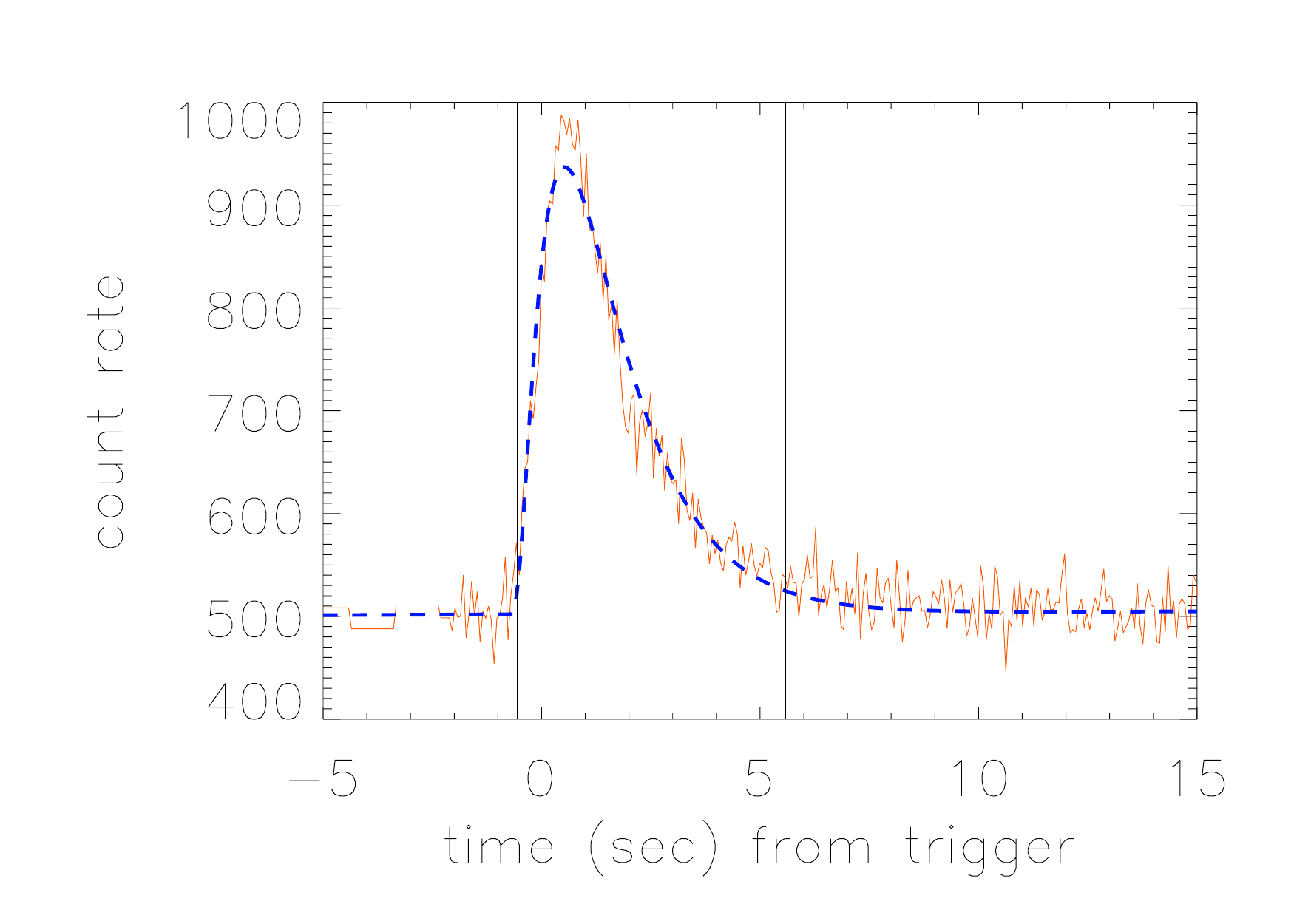} & 
    \includegraphics[width=.25\textwidth]{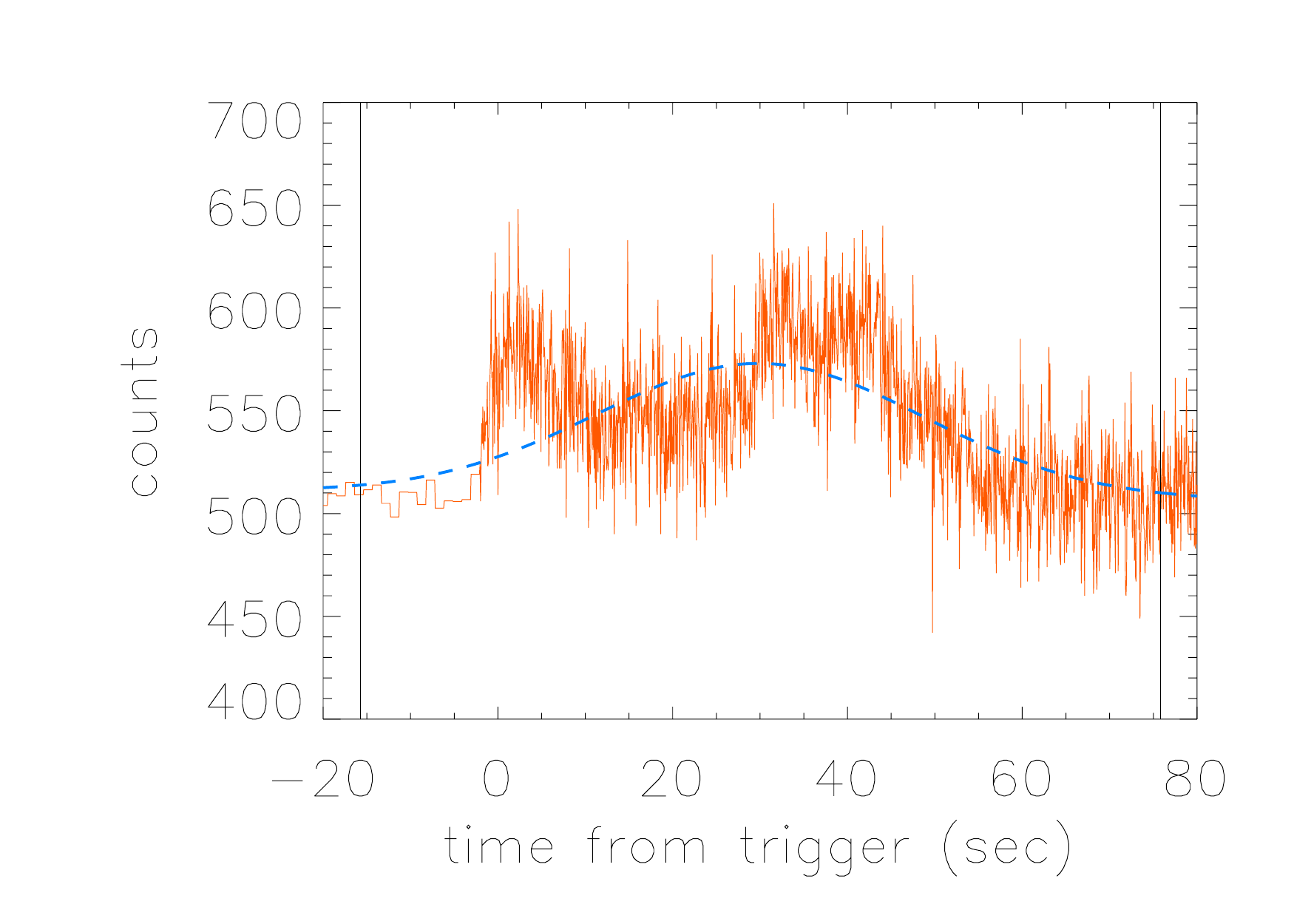} &
    \includegraphics[width=.25\textwidth]{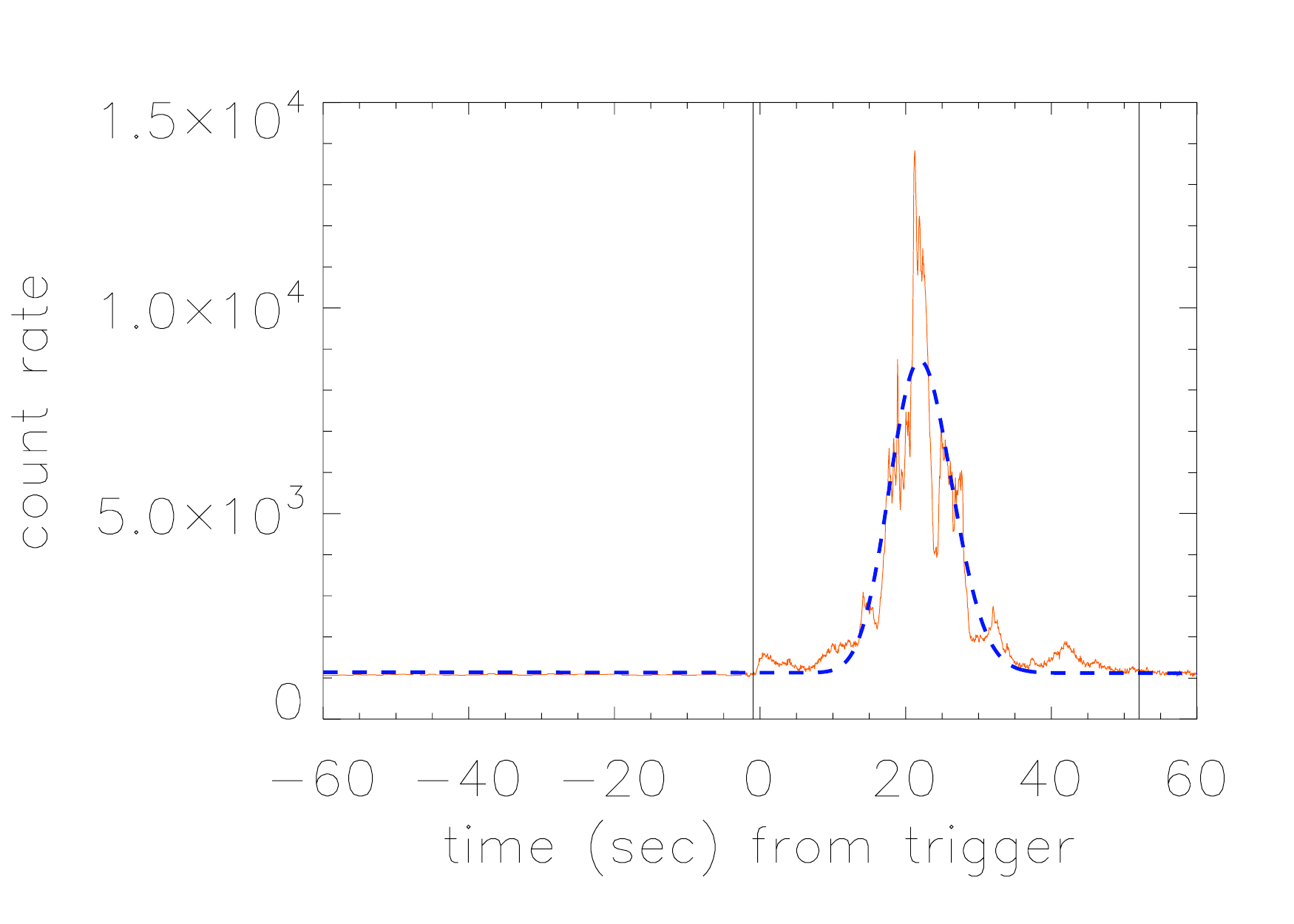} \\
  \end{tabular}
  \caption{Duration windows (dotted vertical lines) for asymmetric BATSE pulse 914 (left panel) and symmetric pulse 1459 (central panel), and structured BATSE pulse 249 (right panel), which has residual structure extending beyond the boundaries of the monotonic pulse. \label{fig:fig3}}
\end{figure}

\paragraph{Pulse durations.} Fits for most pulses are obtained over the pulse duration windows defined by $w$ in Equation \ref{eqn:duration} and Equation \ref{eqn:normdur} (based on the pulse model used). These duration definitions tend to minimize the amount of background included in the fit when pulses are bright, but include larger background intervals (with correspondingly improved goodness-of-fit measures) when pulses are faint. Examples of typical pulse durations windows obtained in this manner are shown in the left and center panels of Figure \ref{fig:fig3}. 
Roughly 15\% of fitted pulses have observed emission that extends beyond $w$; these pulses exhibit residuals that begin before the pulse start-times and/or finish after the pulse end-times. The start- and end-times of these pulses are thus determined by eye to account for extended time-reversed residuals that appear to be associated with the monotonic pulse structure:
\begin{equation}
{\rm duration}=t_{\rm end}-t_{\rm start}.
\end{equation}
An example of one of these extended duration windows is shown in the right panel of Figure \ref{fig:fig3}.

\paragraph{Successfully-fitted temporally symmetric residual structures.} Uncertainties in $t_{\rm 0; mirror}$ and $s_{\rm mirror}$ are obtained through the process of Monte Carlo bootstrapping ({\em e.g.,} \cite{and10}). Random values of $t_{\rm 0; mirror}$ and $s_{\rm mirror}$ are generated for each pulse by adding Poisson variations to the light curve. The Monte Carlo-generated distributions of $t_{\rm 0; mirror}$ and $s_{\rm mirror}$ are used to obtain $\sigma_{\rm t0; mirror}$ and $\sigma_{\rm s~mirror}$. We find that uncertainties larger than $\sigma_{\rm s~mirror} \approx 0.4$ indicate that the $s_{\rm mirror}$ measurement is unreliable. We therefore consider that a pulse has been successfully fitted when the uncertainty in measuring $s_{\rm mirror}$ is $\sigma_{\rm s~mirror} < 0.4$.

\begin{figure}[htb]
\centering
  \begin{tabular}{@{}cccc@{}}
    \includegraphics[width=.33\textwidth]{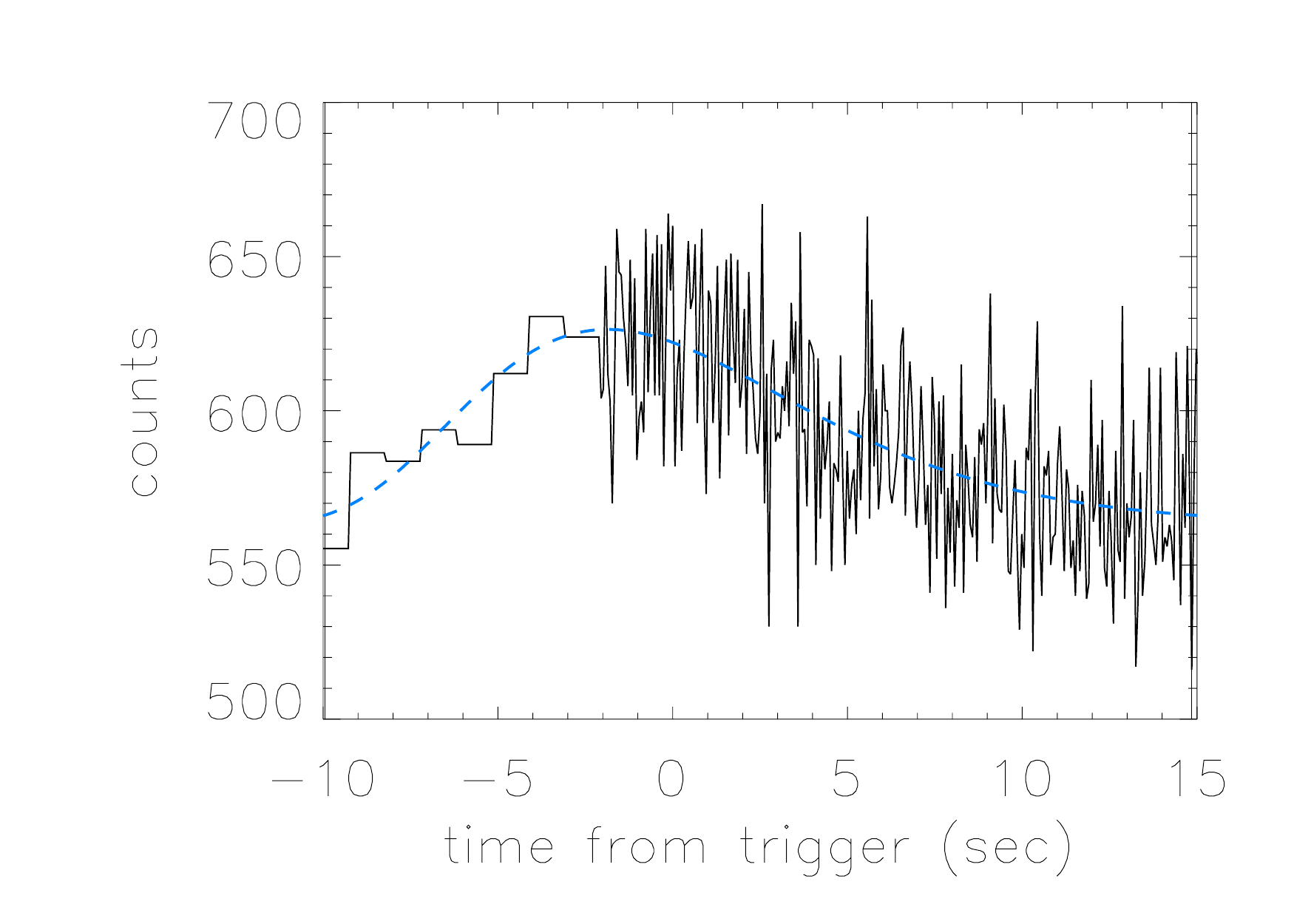} & 
    \includegraphics[width=.33\textwidth]{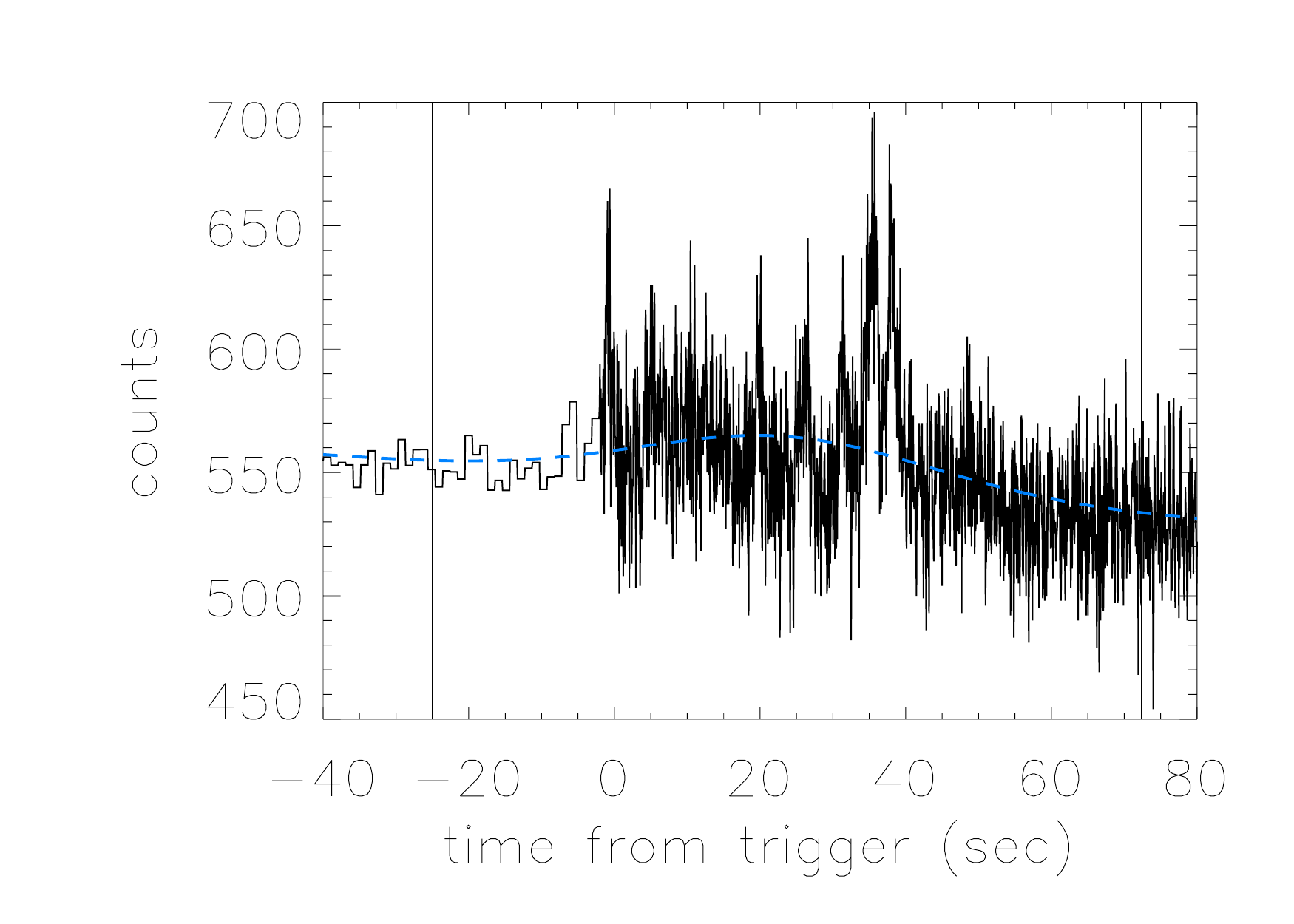} \\
    \includegraphics[width=.33\textwidth]{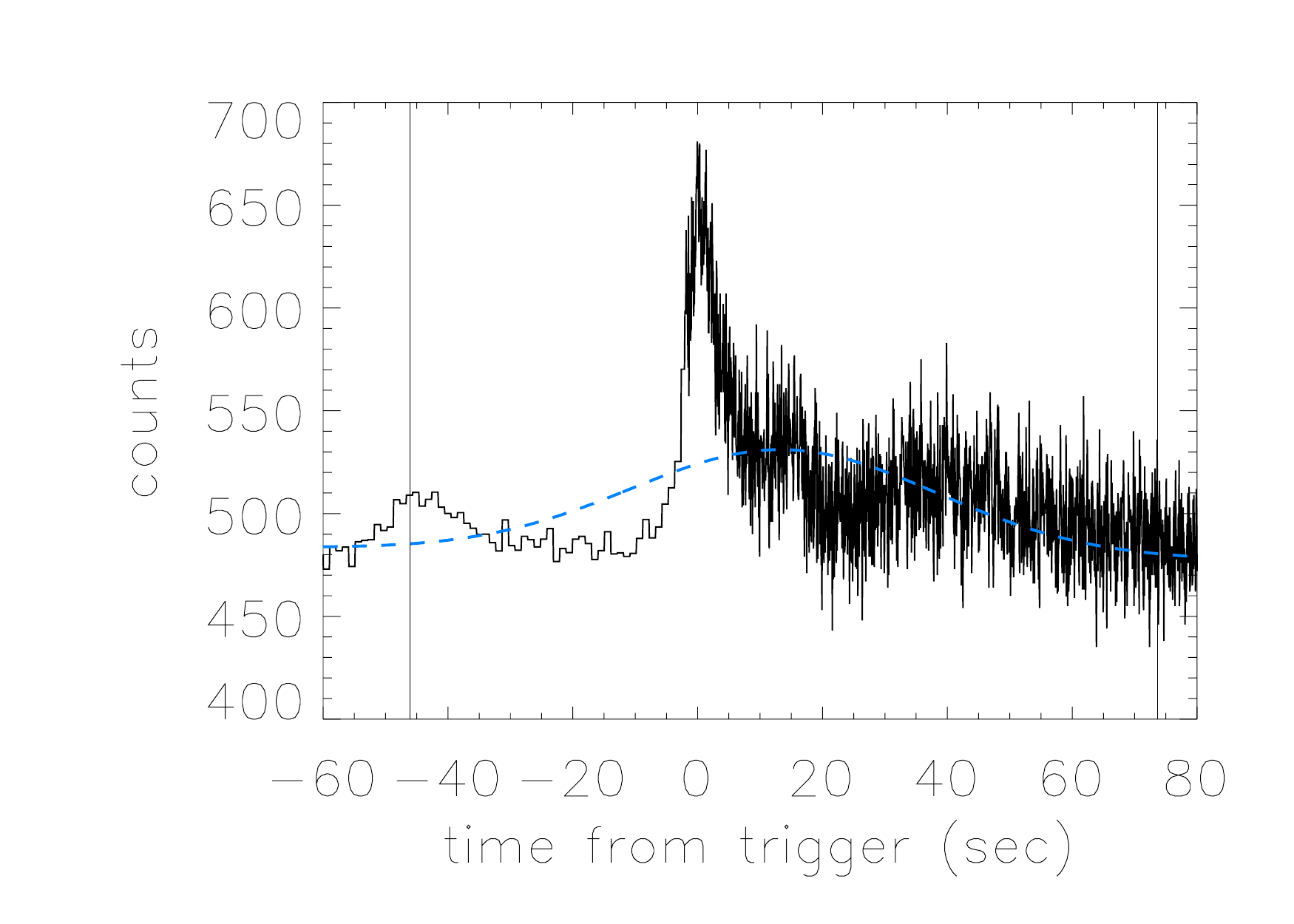} &
    \includegraphics[width=.33\textwidth]{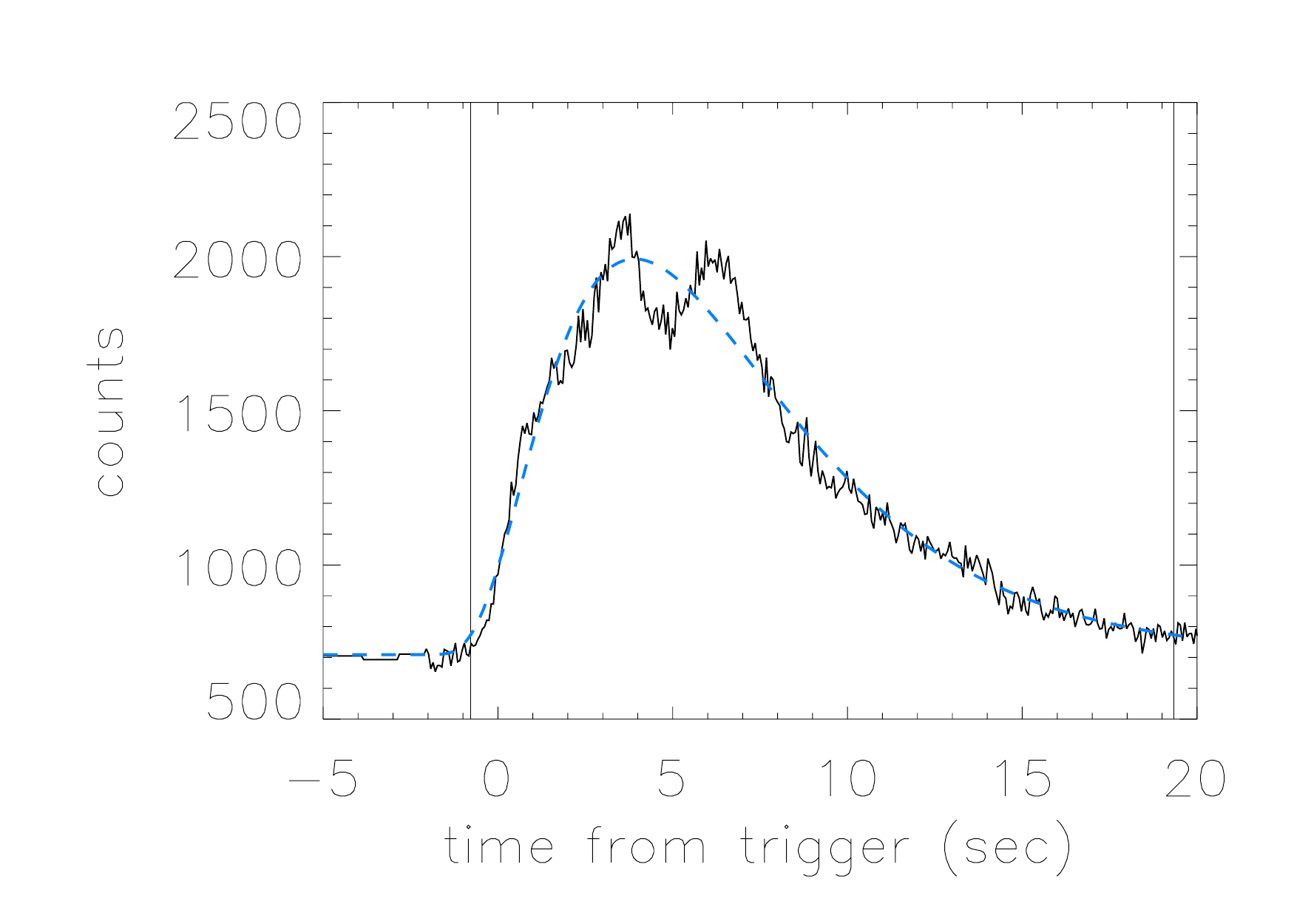} \\ 
  \end{tabular}
  \caption{BATSE pulses with a range of residual significances. BATSE 630 (upper left) has $R=1.09$, BATSE 680 (upper right) has $R=1.26$, BATSE 1046 (lower left) has $R=1.53$, and BATSE 829 (lower right) has $R=3.05$. In each plot, the fitted Norris or Gaussian pulse is indicated by a blue dashed line, and the duration window is identified by solid vertical lines. \label{fig:fig4}}
\end{figure}

\paragraph{Unsuccessfully-fitted temporally symmetric residual structures.} 
Temporally symmetric residual structure is absent when a) residual structures are bright enough to measure, and b) $s_{\rm mirror}$ and $t_{\rm 0; mirror}$ measurements are uncertain. We introduce the {\em residual statistic R} to characterize residual structure brightness, where
\begin{equation}\label{eqn:R}
R = \sqrt{\frac{\sigma_{\rm res}}{\langle B \rangle}}.
\end{equation}
Here, $\langle B \rangle$ is the average number of background counts per bin over the duration of the pulse. $R$ does not generally compare two random stochastic measures: the standard deviation in the residuals $\sigma_{\rm res}$ primarily represents a mean systematic deviation of the pulse fit that is often characterized by wavy features in isolated FRED pulses. $R$ can exceed unity significantly for bright pulses. Still, $R$ can be considered to be a S/N ratio for pulse residuals. Examples of typical $R$ values for BATSE pulses are demonstrated in Figure \ref{fig:fig4}.
From our bootstrapping analysis, we find that $s_{\rm mirror}$ is often unconstrained for faint pulses ($R \le 2$), whereas pulses with bright residuals ($R > 2$) often have $s_{\rm mirror}$ values that can be accurately measured. If pulse residuals are bright and $\sigma_{\rm s~mirror} \ge 0.4$, then the residuals can be measured, but we consider them to be inconsistent with the temporally-symmetric pulse model.

\begin{figure}[htb]
\centering
  \begin{tabular}{@{}ccc@{}}
    \includegraphics[width=.25\textwidth]{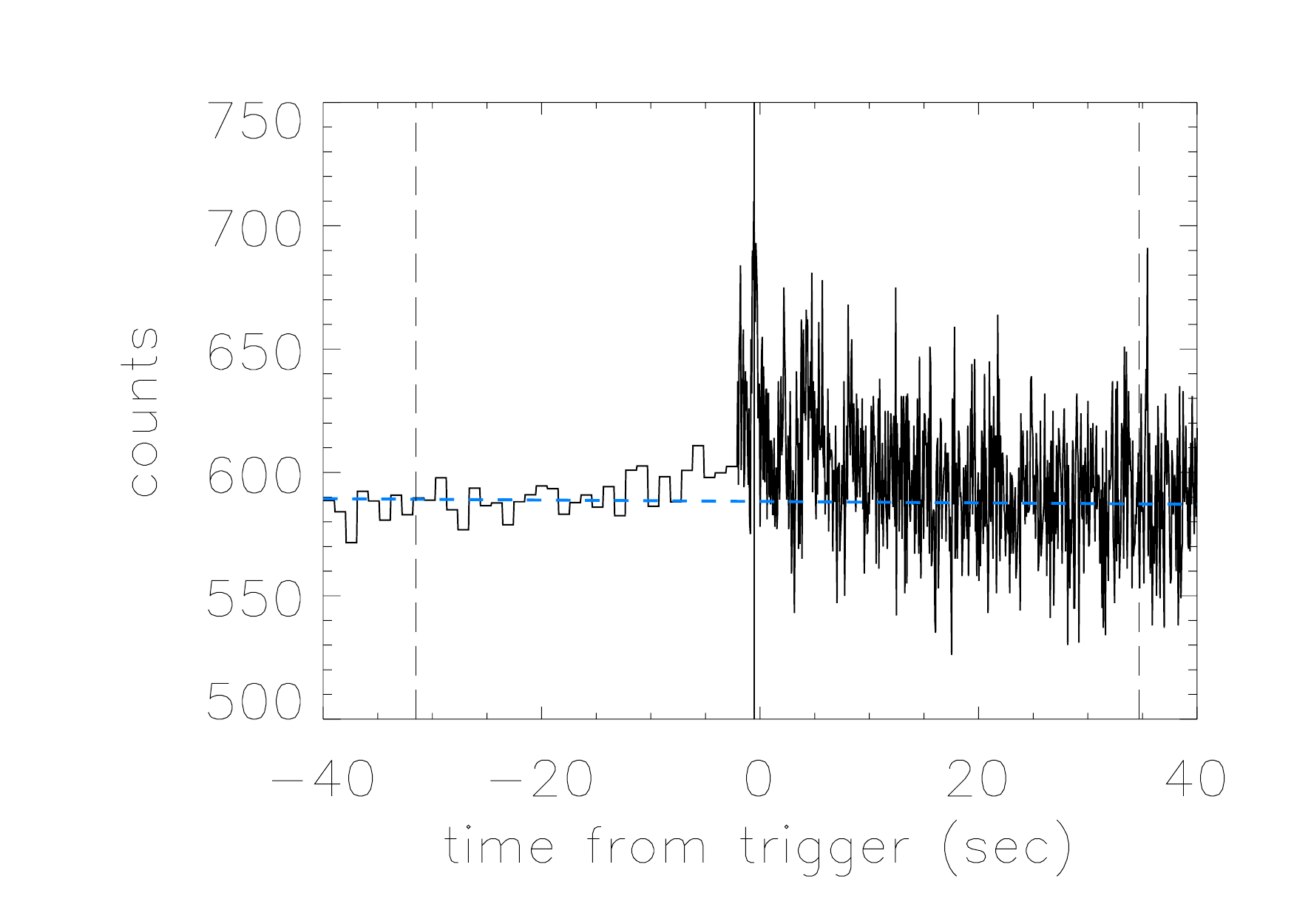} & 
    \includegraphics[width=.25\textwidth]{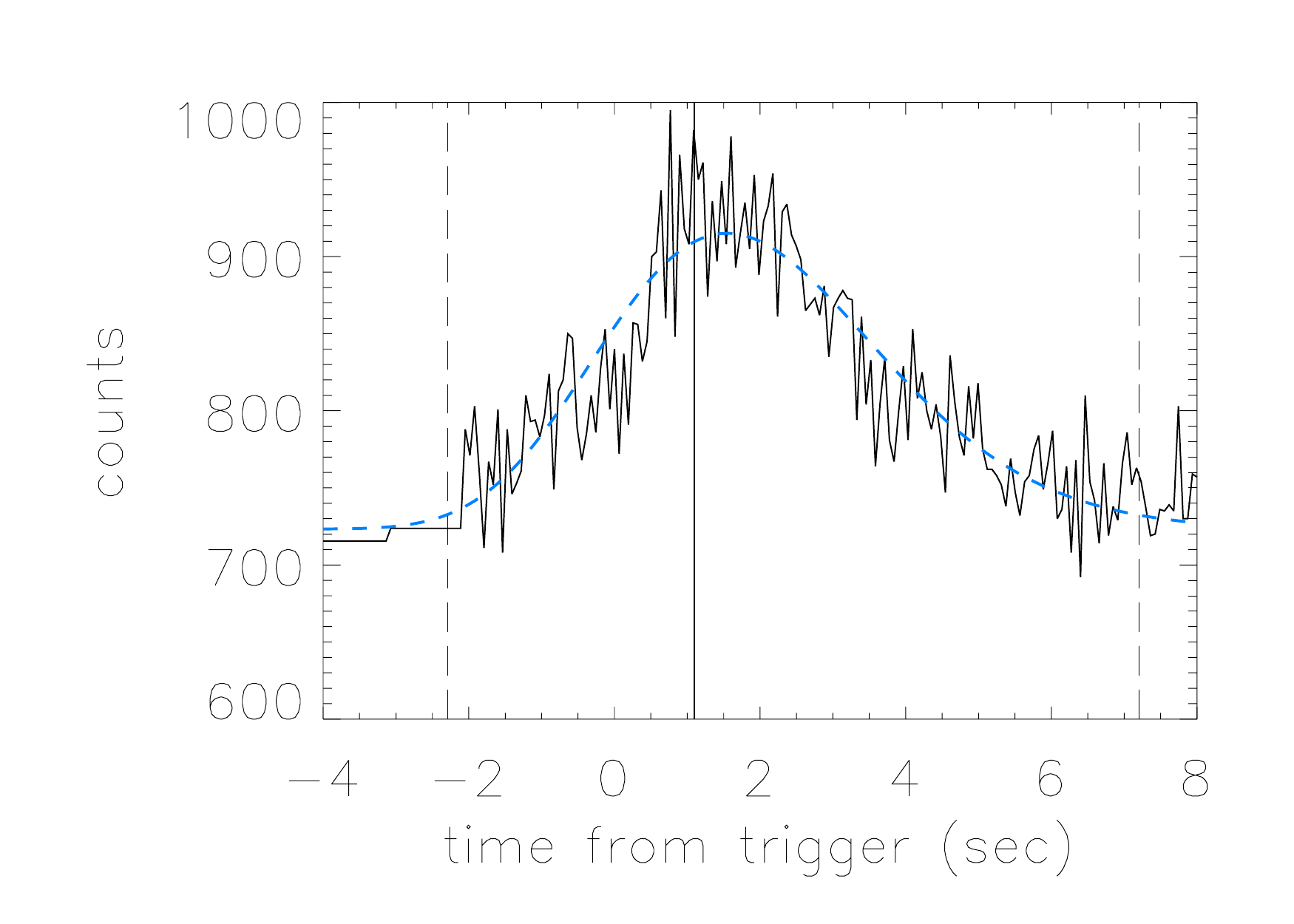} &
    \includegraphics[width=.25\textwidth]{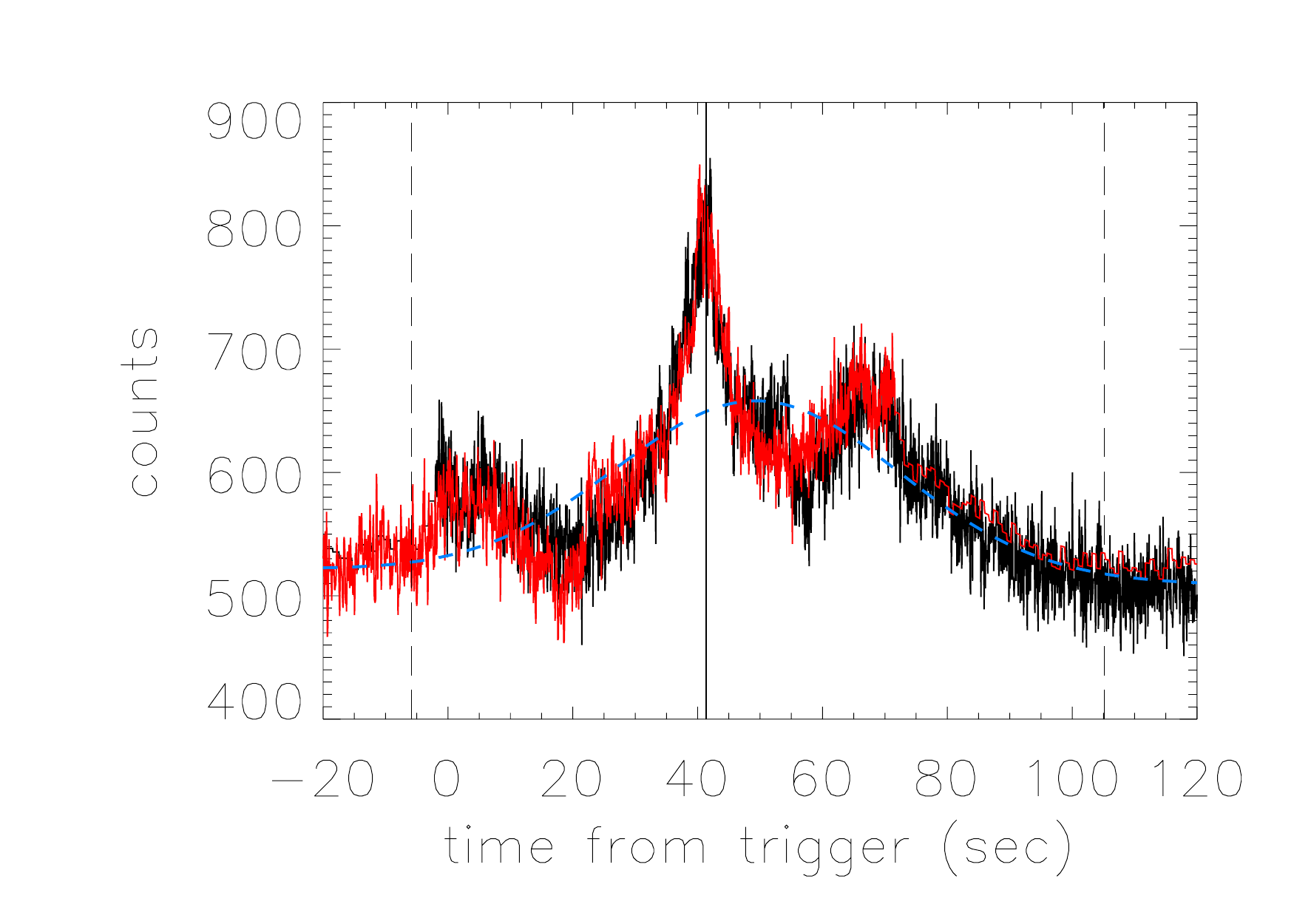} \\
  \end{tabular}
  \caption{Quality factors for various BATSE pulses. Shown are BATSE pulse 717 (left panel) having $q=-1$, BATSE pulse 680 (right panel) having $q=1$, and BATSE pulse 351 (right panel) having $q=2$. BATSE 717 is adequately fitted by background alone (dotted blue line), so neither a pulse fit nor a residual structure fit is needed. BATSE 680 requires an asymmetric monotonic pulse (dotted blue line), so an additional residual structure fit is not needed. BATSE 351 is not adequately fitted by a monotonic pulse model alone (dotted blue line), and requires the additional presence of temporally-reversed structure (solid red line). No example is provided of a pulse having $q=0$, since these pulses cannot be fitted by the models. \label{fig:fig5}}
\end{figure}

\paragraph{Pulse fit quality.} The quality $q$ of a pulse is identified using the aforementioned process. A quality of $q=-1$ indicates that the trigger cannot be fitted by a monotonic pulse model because it is too faint, too short, or has an unusable background. Typically only background can be fitted to $q=-1$ pulses. A quality of $q=1$ indicates that a good monotonic fit can be found to the pulse, or that residuals are insufficiently bright to determine if temporally symmetric pulse structure is present. A quality of $q=2$ indicates that temporally-symmetric residual structure is found to be associated with a fitted pulse. A quality of $q=0$ indicates that the data are inconsistent with the model: either a fit cannot be obtained to the monotonic pulse component, or the residuals are inconsistent with the temporally-symmetric model. These quality descriptions are summarized in Table 1. Examples of quality factors $q=-1$, $q=1$, and $q=2$ are shown in Figure \ref{fig:fig5}.

\paragraph{Alignment between monotonic pulse and temporally-symmetric residual structure.} The temporally-symmetric residual structure generally aligns closely with the monotonic pulse, but there are exceptions. One reason that the $s_{\rm mirror}$ values are so large is that $t_{\rm 0; mirror}$ occurs during the pulse decay and after the monotonic pulse peak time $\tau_{\rm peak}$. We  define the unitless {\em offset} as the delay between the time of reflection and the monotonic pulse peak time, normalized to the pulse duration window, as:
\begin{equation}
{\rm offset}=\frac{\tau_{\rm peak}-t_{\rm 0; mirror}}{t_{\rm end}-t_{\rm start}}.
\end{equation}


\paragraph{Effects of signal-to-noise ratio on pulse analysis.} Faint GRB pulses generally do not exhibit measurable residuals. This appears to be due to the washing-out of these features at low signal-to-noise ratios ({\em S/N}). Appendix A demonstrates and discusses this effect.

\subsection{BATSE Pulse Fits} \label{sec:results}
The aforementioned process has been applied to a sequentially-detected BATSE GRB sample, containing BATSE Triggers 105 to 2068 plus eight bright pulses from later bursts (and discussed in previous papers) for which 64 ms data are available. Approximately 1/4 of the bursts in the sample cannot be fitted due to missing data, nonlinear background rates, and active sources during the trigger. Most of the short GRBs have also been excluded because 64 ms binning has washed out pulse structure and has thus made its characterization unreliable. However, we have augmented the dataset by including some short duration GRBs (described in \cite{hak18a} for which 4 ms TTE data are available. In all, 413 BATSE pulses/GRBs have been characterized by this systematic process. 


\begin{deluxetable*}{lccc}
\tablenum{1}
\tablecaption{Results of Temporal Symmetry Pulse Fitting \label{tab:tab1}}
\tablewidth{0pt}
\tablehead{
\colhead{Fit result}  &  \colhead{quality} &  \colhead{definition} 
&  \colhead{number of events} }
\startdata
Not attempted (poor quality) & -1 &  incomplete data, bad background, OR & 101 \\
& & inadequate temporal resolution \\
Not temporally symmetric  & 0 & Cannot fit OR ($p_{\rm pulse} < 0.05$& 25 \\
& & AND $R \ge 2.0$ AND $\sigma_{\rm s~mirror} \ge 0.4$) \\
Monotonic (indeterminate) & 1 & $p_{\rm pulse} \ge 0.05$ OR & 126 \\
& & ($p_{\rm pulse} < 0.05$ AND $R < 2.0$) \\
Temporally symmetric & 2 & $p_{\rm pulse} < 0.05$ AND $\sigma_{\rm s~mirror} < 0.4$ & 161\\
\enddata
\end{deluxetable*}

The temporal symmetry properties of each GRB pulse in the sample having $q \ge 0$, along with associated pulse morphology characteristics, are summarized in Table \ref{tab:tab1}. Additional pulse properties can be found in Table \ref{tab:tab2}, which contains monotonic pulse fitting parameters, as well as the time of reflection for the residuals (when available). Tables \ref{tab:tab1} and \ref{tab:tab2} are only partially presented here due to their large sizes; the complete machine-readable tables are available from the publisher. 

\movetabledown=5.cm
\begin{rotatetable}
\begin{deluxetable}{lcccccccccccccc}
\tablenum{2}
\tablecaption{BATSE GRB Time-Reversed Pulse and Morphology Catalog \label{tab:tab2}}
\tablewidth{0pt}
\tablehead{
\colhead{Pulse}  &  \colhead{res (s)}   &  \colhead{quality} &  \colhead{morphology} &  \colhead{S/N}   &  \colhead{R}  &  \colhead{CCF ratio}  &  \colhead{$\kappa$} &  \colhead{$\kappa$ error}  &  \colhead{$s_{\rm mirror}$}  &  \colhead{$s_{\rm mirror}$ error}  &  \colhead{offset} & \colhead{offset error} &  \colhead{start (s)} & \colhead{end (s)} \
}
\startdata
575p1 & 0.004 & 2 & FRED & 5.1 & 1.38 & 0.523 & 0.304 &1.000 & 0.46 & 0.33 & 0.01 & 0.67 & 0.12 & 0.33\\
575p2 & 0.004 & 1 & FRED & 6.1 & 1.47 & 0.348 & 0.728 & 0.044 & & & & & 0.39 & 0.78\\
577 & 0.064 & 1 & FRED & 6.9 & 1.07 & 0.128 & 0.683 & 0.038 &  &  &  & & -2.60 & 34.27\\
591 & 0.064 & 2 & u-pulse & 8.2 & 1.60 & 0.881 & 0.000 & 0.000 & 1.01 & 0.30 & -0.03 & 0.02 & -32.87 & 72.86\\
593 & 0.064 & 2 & no pulse u-pulse & 6.7 & 1.36 & 0.993 & -99 & -99 & 0.97 & 0.34 & -99 &-99 & -106.40 & 50.40\\
594 & 0.064 & 0 & u-pulse & 10.4 & 2.39 & 0.713 & 0.000 & 0.000 & & & & & -11.74 & 33.68\\
603 & 0.064 & -1 &  &  &  &  &  &  &  &  &  &  &  & \\
606 & 0.064 & 1 & rollercoaster & 9.4 & 1.44 & 0.284 & 0.779 & 0.025 & &  & & & -1.36 & 26.50\\
\enddata
\tablecomments{Table \ref{tab:tab2} is published in its entirety in the machine-readable format.
      A portion is shown here for guidance regarding its form and content. Each pulse is identified by its number (Column 1) and its temporal resolution (Column 2). The resolution is 4 ms for TTE pulses in the sample, and 64 ms otherwise. The pulse quality (2=temporally symmetric, 1=monotonic, 0=not temporally symmetric, and -1=not attempted) is listed in Column 3. The pulse morphology (see Section \ref{sec:morphology} lists the associated pulse type (Column 4). Parameters describing pulse and structural brightness are $S/N$ (Column 5, see Equation \ref{eqn:SN}), $R$ (Column 6, see Equation \ref{eqn:R}), and CCF ratio (Column 7, see Equation \ref{eqn:CCF}). The monotonic pulse asymmetry $\kappa$ is given in Column 8, and its error in Column 9, the stretching parameter $s_{\rm mirror}$ is provided in Column 10 along with its error in Column 11, and the offset is given in Column 12 along with its error in Column 13. The start and end times of the pulse are provided in Column 14 and Column 15, respectively. Note that values of $-99$ have been assigned to the $\kappa$ values, $\kappa$ errors), offset values, and offset errors of no-pulse u-pulses, for which these properties could not be measured.}
\end{deluxetable}
\end{rotatetable}

\movetabledown=6.25cm
\begin{rotatetable}
\begin{deluxetable}{lcccccccccccccccccccc}
\tablenum{3}
\tablecaption{Additional BATSE GRB Pulse Properties \label{tab:tab3}}
\tablewidth{0pt}
\tablehead{
\colhead{Pulse}  &  \colhead{res (s)}   &  \colhead{$B_0$} &  \colhead{$B_0$ err} &  \colhead{BS}   &  \colhead{BS err}  &  \colhead{A}  &  \colhead{A err}  &  \colhead{$t_s$}   &  \colhead{$t_s$ err} &  \colhead{$\tau_1$} & \colhead{$\tau_1$ err} & \colhead{$\tau_2$} & \colhead{$\tau_2$ err} &  \colhead{$\tau_{\rm peak}$} & \colhead{$\tau_{\rm peak}$ err}  & \colhead{$t_0$}   &  \colhead{$t_0$ err} & \colhead{$\sigma$}  & \colhead{$\sigma$ err} &  \colhead{$t_{\rm 0; mirror}$} \\
}
\startdata
107 & 0.064 & 546.2 & 0.4 & 0.02 & 0.00 & 37.0 & 0.9 & & & & & & & & & -149.67 &  1.19 & 42.81 & 1.30 & -158.69 \\
109 & 0.064 & 640.9 & 1.0 & 0.09 & 0.00 & 527.7 & 3.9 & -1.35 & 0.19 & 4.94 & 0.35 & 40.67 & 0.63 & 12.83 & 0.54 & & & & & 49.60 \\
110p1 & 0.064 & 518.0 & 1.6 & -0.27 & 0.01 & 24.2 & 1.8 & & & & & & & & &  -251.00 & 0.95 & 11.31 & 1.07 & -264.87 \\
110p2 & 0.064 & 530.9 & 0.6 & -0.2 & 0.01 & 73.0 & 3.3 & & & & & & & & &  -0.64 & 0.16 & 2.96 & 0.17 & -4.54 \\
111 & 0.064 & 541.4 & 0.4 & 0.00 & 0.00 & 86.4 & 1.4 & -16.96 & 0.92 & 16.16 & 2.71 & 26.99 & 1.26 & 3.93 & 2.04 & & & & & 7.41 \\
114 & 0.064 & 549.3 & 0.3 & -0.08 & 0.00 & 115.2 & 3.9 & & & & & & & & & -0.42 & 0.09 & 2.27 & 0.09 & -0.18 \\
121 & 0.064 & 480.8 & 0.5 & 0.08 & 0.00 & 109.0 & 1.0 & & & & & & & & & 29.70 & 0.35 &  34.50 & 0.40 &  39.49 \\
130 & 0.064 & 474.9 & 0.3 & -0.11 & 0.00 & 286.7 & 1.3 & & & & & & & & & 31.22 & 0.12 & 23.84 & 0.11 & 27.04 \\
\enddata
\tablecomments{Table 3 is published in its entirety in the machine-readable format. A portion is shown here for guidance regarding its form and content. Each pulse is listed by name (Column 1) and associated resolution (Column 2). All pulse fits include a linear background model characterized by $B_0$ (Column 3) and BS (Column 5) and described in Equation \ref{eqn:normfunction}, with uncertainties $B_0$ err (Column 4) and BS err (Column 6), respectively. Each pulse has been fitted using either the monotonic asymmetric \cite{nor05} pulse model (Equation \ref{eqn:function}) or the monotonic symmetric Gaussian pulse model (Equation \ref{eqn:normfunction}). The fit parameters of the \cite{nor05} model are A (Column 7), $t_s$ (Column 9), $\tau_1$ (Column 11), and $\tau_2$ (Column 13). When fitted along with the background, these parameters result in the corresponding uncertainties A err (Column 8), $t_s$ err (Column 10), $\tau_1$ err (Column 12), and $\tau_2$ err (Column 14).  The peak of the \cite{nor05} function occurs at time $\tau_{\rm peak}$ (Column 15), and the propagated uncertainties in the value are $\tau_{\rm peak}$ err (Column 16). The fit parameters of the Gaussian model are A (Column 7; this variable is C in equation \ref{eqn:normfunction}), $t_0$ (Column 17), and $\sigma$ (Column 19). When fitted along with the background, these parameters result in the corresponding uncertainties A err (Column 8), $t_0$ err (Column 18), and $\sigma$ err (Column 20). The time of reflection for each fitted pulse is $t_{\rm 0; mirror}$ (Column 21). }
\end{deluxetable}
\end{rotatetable}

Table \ref{tab:tab4} describes the fitting process used for assigning a quality value for each attempted pulse, as well as the resulting distribution of fits. Excluding monotonic pulses, $87\%$ of the pulses (162/187) successfully fit the model described by a monotonic pulse augmented by time-symmetric residuals. The modeled light curves of the quality factor 2 pulses referenced in Table \ref{tab:tab4} can be found in 
\figsetstart
\figsetnum{6}
\figsettitle{Fitted GRB pulse light curves.}

\figsetgrpstart
\figsetgrpnum{6.1}
\figsetgrptitle{Model pulse fit for BATSE pulse 107}
\figsetplot{f6_1.eps}
\figsetgrpnote{Temporally-symmetric model fits to GRB pulse light curves. Shown are the counts data (black), the fit to the Norris/Gaussian model (blue dashed line), the time-reversed model (red), the duration window (vertical dashed lines), and the time of reflection (vertical solid line).The complete figure set (298 images) is available in the online journal.}
\figsetgrpend

\figsetend

\begin{figure}
\plotone{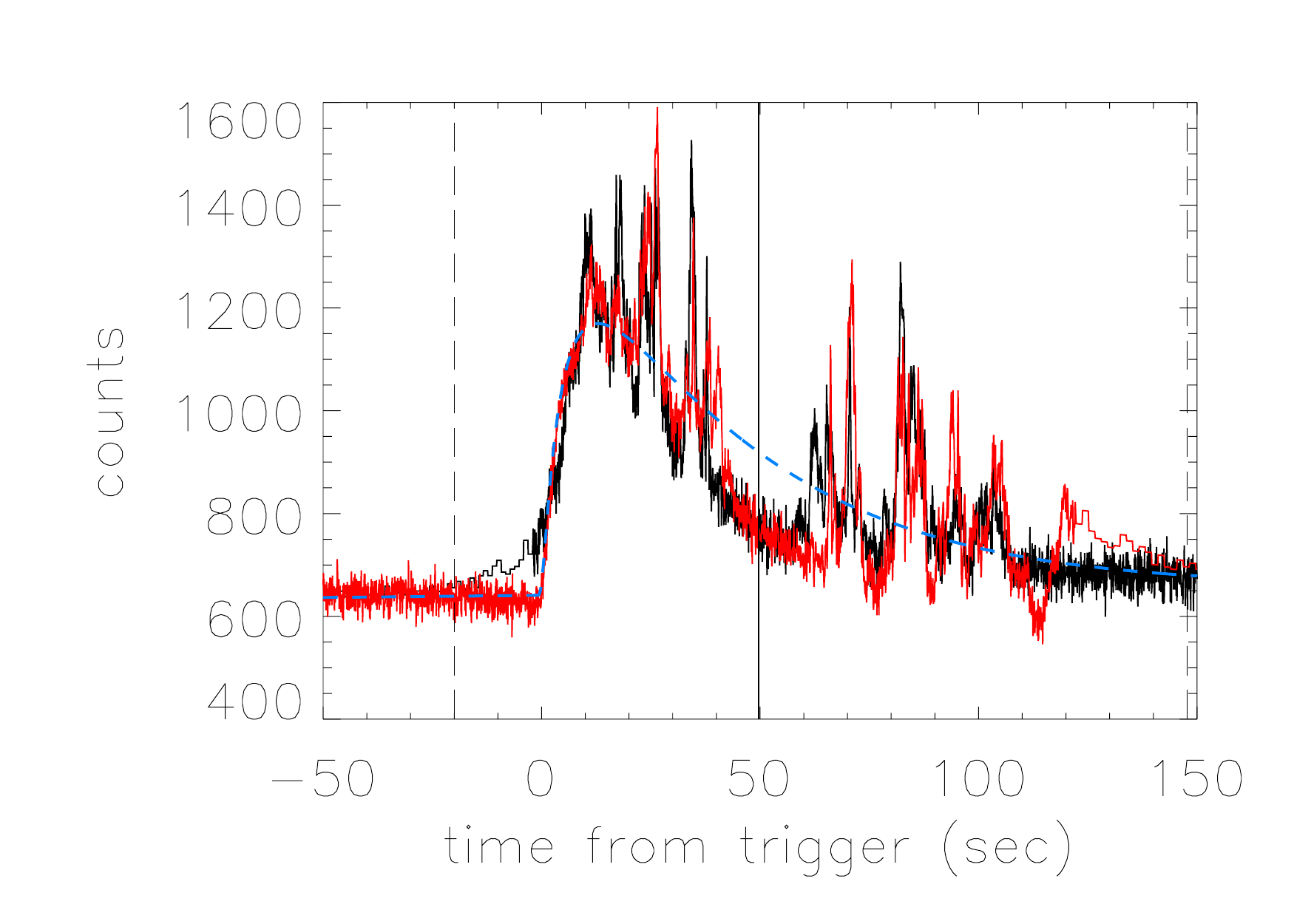}
\caption{Temporally-symmetric model fits to GRB pulse light curves. Shown are the counts data (black), the fit to the Norris/Gaussian model (blue dashed line), the time-reversed model (red), the duration window (vertical dashed lines), and the time of reflection (vertical solid line).The complete figure set (298 images) is available in the online journal.}
\end{figure}

We note that the sample contains 26 double-pulsed events (9\%) and 2 triple-pulsed events (1\%).

In order to more accurately study the $s_{\rm mirror}$ vs.~$\kappa$ relationship, we identify a more reliable subset of BATSE GRB pulses having $s_{\rm mirror} \le 0.3$ and $\kappa \le 0.3$. Table  \ref{tab:tab5} summarizes the distribution of pulses in this reduced dataset.

\begin{deluxetable*}{lcccc}
\tablenum{4}
\tablecaption{Fitted BATSE GRB pulses \label{tab:tab4}}
\tablewidth{0pt}
\tablehead{
\colhead{{\bf Quality factor}} &  \colhead{{\bf 0}}  &  \colhead{{\bf 1}}  &  \colhead{{\bf 2}} &  \colhead{{\bf Total}} 
}
\startdata
TTE Gaussian  &  0 &  4  &  1 &  {\bf 6} \\
TTE Norris  & 3  & 14  & 10  & {\bf 27} \\
Gaussian  &  15  &  31  &  66    &  {\bf 112} \\
Norris  & 7  & 77  & 84  & {\bf 168} \\
{\bf Total}  &  {\bf 25} &  {\bf 126}  &  {\bf 161} & {\bf 312} \\
\enddata
\end{deluxetable*}

\begin{deluxetable*}{ccccc}
\tablenum{5}
\tablecaption{Temporally symmetric pulses ($q=2$) with small measurement uncertainties ($\sigma_{\rm s~mirror} \le 0.3$ and $\kappa \le 0.3$). \label{tab:tab5}}
\tablewidth{0pt}
\tablehead{
\colhead{{\bf TTE Gaussian}} &  \colhead{{\bf TTE Norris}}  &  \colhead{{\bf Gaussian}}  &  \colhead{{\bf Norris}} &  \colhead{{\bf Total}} 
}
\startdata
1 &  1  &  51 &  68 & {\bf 121} \\
\enddata
\end{deluxetable*}

\subsection{The $s_{\rm mirror}$ vs.~Asymmetry Pulse Relationship}
 
The key goal of our study is to reexamine the relationship between pulse asymmetry ($\kappa$) and stretching in the time-reversed residuals ($s_{\rm mirror}$). From a much smaller sample of selected bright GRB pulses, \cite{hak19} demonstrated that this relationship was real. The existence of such a relationship has two immediate repercussions:
\begin{enumerate}
\item The alignment of structure with each pulse indicates that structure is part of each pulse, rather than unrelated to it. By redefining pulses rather than treating each structural variation as a separate pulse, we could say that GRB pulses must be rare rather than common events.
\item Being able to fit a large percentage of GRB pulses with the temporally-reversed model indicates that temporal reversibility is a strange yet defining characteristic of GRB pulse structure.
\end{enumerate}

\begin{figure}
\plottwo{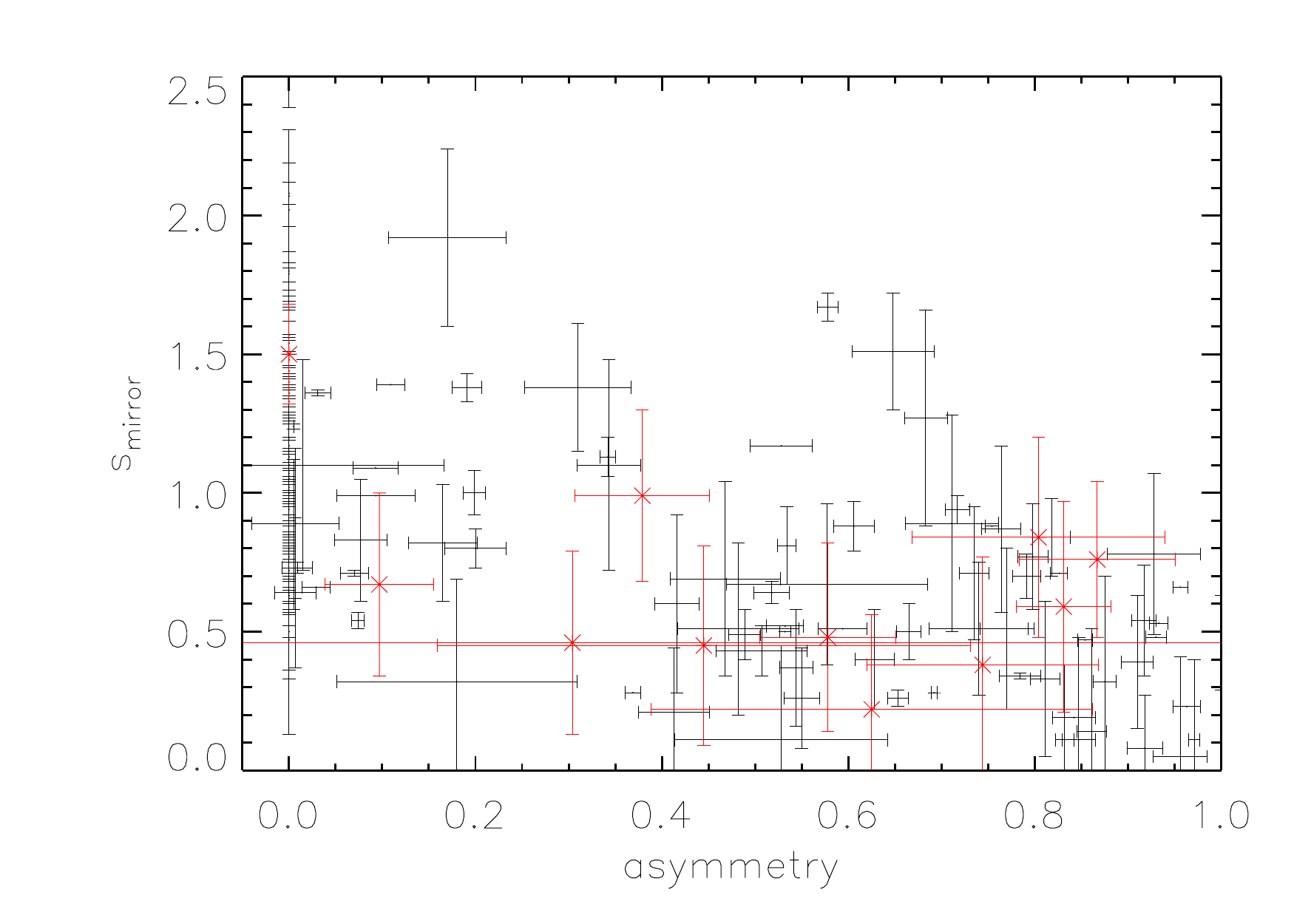}{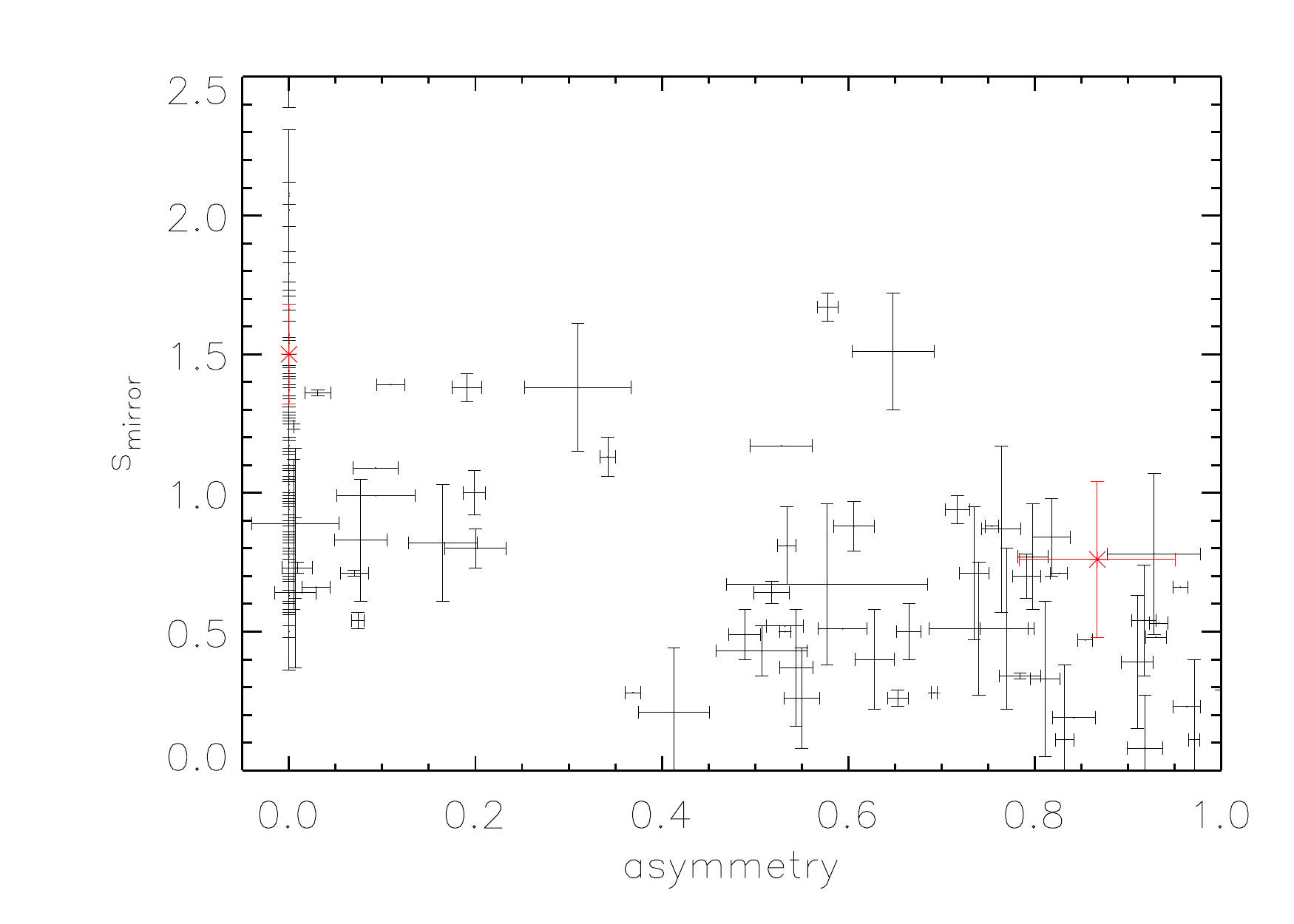}
\caption{Observed values of $s_{\rm mirror}$ vs.~asymmetry ($\kappa$) for the GRB pulses having measurable temporally-symmetric residuals. The left panel demonstrates the relation for the 160 GRB pulses in Table 3 for which $s_{\rm mirror}$ values could be measured, while the right panel shows the relation for the 122 GRB pulses in Table 4 having $\sigma_{\rm s~mirror} \le 0.3$. Black error bars indicate 64 ms pulse fits and red error bars indicate 4 ms-binned TTE pulse fits. Spearman Rank Order correlation tests find an anti-correlation of $-0.64$ between the parameters in the left panel with a $p-$value of $4.7 \times 10^{-20}$, and an anti-correlation of $-0.62$ between the parameters in the right panel with a $p-$value of $1.5 \times 10^{-14}$.\label{fig:fig7}}
\end{figure}

The temporally-symmetric characteristics of the 162 BATSE GRB pulses having measurable residuals are shown in the left panel of Figure \ref{fig:fig7}, while the characteristics of the 121 bight pulses described in Table \ref{tab:tab5} are shown in the right panel of Figure \ref{fig:fig7}. Each fitted pulse is identified by the error bars $\sigma_{\rm s~mirror}$ and $\sigma_\kappa$ in measuring $s_{\rm mirror}$ and $\kappa$, respectively. The plots demonstrate that fewer TTE pulses (red error bars) have measurable residuals than 64 ms pulses (black error bars), because structure has been washed out by small-number photon counting statistics. 

A strong anti-correlation between $\kappa$ and $s_{\rm mirror}$ and $\kappa$ is present in both plots, with Spearman Rank Order correlation tests indicating that both relationships are highly unlikely to have occurred by chance. Thus, the temporally symmetric residual structures do indeed appear to be tied to the underlying monotonic pulse structures. We look to the light curves of pulses shown in this Figure \ref{fig:fig7} to provide us with greater insights regarding the large intrinsic scatter between $s_{\rm mirror}$ vs.~$\kappa$.

\subsection{GRB Pulse Morphology} \label{sec:morphology}

{\em Visual inspection unexpectedly shows us that similarities among the shapes of individual pulse light curve are responsible for the large dispersion between $s_{\rm mirror}$ vs.~$\kappa$.} This is surprising because, up until now, the only formally-recognized pulse morphology has been the FRED pulse type. Although we have used visual classification to aid us with our understanding, we note that automated and machine learning classification are also possible using techniques such as using agglomerative clustering to identify light curve similarities ({\em e.g.,} \cite{can20}). Figure \ref{fig:fig8} demonstrates that different pulse morphologies are confined to specific regions of the $s_{\rm mirror}$ vs.~$\kappa$ parameter space. Besides FREDs, we classify these new pulse types as ``rollercoaster pulses", ``crowns", ``u-pulses", and ``asymmetric u-pulses." The diagram also contains a small number of unclassified pulses. Unfitted pulses and a small number of ``no pulse u-pulses'' are excluded from this diagram because they are suspected of being single structured pulses whose monotonic components cannot be identified.

\subsubsection{FRED pulses}

\begin{figure}
\epsscale{0.80}
\plotone{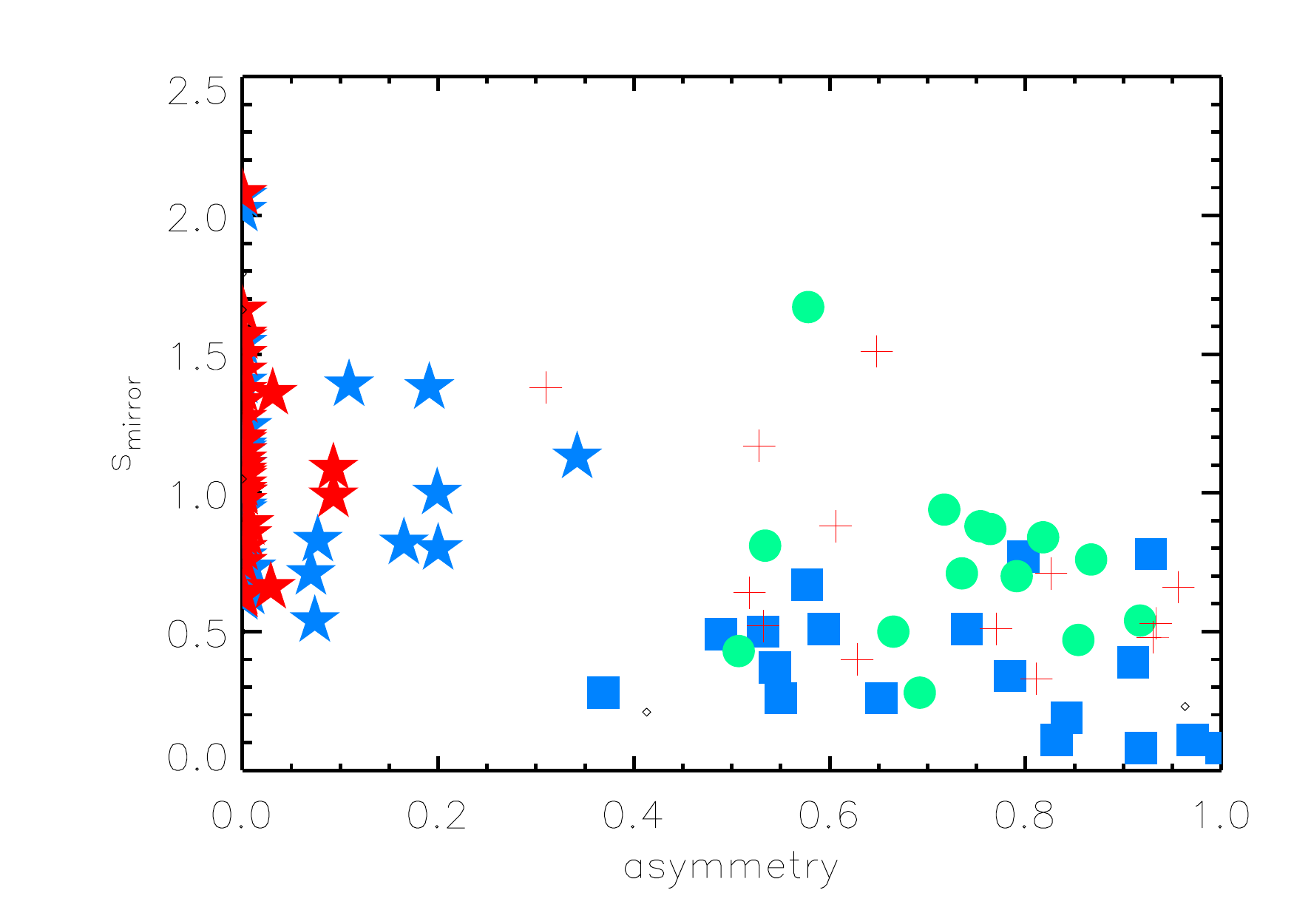}
\caption{Observed values of $s_{\rm mirror}$ vs.~asymmetry ($\kappa$) for the 122 GRB pulses described in Table 4, labeled by pulse morphology. Blue squares indicate FRED pulses, red stars indicate u-pulses, blue stars indicate crown pulses, green circles indicate rollercoaster pulses, and red plus signs indicate asymmetric u-pulses. No-pulse u-pulses are not included in the plot, and unclassified pulses are shown as small open black diamonds. \label{fig:fig8}}\end{figure}

\begin{figure}[htb]
\centering
  \begin{tabular}{@{}cc@{}}
    \includegraphics[width=.5\textwidth]{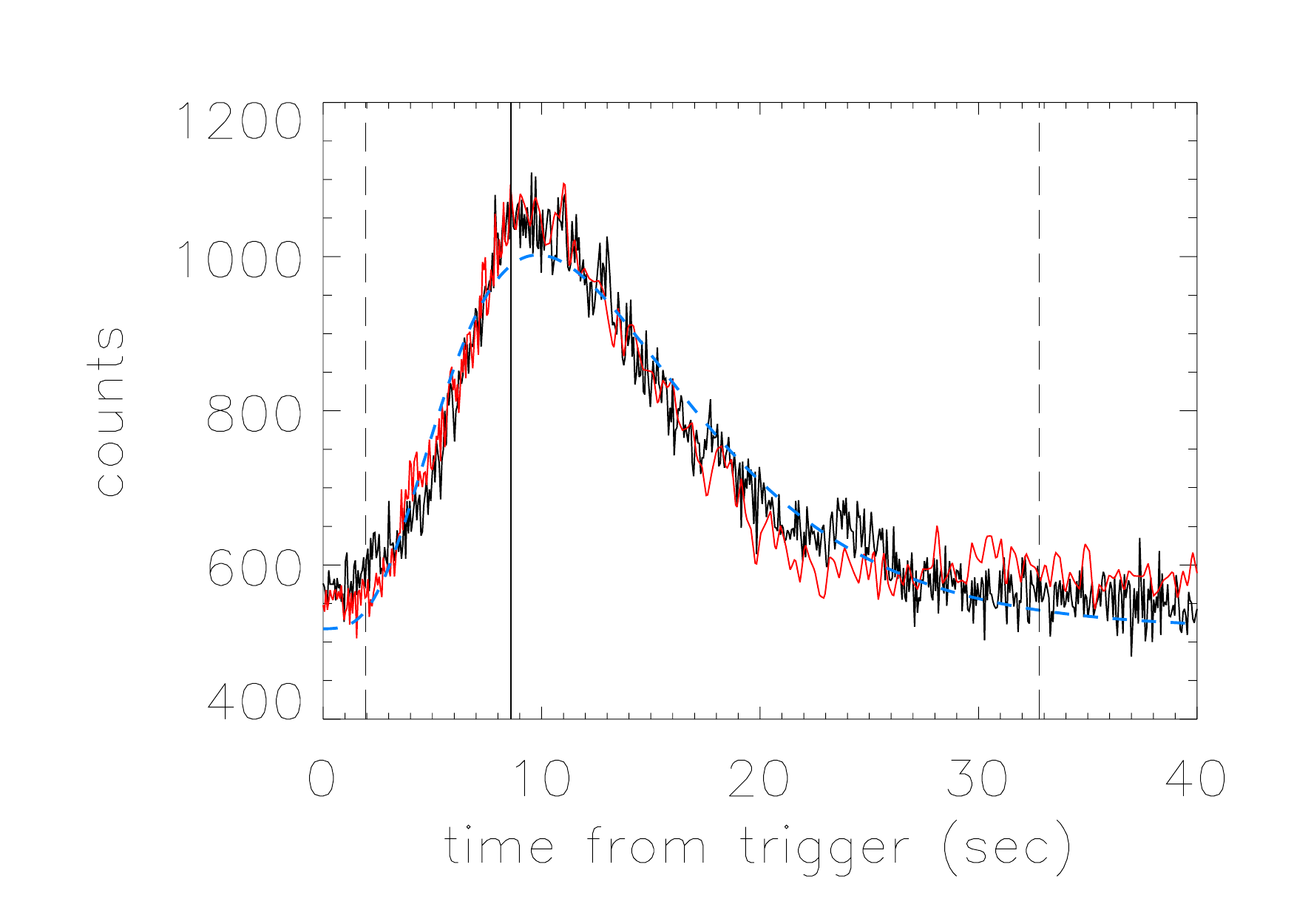} & 
    \includegraphics[width=.5\textwidth]{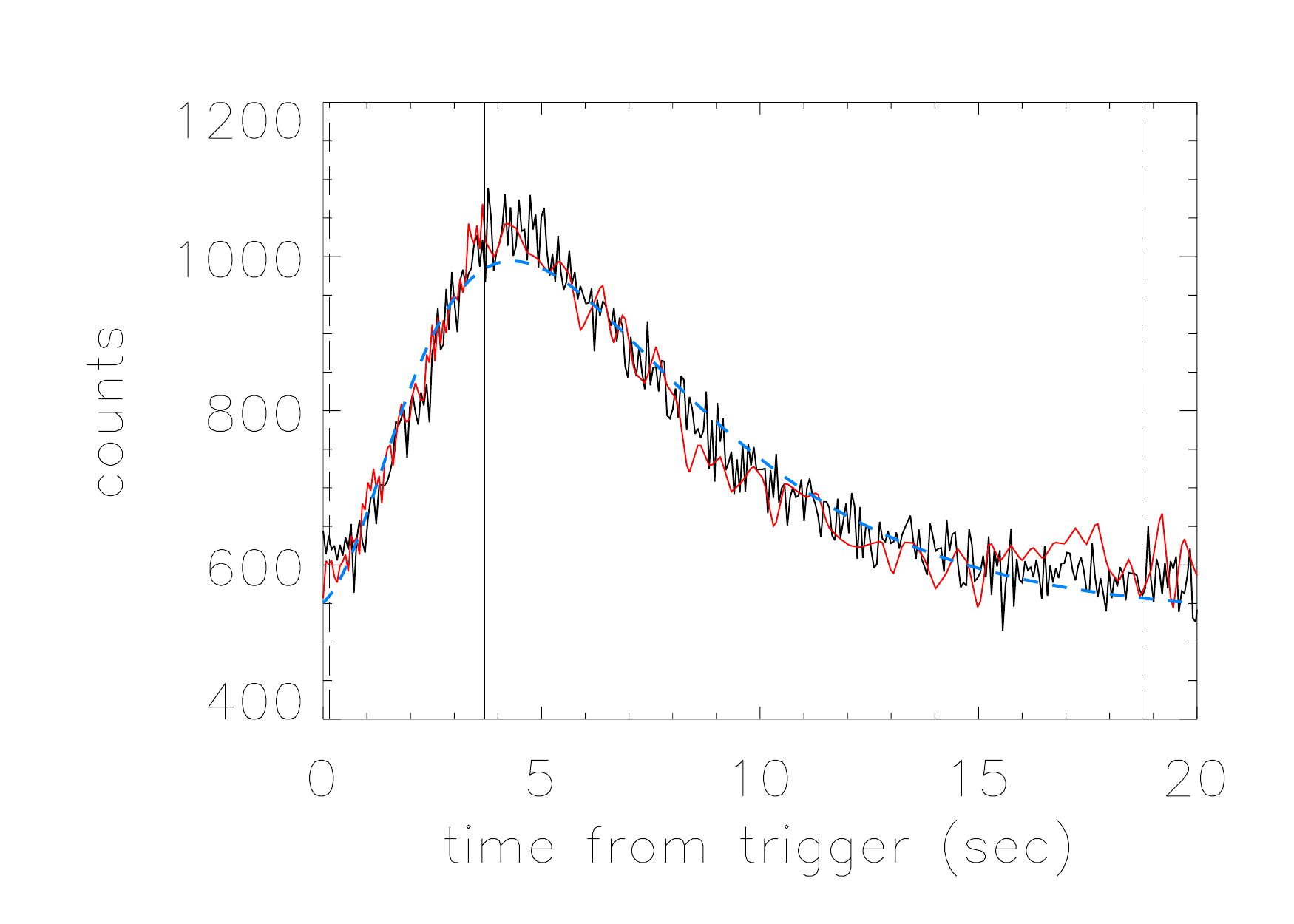} \\
    \includegraphics[width=.5\textwidth]{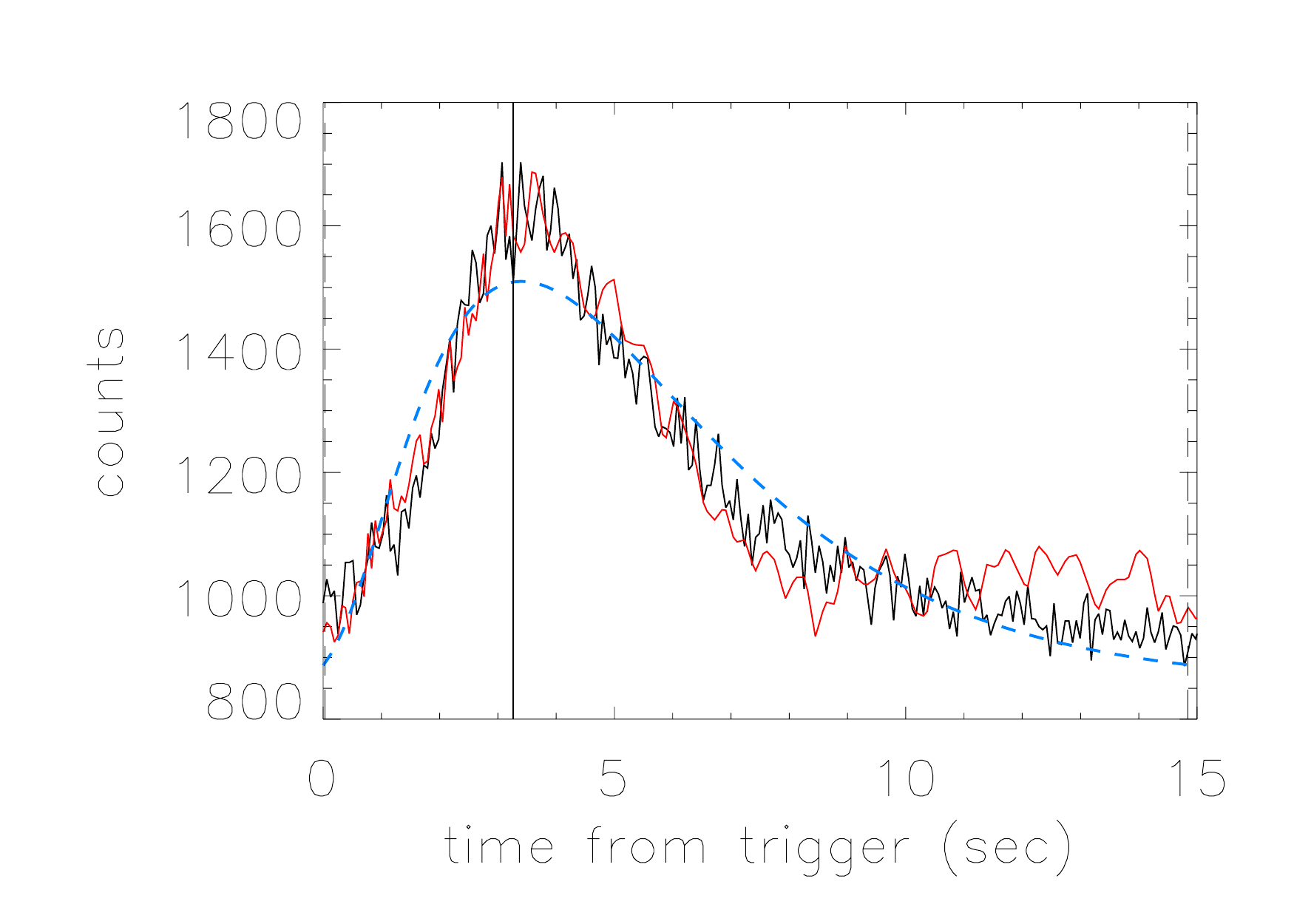}  & 
    \includegraphics[width=.5\textwidth]{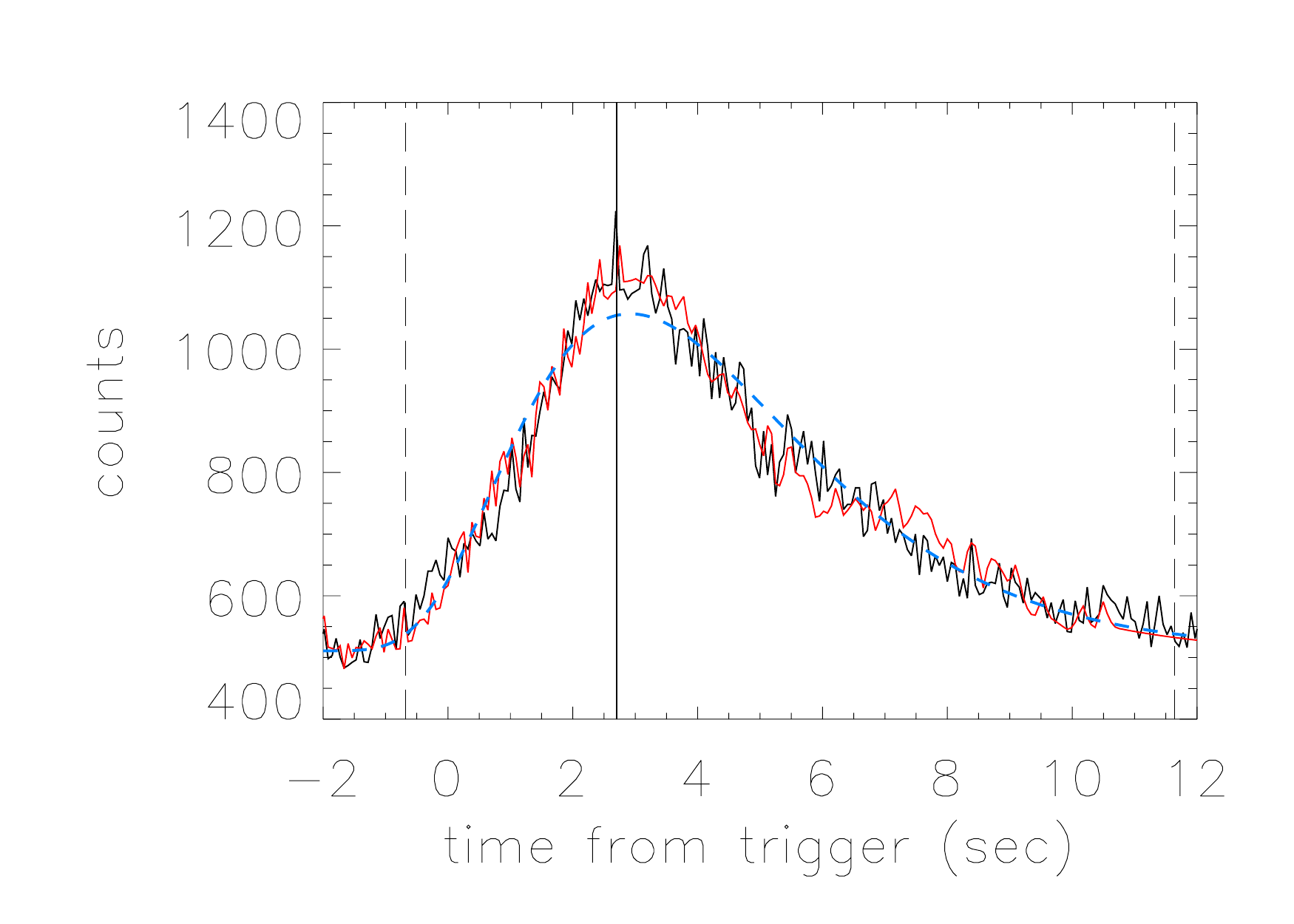}   \\
  \end{tabular}
  \caption{Time-reversed models for FRED pulses BATSE 3003 (upper left), BATSE 1467 (upper right), BATSE 1733 (lower left) and BATSE 1956 (lower right). Shown are the counts data (black), the fit to the Norris/Gaussian model (blue dashed line), the time-reversed model (red), the duration window (vertical dashed lines), and the time of reflection (vertical solid line). \label{fig:fig9}}
\end{figure}

FRED pulses are not always the monotonic bumps they have been portrayed to be. Brighter FREDs have smoothly varying non-monotonic structure that can be explained by the ``triple-peaked residuals" described by \cite{hak14}. This structure appears to be washed out in faint FRED pulses, as these exhibit insignificant residuals and cannot be included in Figure \ref{fig:fig8}. FRED pulses with temporally-symmetric structures are plotted in Figure \ref{fig:fig8} as blue squares. These pulses follow a much tighter anti-correlation between $s_{\rm mirror}$ vs.~$\kappa$ than that of the overall distribution. Examples of the temporally-symmetric pulse models of FREDs are 
the pulses shown in Figure \ref{fig:fig9}.

\subsubsection{Rollercoaster pulses}

\begin{figure}[htb]
\centering
  \begin{tabular}{@{}cc@{}}
    \includegraphics[width=.5\textwidth]{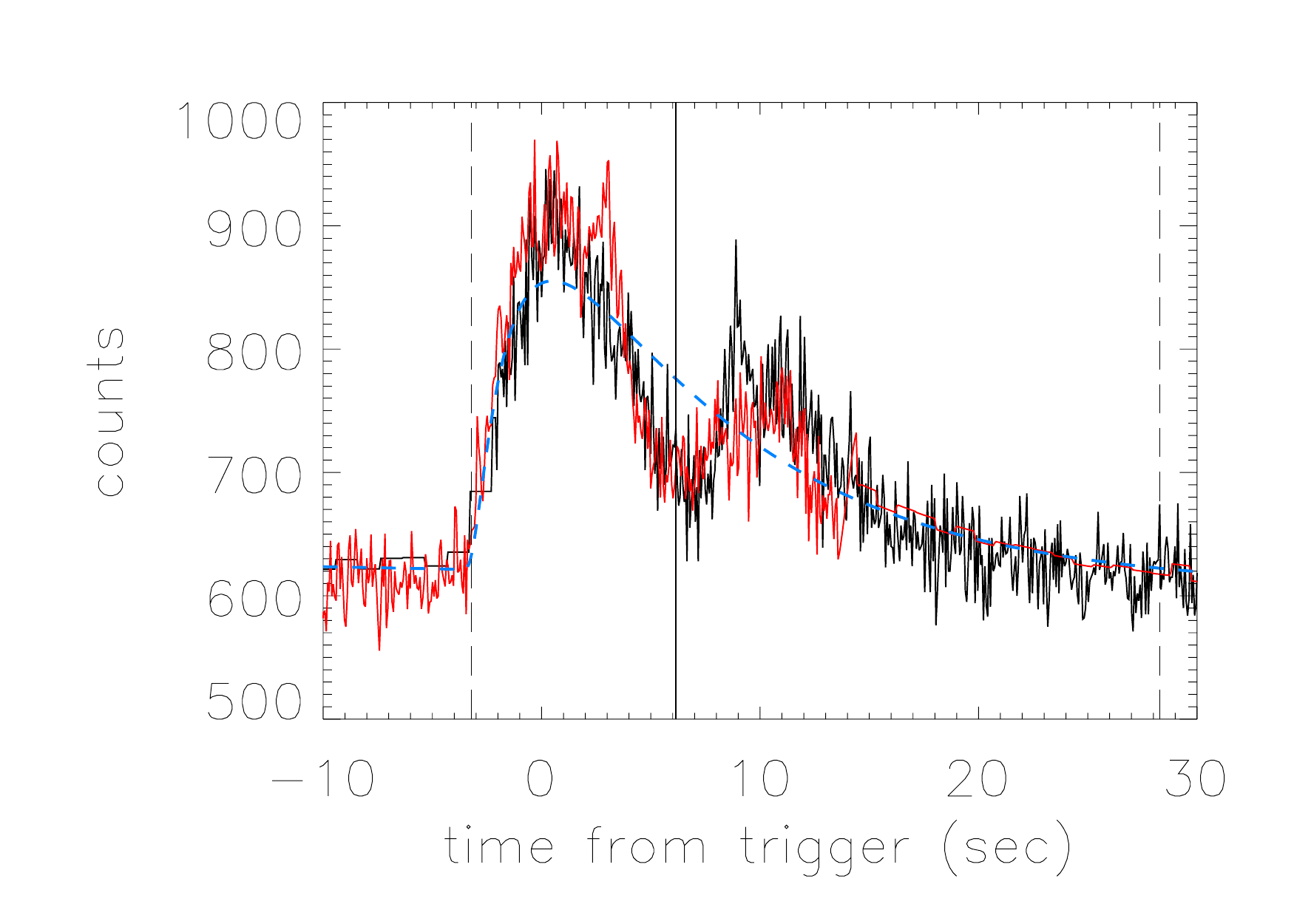}  & 
    \includegraphics[width=.5\textwidth]{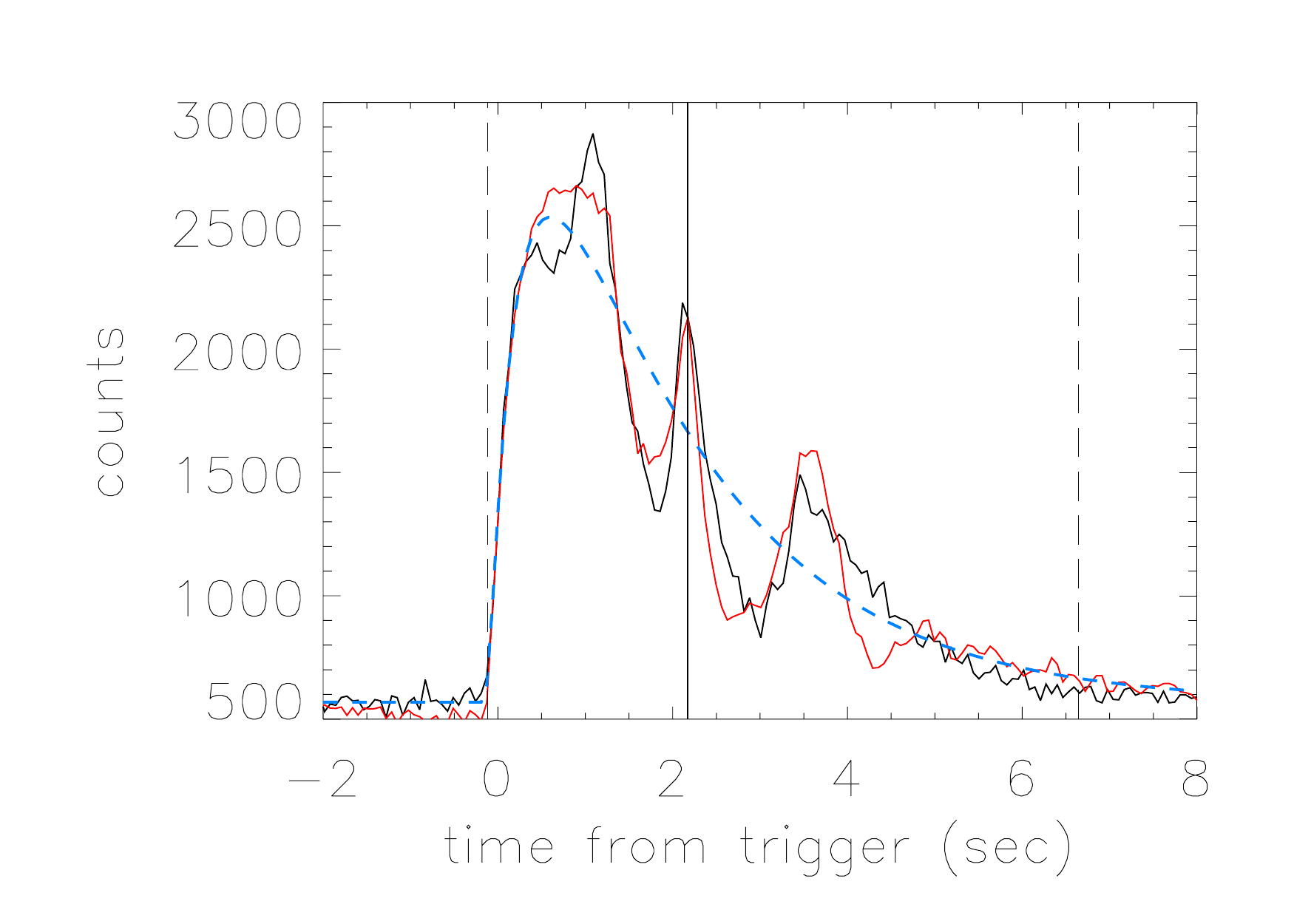} \\
    \includegraphics[width=.5\textwidth]{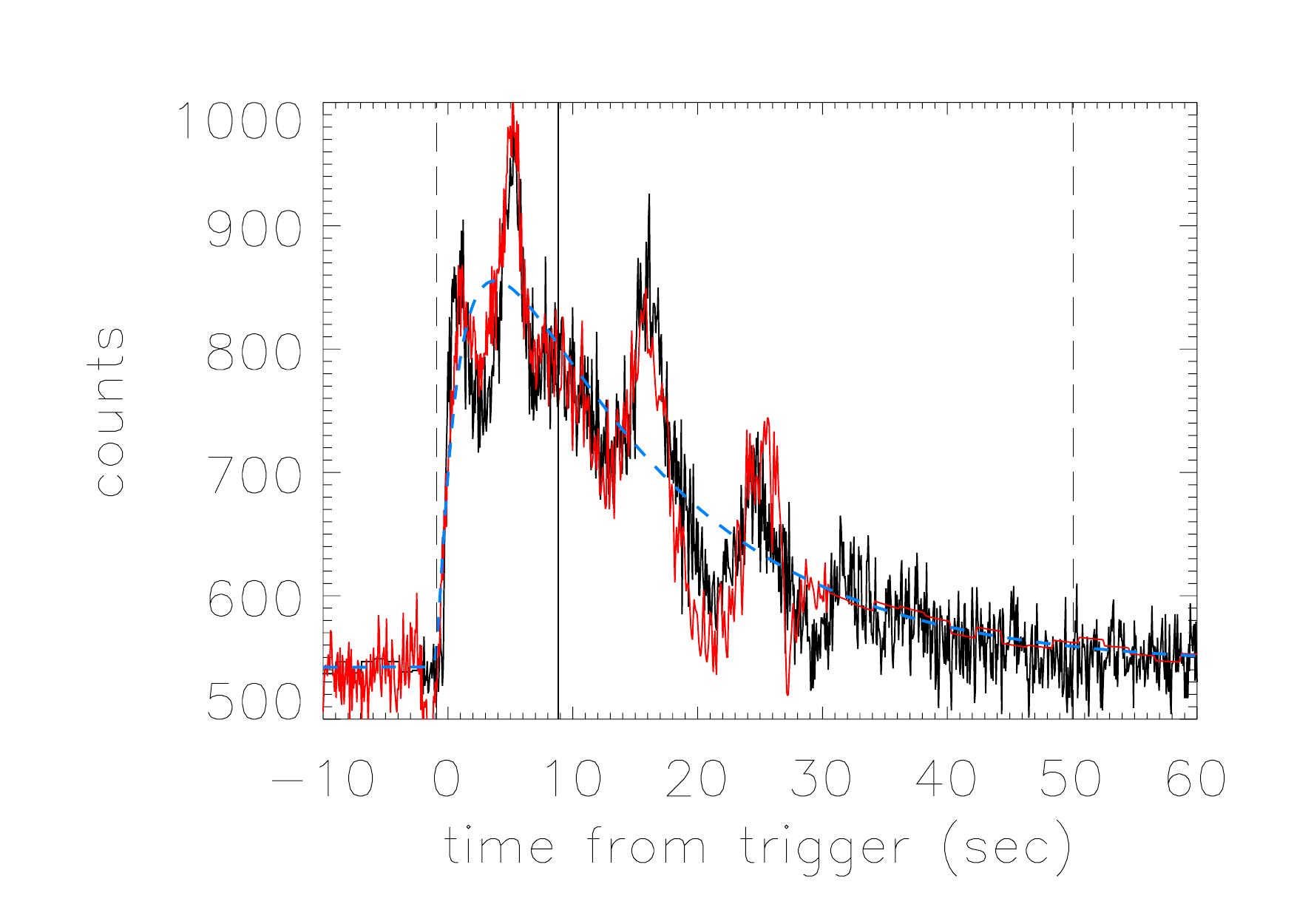}  & 
    \includegraphics[width=.5\textwidth]{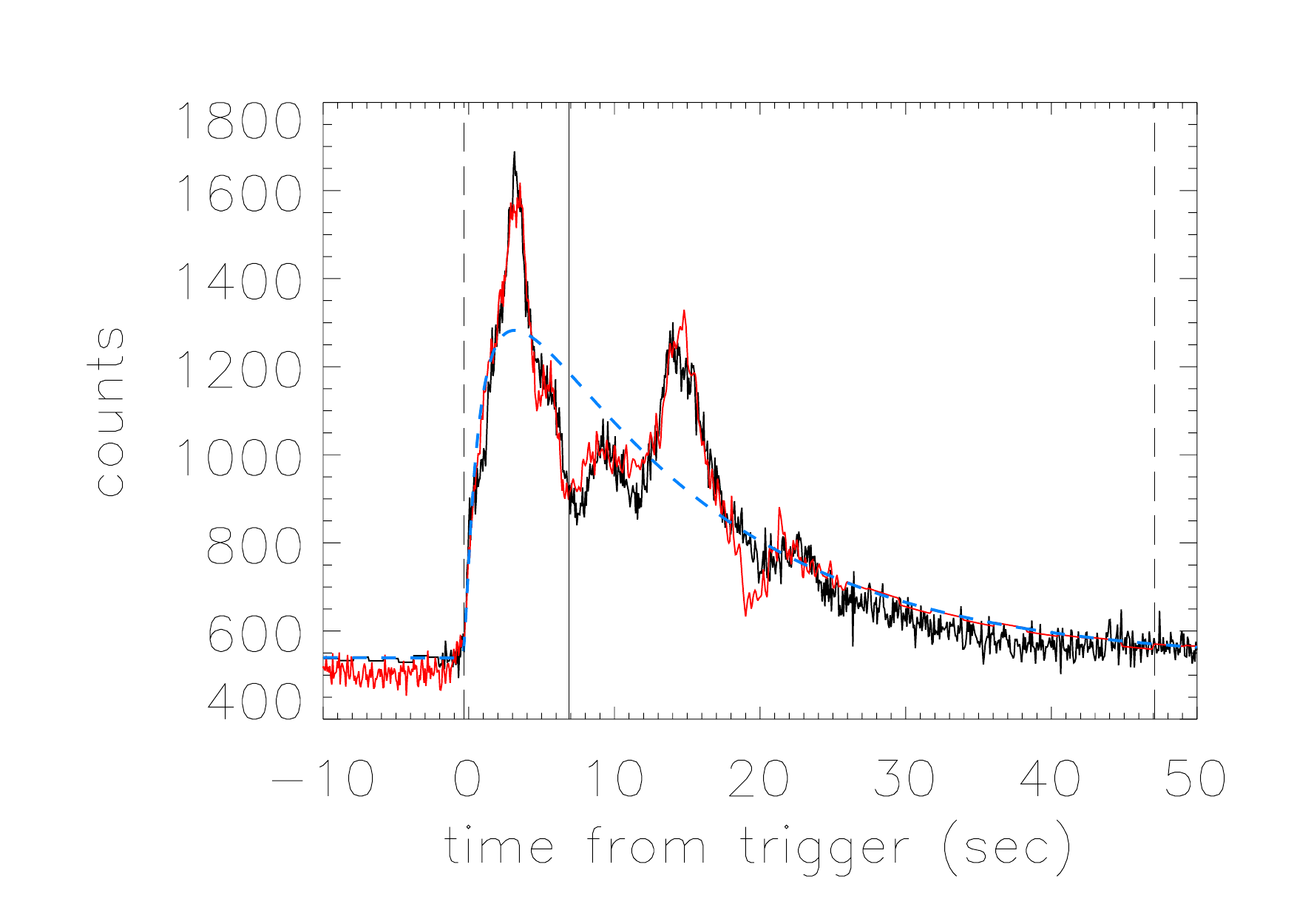}   \\
  \end{tabular}
  \caption{Temporally-symmetric models of rollercoaster pulses BATSE 398 (upper left), BATSE 543 (upper right), BATSE 548 (lower left) and BATSE 647 (lower right). Shown are the counts data (black), the fit to the Norris/Gaussian model (blue dashed line), the time-reversed model (red), the duration window (vertical dashed lines), and the time of reflection (vertical solid line). \label{fig:fig10}}
\end{figure}

Many pulses with FRED-like asymmetries but with substantially larger $s_{\rm mirror}$ values are found to have similar light curve morphologies. These pulses, which we call {\em rollercoaster} pulses, have the appearance of FREDs with smooth wavy structures appearing during the pulse decay phase. These wavy structures are themselves temporally-symmetric. Examples of temporally-symmetric models of rollercoaster pulses are 
the pulses shown in Figure \ref{fig:fig10}.

\begin{figure}
\epsscale{0.80}
\plotone{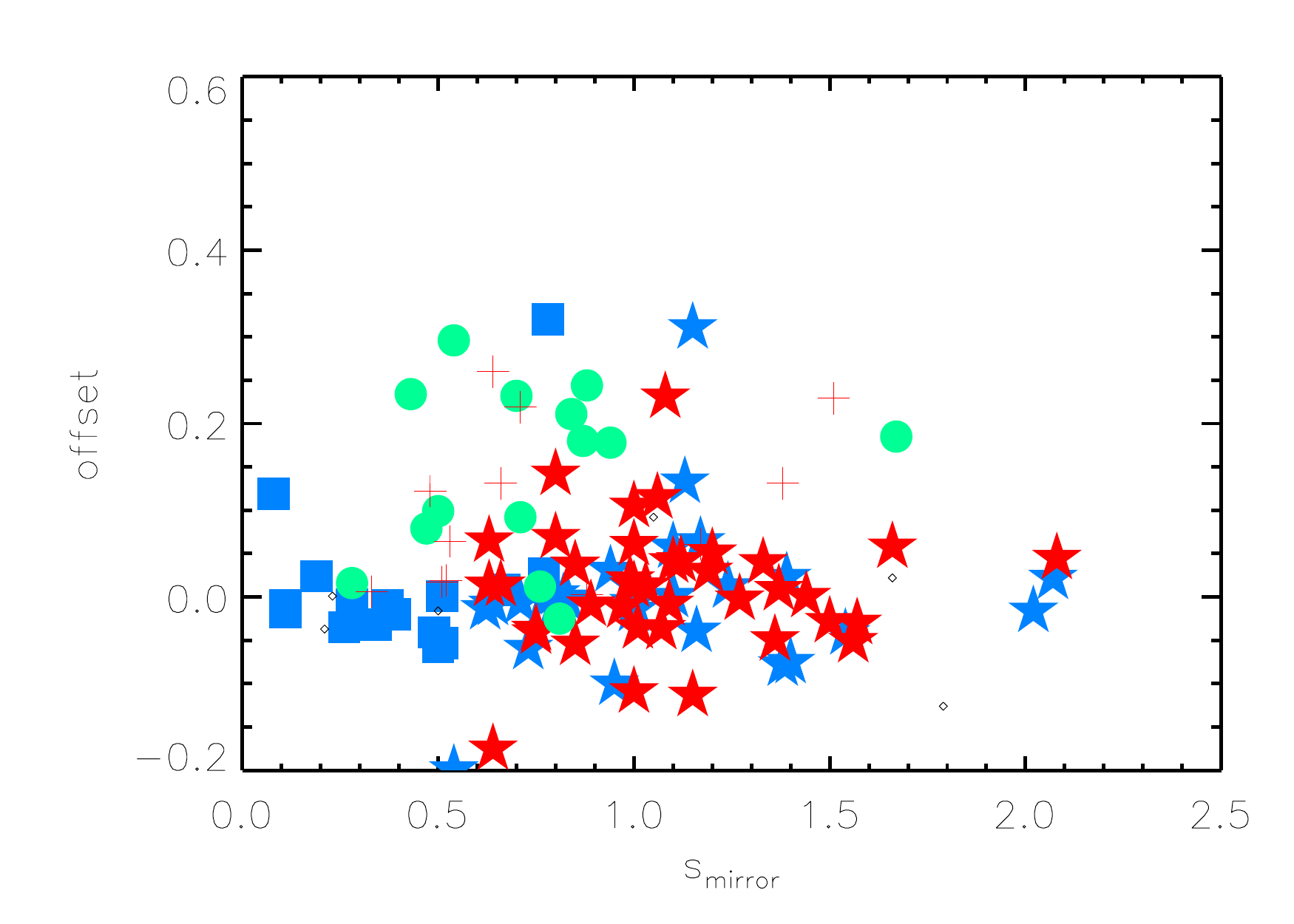}
\caption{Offset vs.~$s_{\rm mirror}$ for the pulses in this GRB sample. Rollercoaster pulses (green circles) and asymmetric u-pulses (red crosses) have larger offsets than FREDS (blue squares), crowns (blue stars), u-pulses (red stars), and unclassified pulses (small open black diamonds). \label{fig:fig11}}\end{figure}

Both the $s_{\rm mirror}$ and $\kappa$ values of rollercoaster pulses are typically between 0.5 and 1.0. These are larger $s_{\rm mirror}$ values than those found for FREDs having similar $\kappa$ values. Rollercoaster pulses accommodate these larger $s_{\rm mirror}$ values by having the time of reflection delayed relative to the time of the pulse peak time. Figure \ref{fig:fig11} demonstrates the additional role that the offset plays in delineating rollercoaster pulses (green circles) from FREDs and other pulse types. 

\subsubsection{Crown pulses} 

Symmetric FRED-like pulses differ from asymmetric FRED pulses in that they have enhanced structure and large $s_{\rm mirror}$ values (typically $s_{\rm mirror} \ge 1$). We call pulses with these characteristics {\em crown} pulses. When $s_{\rm mirror}$ exceeds unity, the decay portions of the residuals are compressed relative to the rise portion rather than stretched, changing the overall pulse appearance.

\begin{figure}
\epsscale{0.80}
\plotone{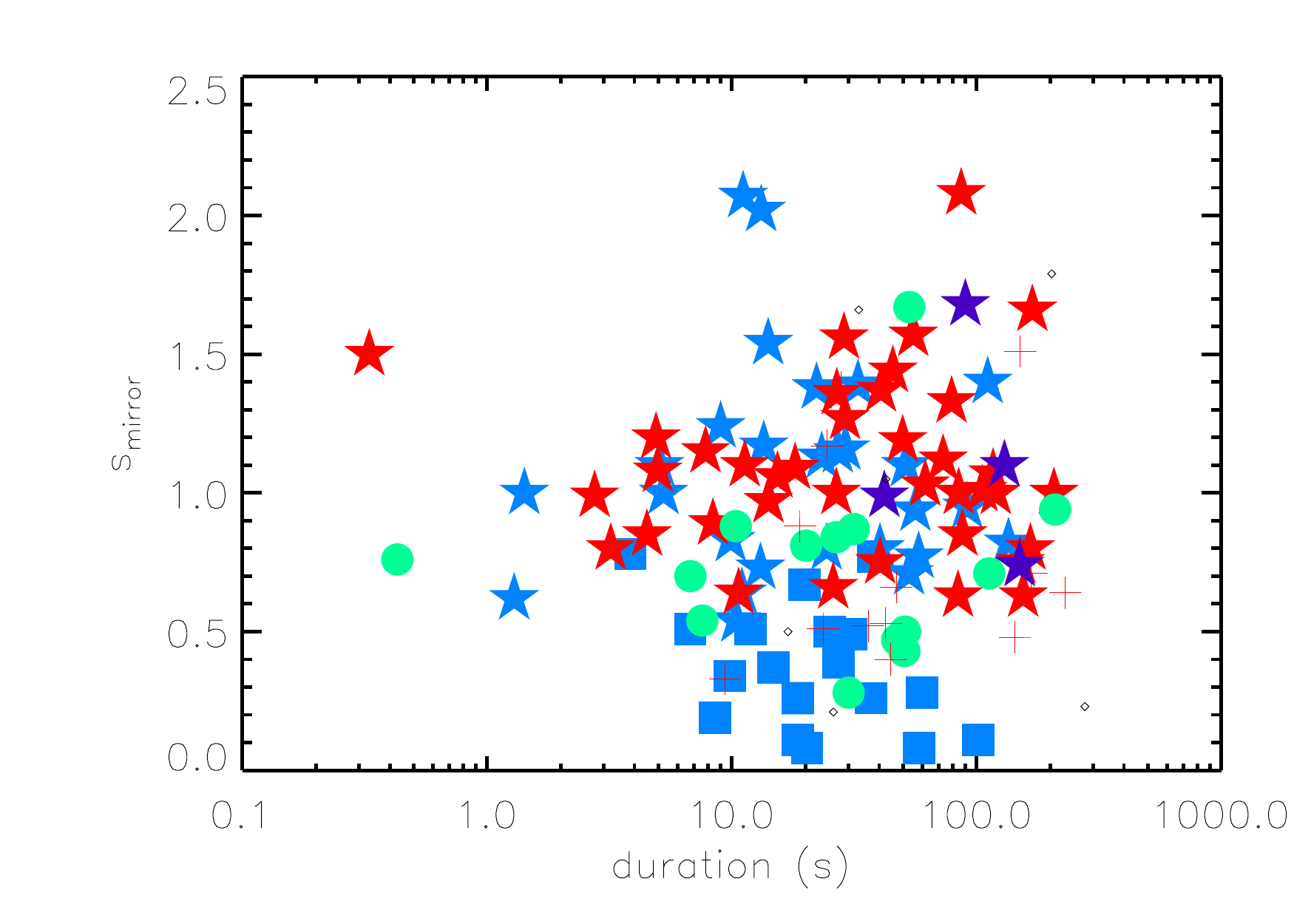}
\caption{$s_{\rm mirror}$ vs.~pulse duration for the pulses in this GRB sample. The longest-duration pulses are u-pulses (red stars), no-pulse u-pulses (purple stars), and crowns (blue stars). FREDS (blue squares), rollercoaster pulses (green circles), and asymmetric u-pulses (red crosses) are shorter. Also plotted are unclassified pulses (small open black diamonds). Short GRB pulses (durations shorter than 1 s) comprise what appears to be a separate but parallel pulse morphology distribution to that of long GRB pulses. \label{fig:fig12}}\end{figure}

\begin{figure}[htb]
\centering
  \begin{tabular}{@{}cc@{}}
    \includegraphics[width=.5\textwidth]{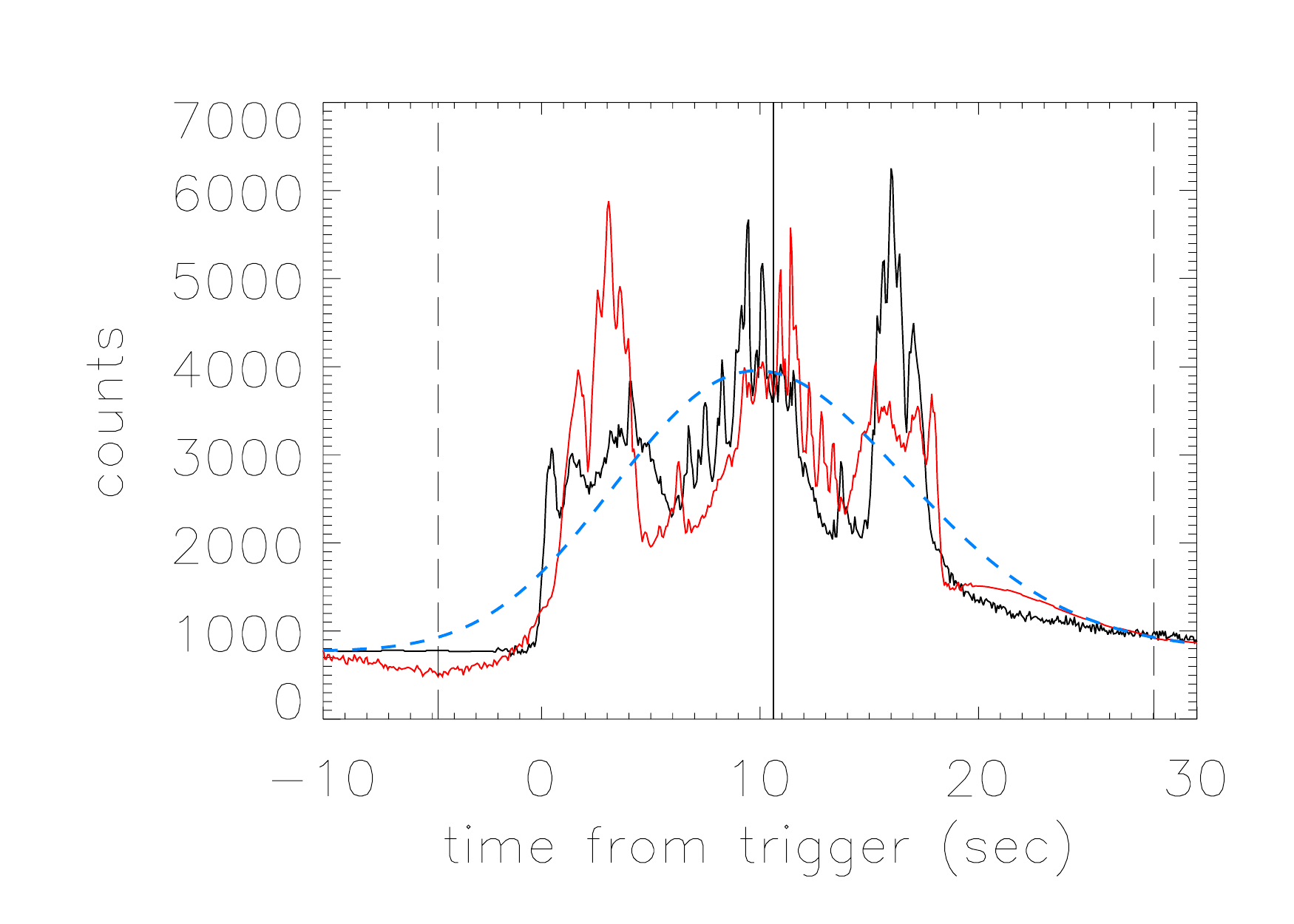}  & 
    \includegraphics[width=.5\textwidth]{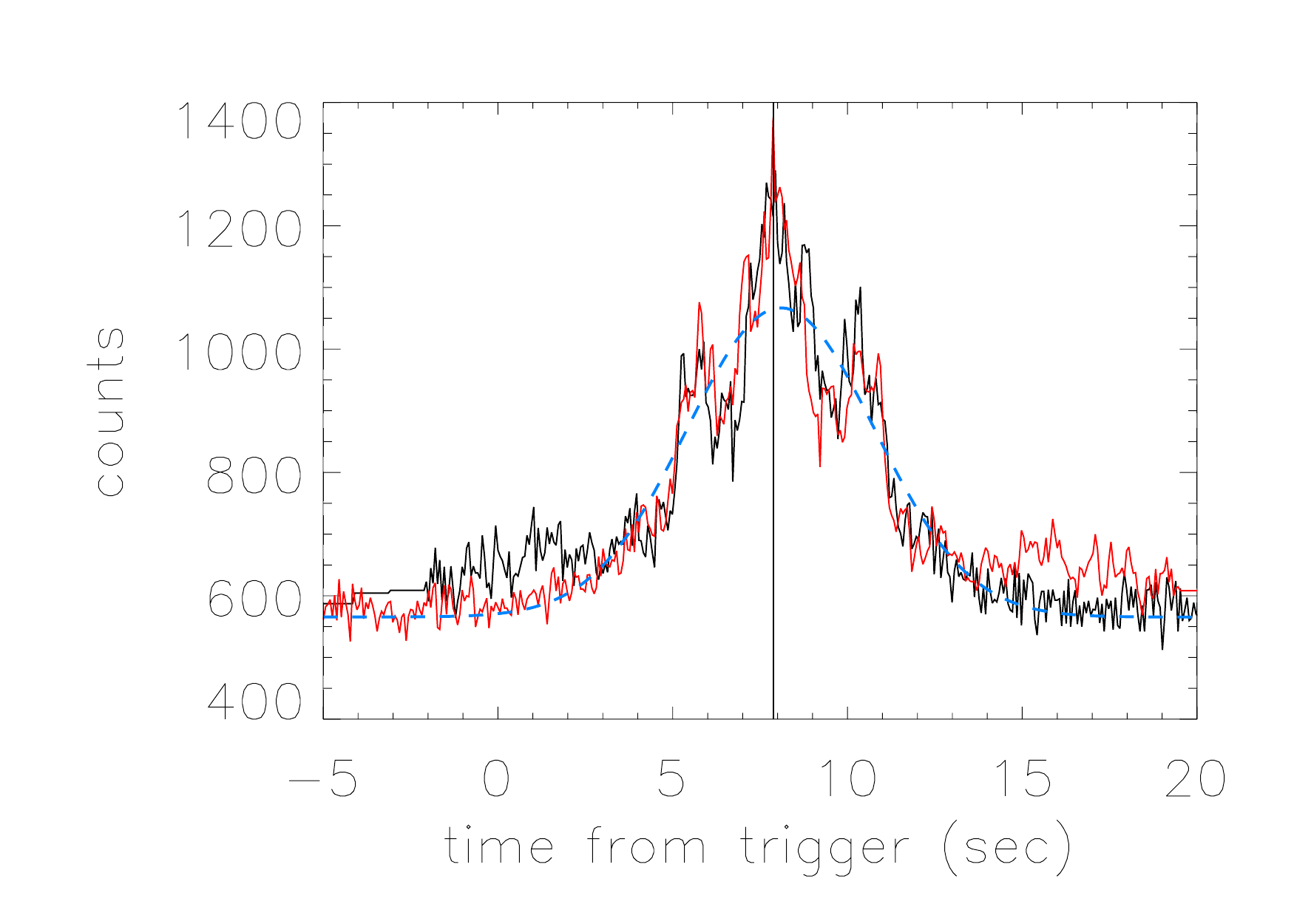} \\
    \includegraphics[width=.5\textwidth]{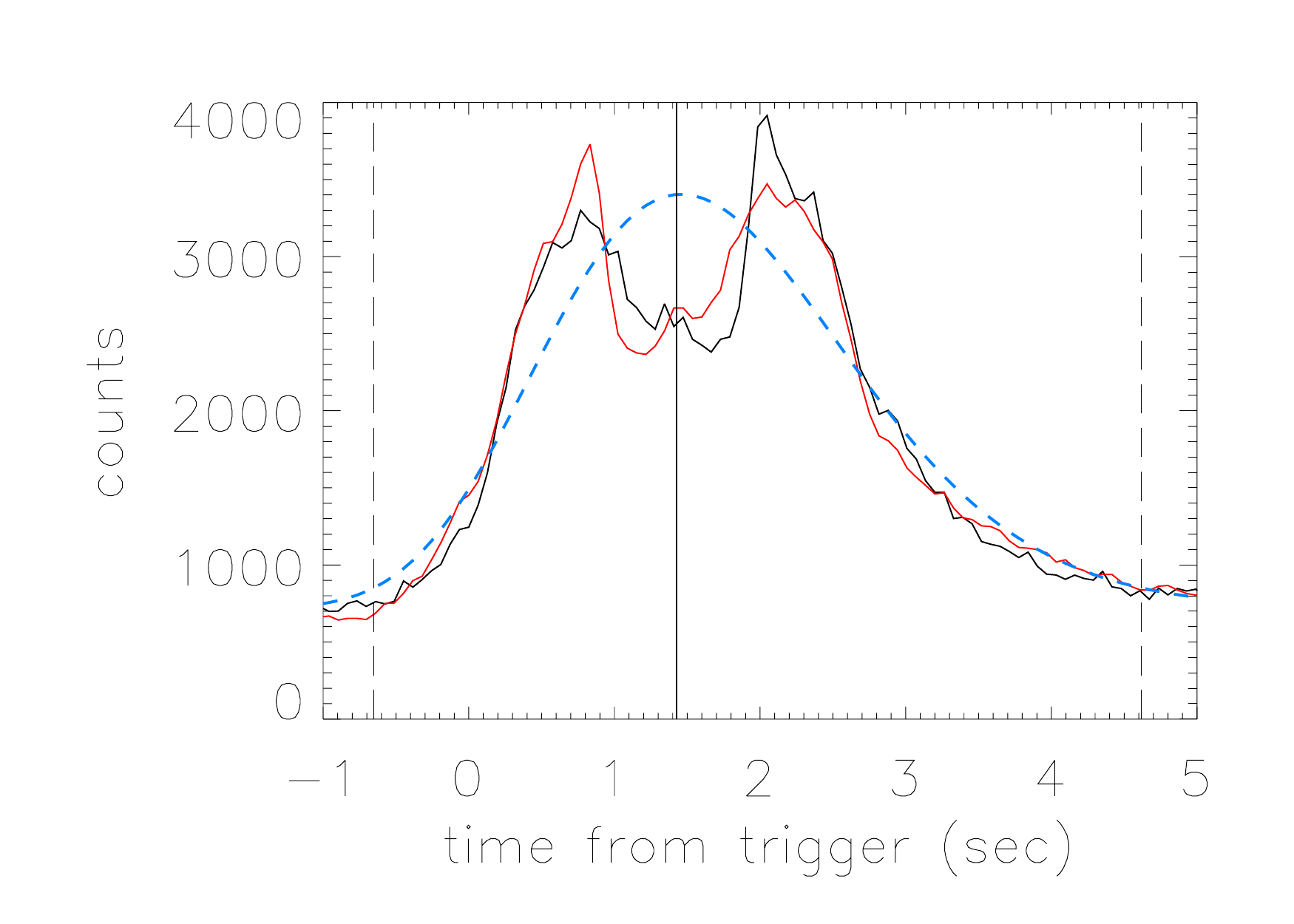}  & 
    \includegraphics[width=.5\textwidth]{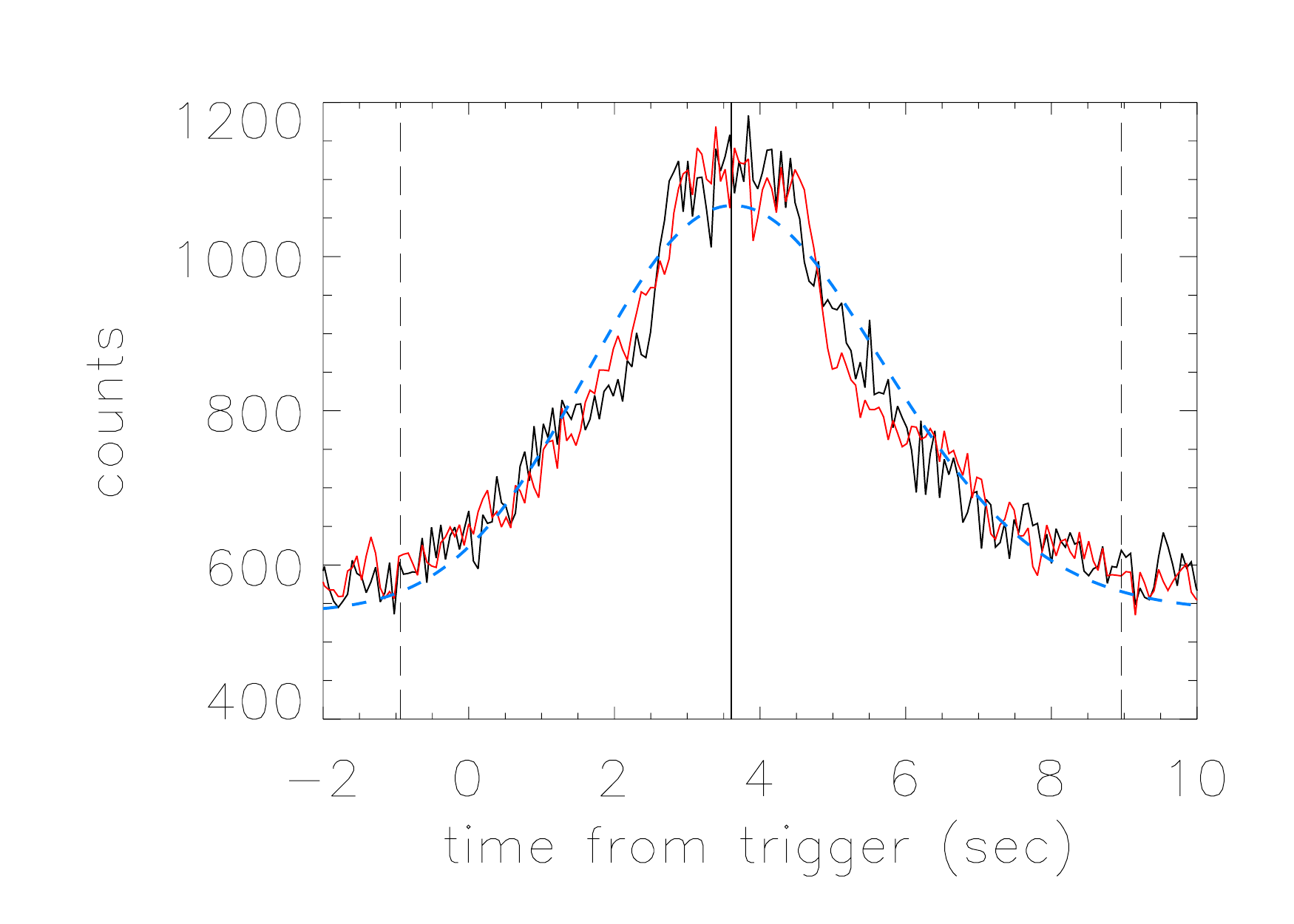}   \\
  \end{tabular}
  \caption{Temporally-symmetric models of crown pulses BATSE 1663 (upper left), BATSE 612 (upper right), BATSE 1709 (lower left) and BATSE 1717 (lower right). Shown are the counts data (black), the fit to the Norris/Gaussian model (blue dashed line), the time-reversed model (red), the duration window (vertical dashed lines), and the time of reflection (vertical solid line). \label{fig:fig13}}
\end{figure}

In general, crowns have offset times similar to those of FREDs (Figure \ref{fig:fig11}) and durations that are larger than FREDs and rollercoaster pulses (Figure \ref{fig:fig12}). Examples of temporally-symmetric models of crowns are shown in Figure \ref{fig:fig13}.

\subsubsection{U-pulses} 

{\em U-pulses} were identified by \citep{hak18a} to describe a subset of TTE GRB pulses characterized by bright time-symmetric intensity structures that drop down to near the background rate around the time of the monotonic pulse peak (they look like the letter "u"). U-pulses make up a surprisingly large fraction of GRB pulses -- there are more u-pulses than any other defined pulse morphological type (see Table \ref{tab:tab6}). U-pulses are more structured than FREDs or rollercoaster pulses, and the structure they exhibit undergoes rapid time variations. This gives u-pulses more chaotic appearances than are found in the smoother FRED or rollercoaster light curves. In many ways, u-pulses look like the residuals of crowns but with much weaker underlying monotonic pulse components. 


\begin{figure}[htb]
\centering
  \begin{tabular}{@{}cc@{}}
    \includegraphics[width=.5\textwidth]{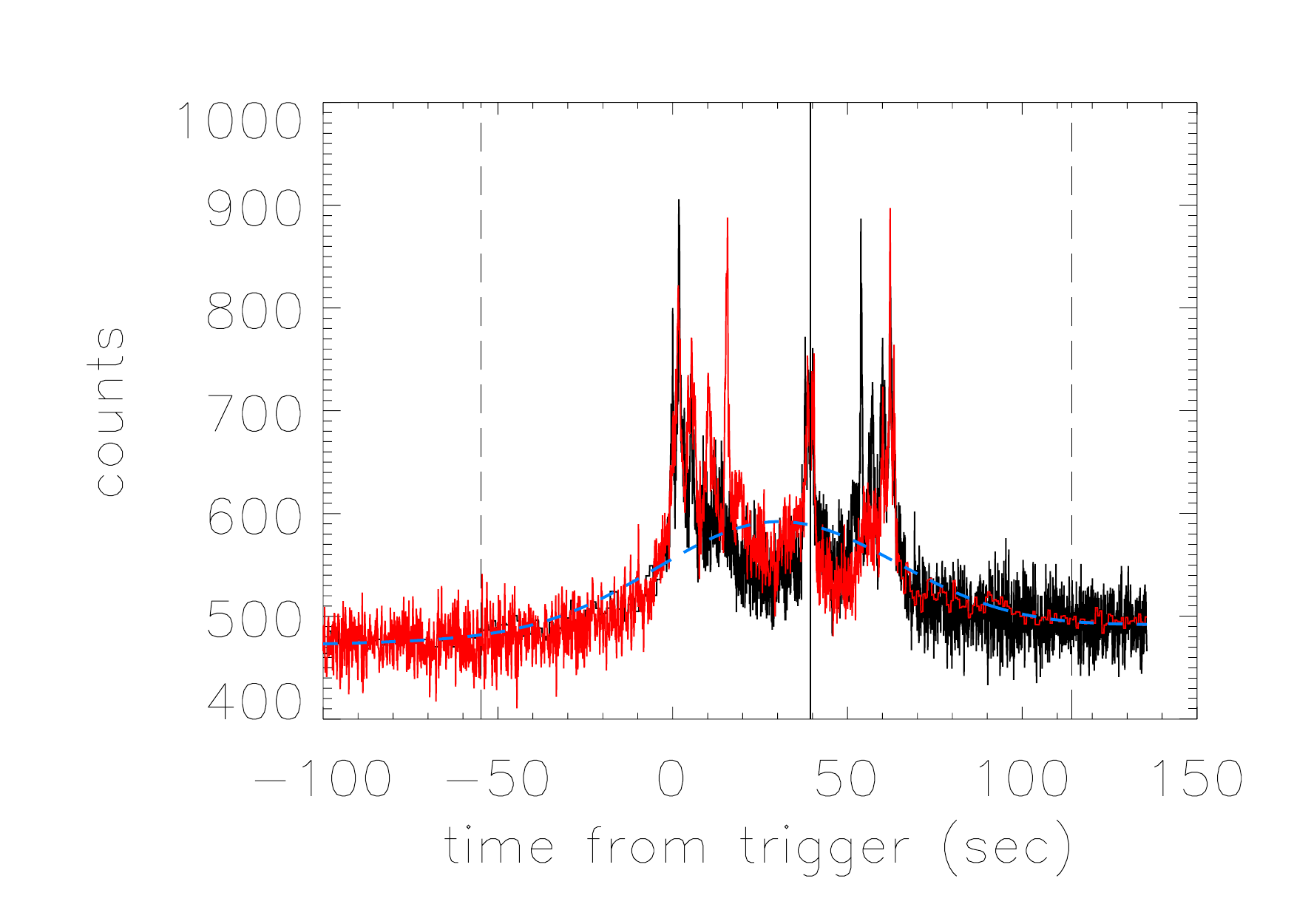}  & 
    \includegraphics[width=.5\textwidth]{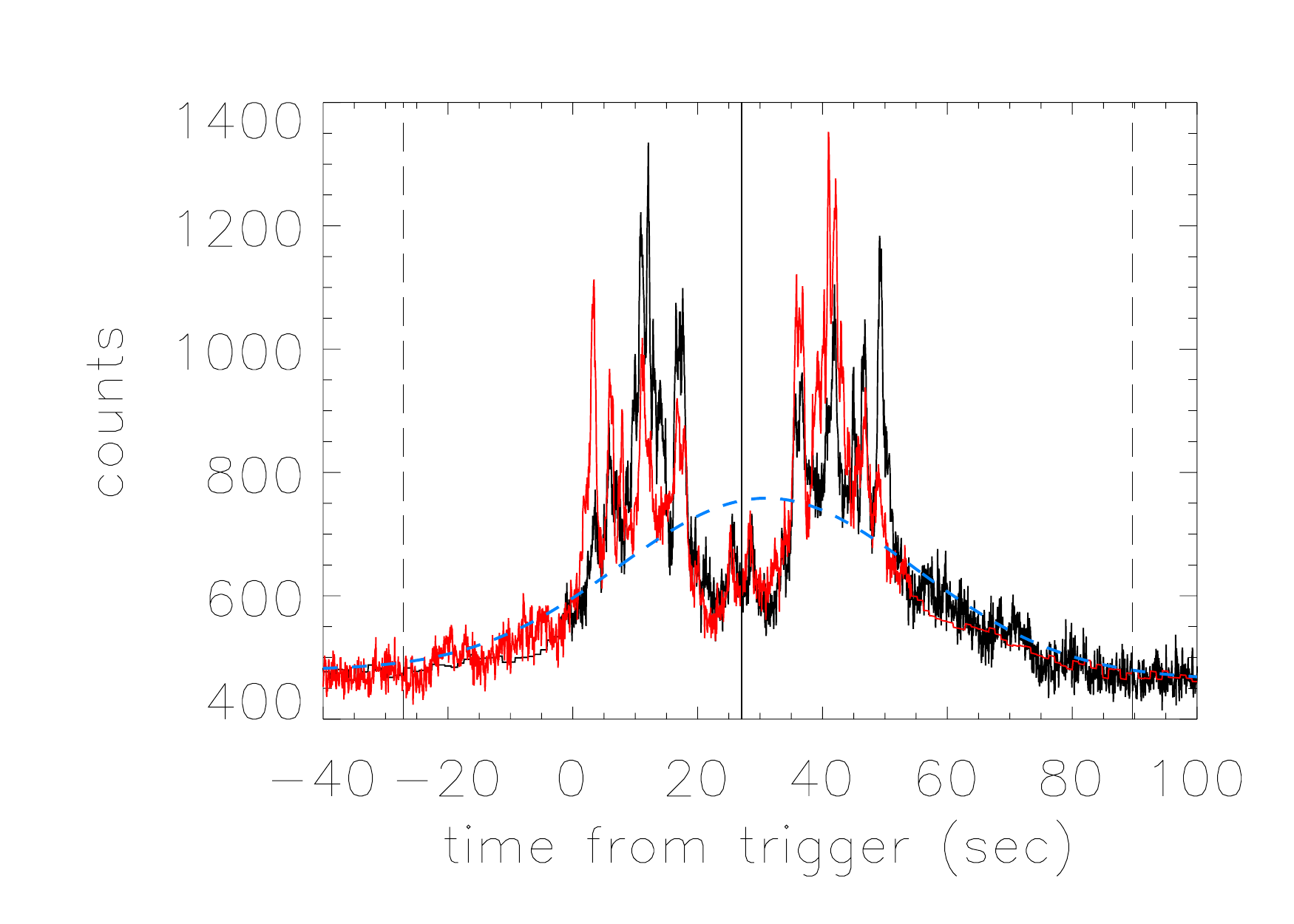} \\
    \includegraphics[width=.5\textwidth]{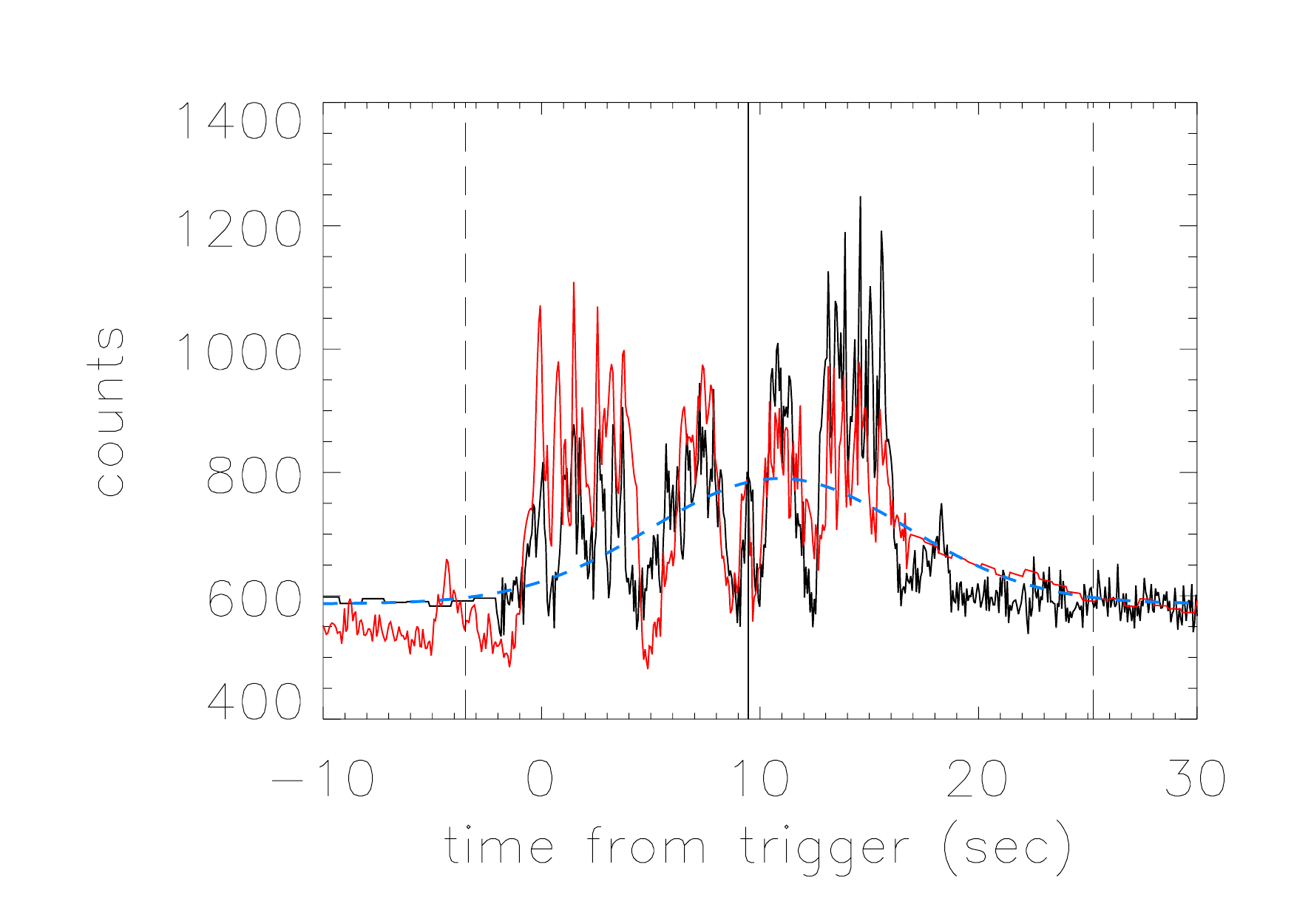}  & 
    \includegraphics[width=.5\textwidth]{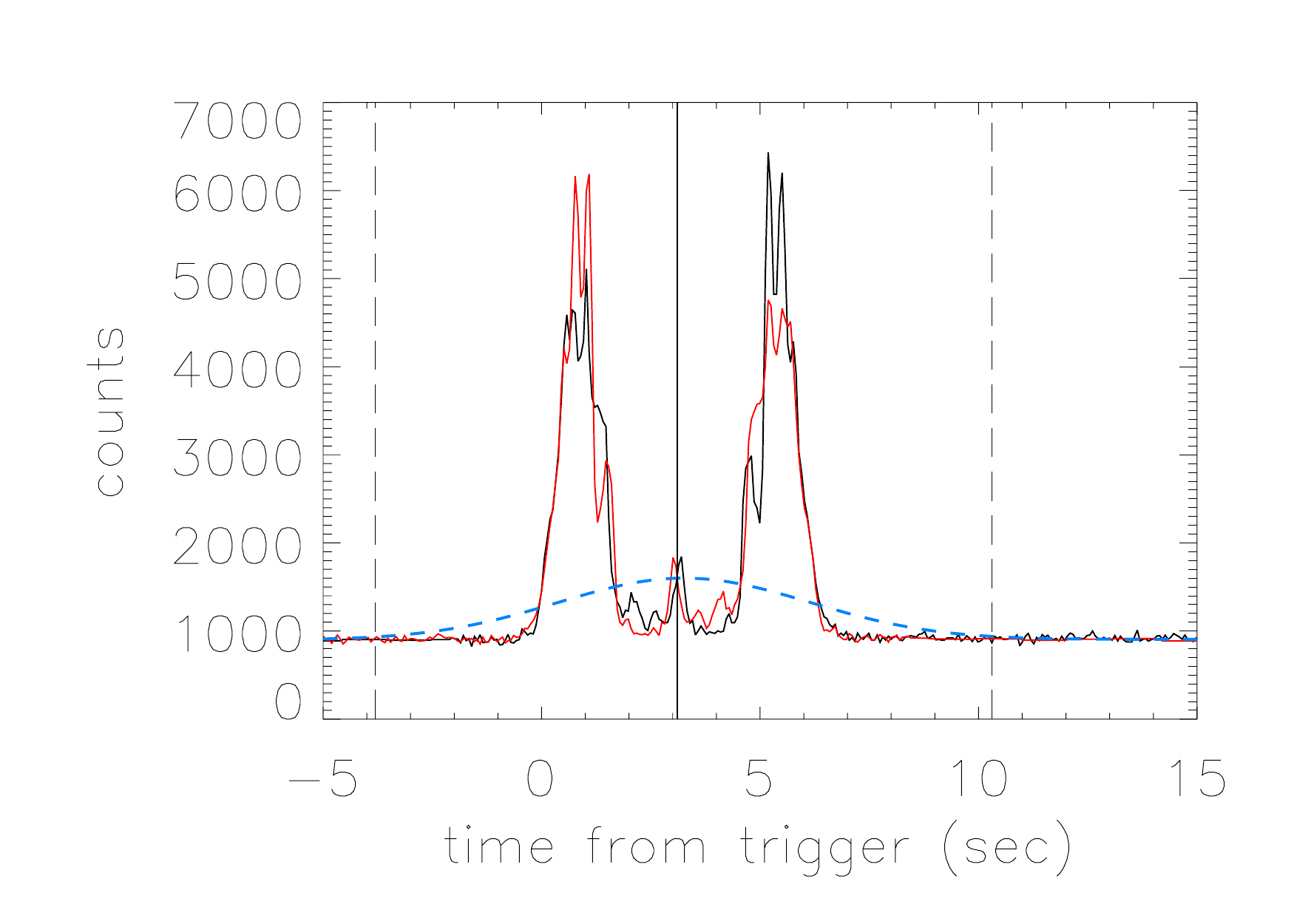}   \\
  \end{tabular}
  \caption{Temporally-symmetric models of u-pulses BATSE 121 (upper left), BATSE 130 (upper right), BATSE 160 (lower left) and BATSE 1711 (lower right). Shown are the counts data (black), the fit to the Norris/Gaussian model (blue dashed line), the time-reversed model (red), the duration window (vertical dashed lines), and the time of reflection (vertical solid line). \label{fig:fig14}}
\end{figure}

\begin{deluxetable*}{ccccccc}
\tablenum{6}
\tablecaption{Morphology distribution of BATSE GRB pulses having temporally-symmetric structure \label{tab:tab6}}
\tablewidth{0pt}
\tablehead{
\colhead{FRED} &  \colhead{rollercoaster}  &  \colhead{crown}  &  \colhead{u-pulse}  &  \colhead{asymmetric u-pulse} & \colhead{no pulse u-pulse} & \colhead{unclassified} 
}
\startdata
23\%  &  10\%  &  22\%  &  29\%  &  8\% &  3\% &  5\% \\
\enddata
\tablecomments{Percentage of GRB pulses with quality 2 described in Table \ref{tab:tab4}, characterized by morphological type.} 
\end{deluxetable*}

U-pulses exhibit offset times that are similar to those of FREDs and crowns (Figure \ref{fig:fig11}), but their durations are among the longest of all pulse types (Figure \ref{fig:fig12}). This perhaps accounts for why they have not previously been identified as a separate pulse class: their long durations, faint underlying pulses, and heavily-structured residuals make them potentially appear to observers as many short, spiky pulses. Their single-pulsed nature is only revealed by viewing them through the lens of temporal symmetry, and by binning their light curves at lower temporal resolutions. Examples of time-reversed models of u-pulses are shown in Figure \ref{fig:fig14}.

\subsubsection{Asymmetric u-pulses.}
A small fraction of asymmetric pulses exhibit rapidly-varying, time-symmetric residual structures. In some ways, these {\em asymmetric u-pulses} are like rollercoaster pulses in that the time of reflection occurs during the pulse decay, offset from the pulse peak. In other ways, they are like u-pulses, in that the flux is highly variable and drops close to the background rate near $t_{\rm 0; mirror}$.
    
\begin{figure}[htb]
\centering
  \begin{tabular}{@{}cc@{}}
    \includegraphics[width=.5\textwidth]{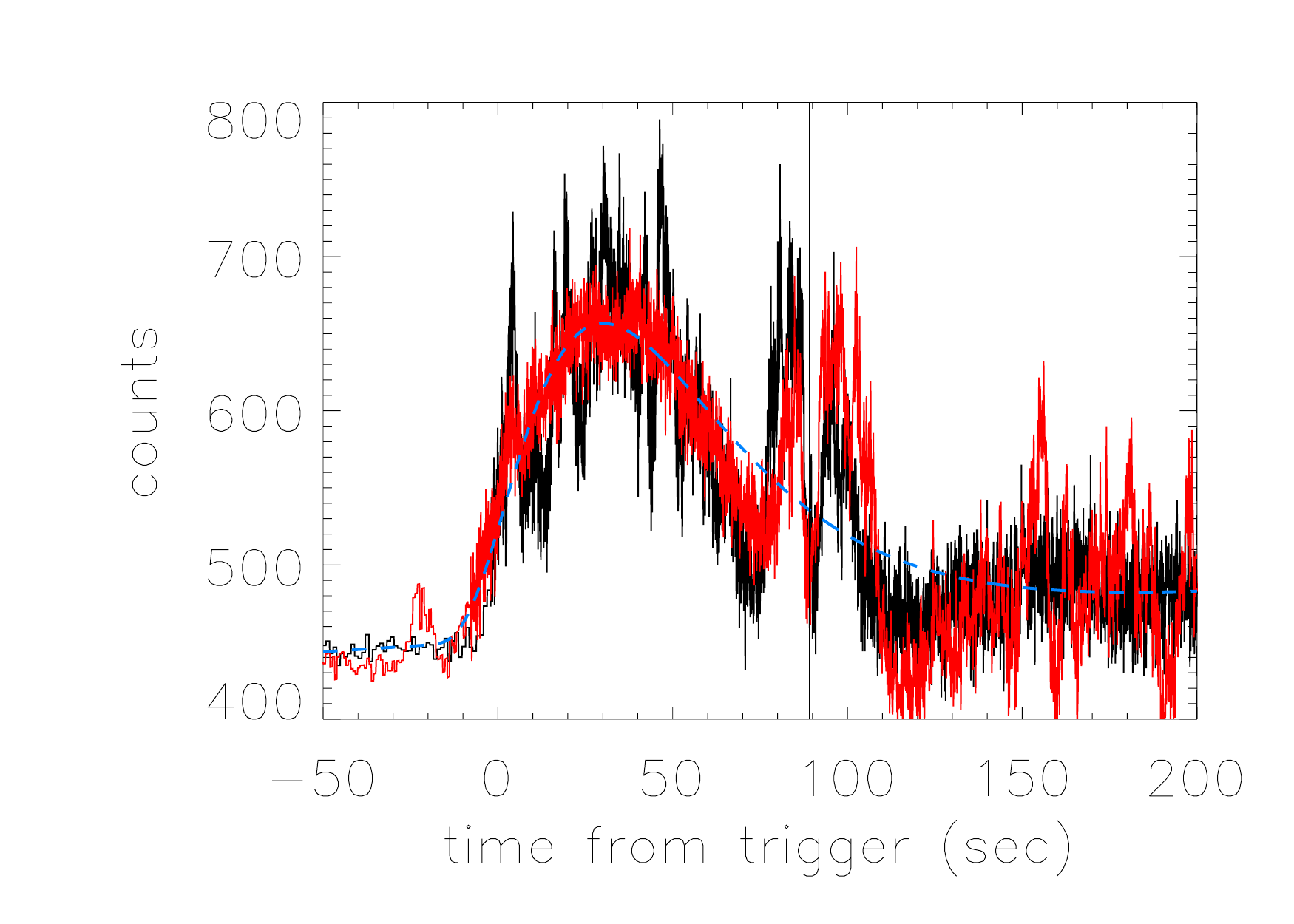}  & 
    \includegraphics[width=.5\textwidth]{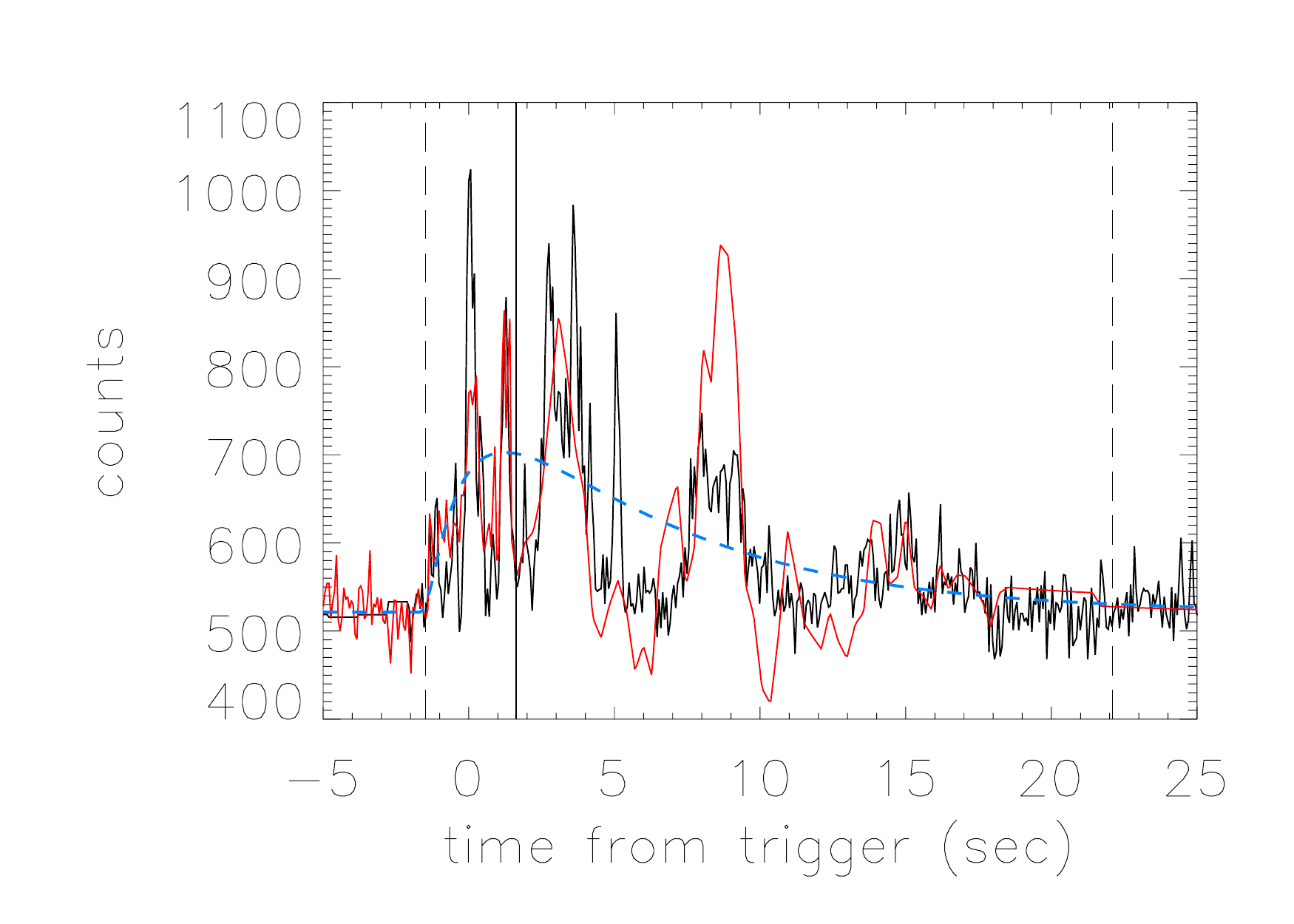} \\
    \includegraphics[width=.5\textwidth]{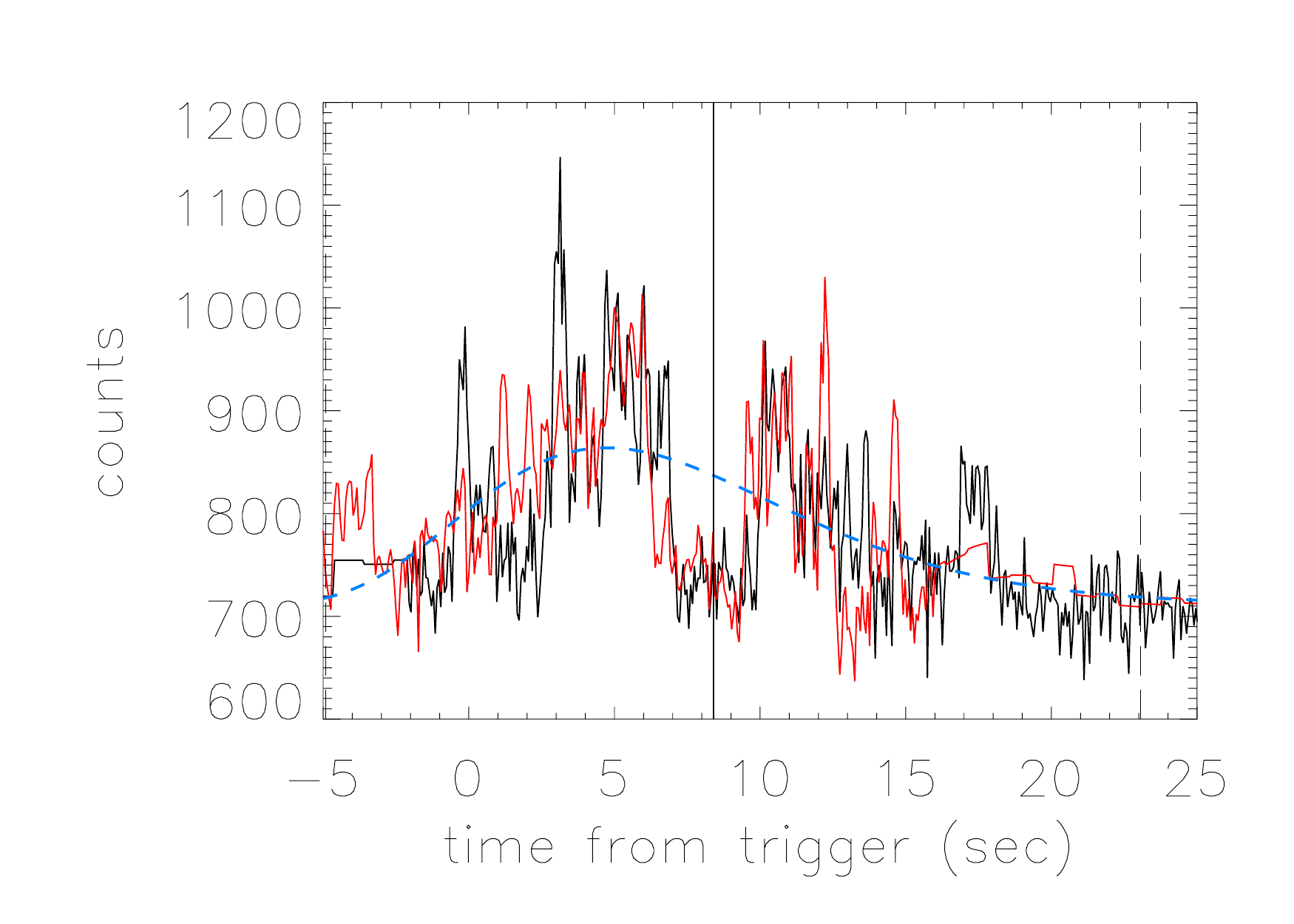}  & 
    \includegraphics[width=.5\textwidth]{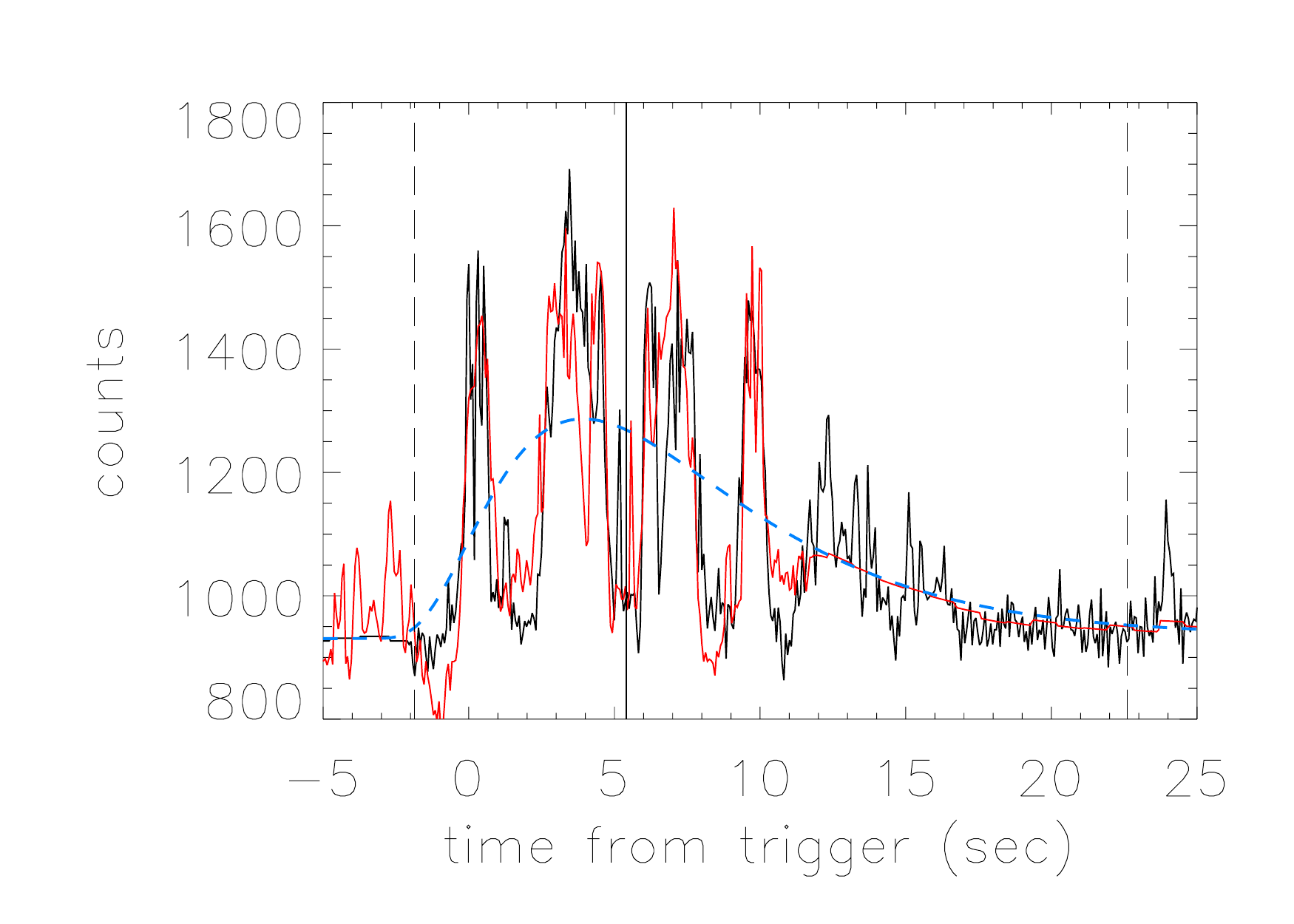}   \\
  \end{tabular}
  \caption{Temporally-symmetric models of asymmetric u-pulses BATSE 226 (upper left), BATSE 1291 (upper right), BATSE 1303 (lower left) and BATSE 1601 (lower right). Shown are the counts data (black), the fit to the Norris/Gaussian model (blue dashed line), the time-reversed model (red), the duration window (vertical dashed lines), and the time of reflection (vertical solid line).\label{fig:fig15}}
\end{figure}

The properties of asymmetric u-pulses are intermediary between those of rollercoaster pulses and those of FREDs. In the $s_{\rm mirror}$ vs.~$\kappa$ parameter space, asymmetric u-pulses are positioned between FREDs and rollercoaster pulses (Figure \ref{fig:fig8}). Here, the anti-correlation between  $s_{\rm mirror}$ and $\kappa$ is more pronounced than that found for FREDs and more similar to that found for rollercoaster pulses. Like rollercoaster pulses, asymmetric u-pulse offsets are greater than zero, but these offsets do not generally appear to be as large as those of rollercoaster pulses (Figure \ref{fig:fig11}). The durations of asymmetric u-pulses are among the longest of GRB pulse types (Figure \ref{fig:fig12}). Examples of time-reversed models of u-pulses are shown in Figure \ref{fig:fig15}.

\subsubsection{No-pulse u-pulses.}

As mentioned previously, it is difficult to identify the number of pulses contained in a structured GRB because bright structure is often indistinguishable from individual pulses. Examples of this can be seen in the early and late structure seen in BATSE pulses 7301p1 and 7301p2 (see figures 15 and 19 of \cite{hak18b}). Furthermore, it is difficult to uniquely identify the numbers of pulses in GRBs where separations between and durations of multiple emission episodes make the emission episodes look like structure instead of like pulses. However, some multiple-episodic structures do appear to comprise individual pulses. We call these episodes {\em no-pulse u-pulses} because they look like u-pulses for which the underlying monotonic component may be present but is too faint to measure. We have tentatively accepted these events as a separate pulse class even though they do not have underlying monotonic pulses.

\begin{figure}[htb]
\centering
  \begin{tabular}{@{}cc@{}}
    \includegraphics[width=.5\textwidth]{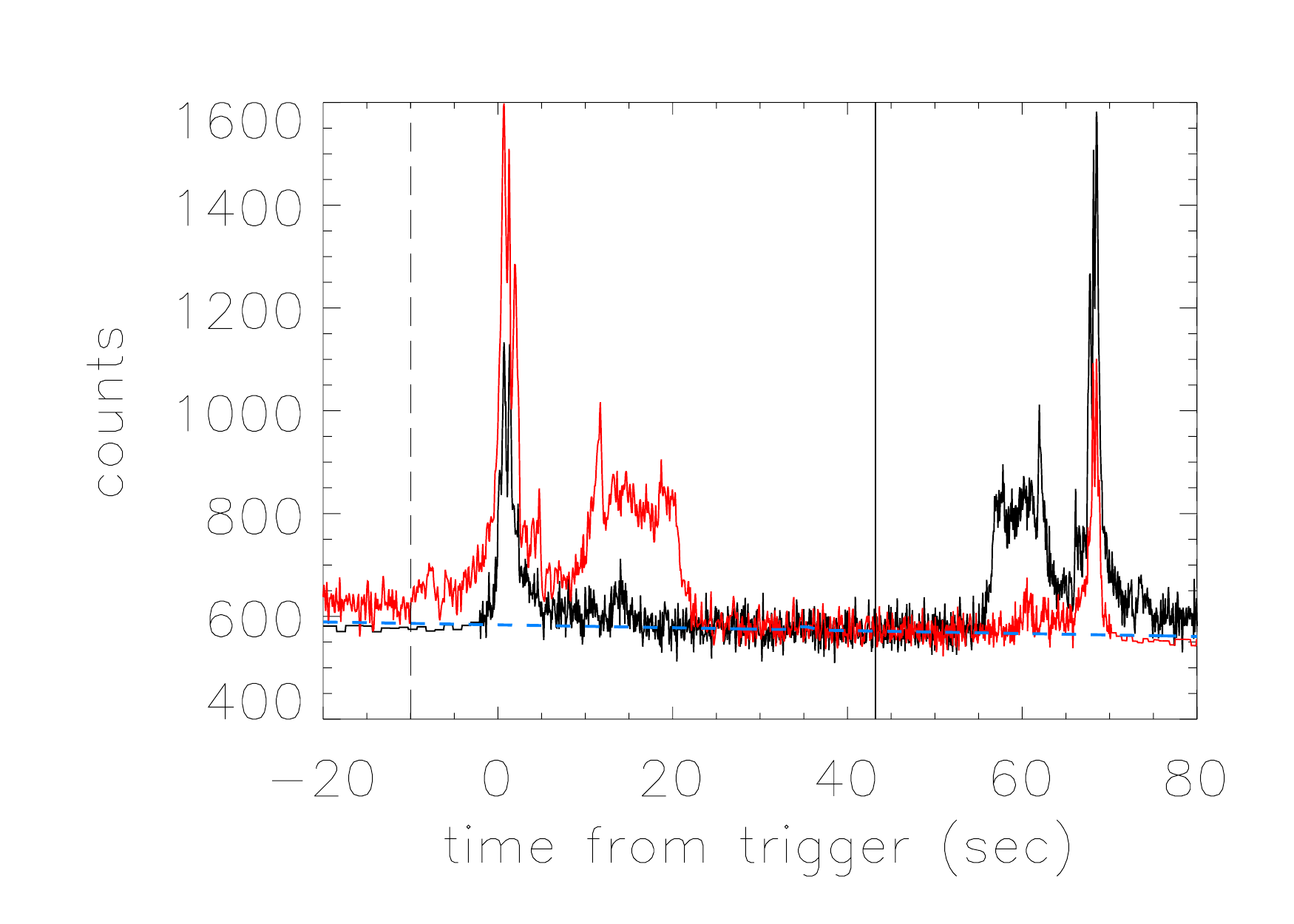}  & 
    \includegraphics[width=.5\textwidth]{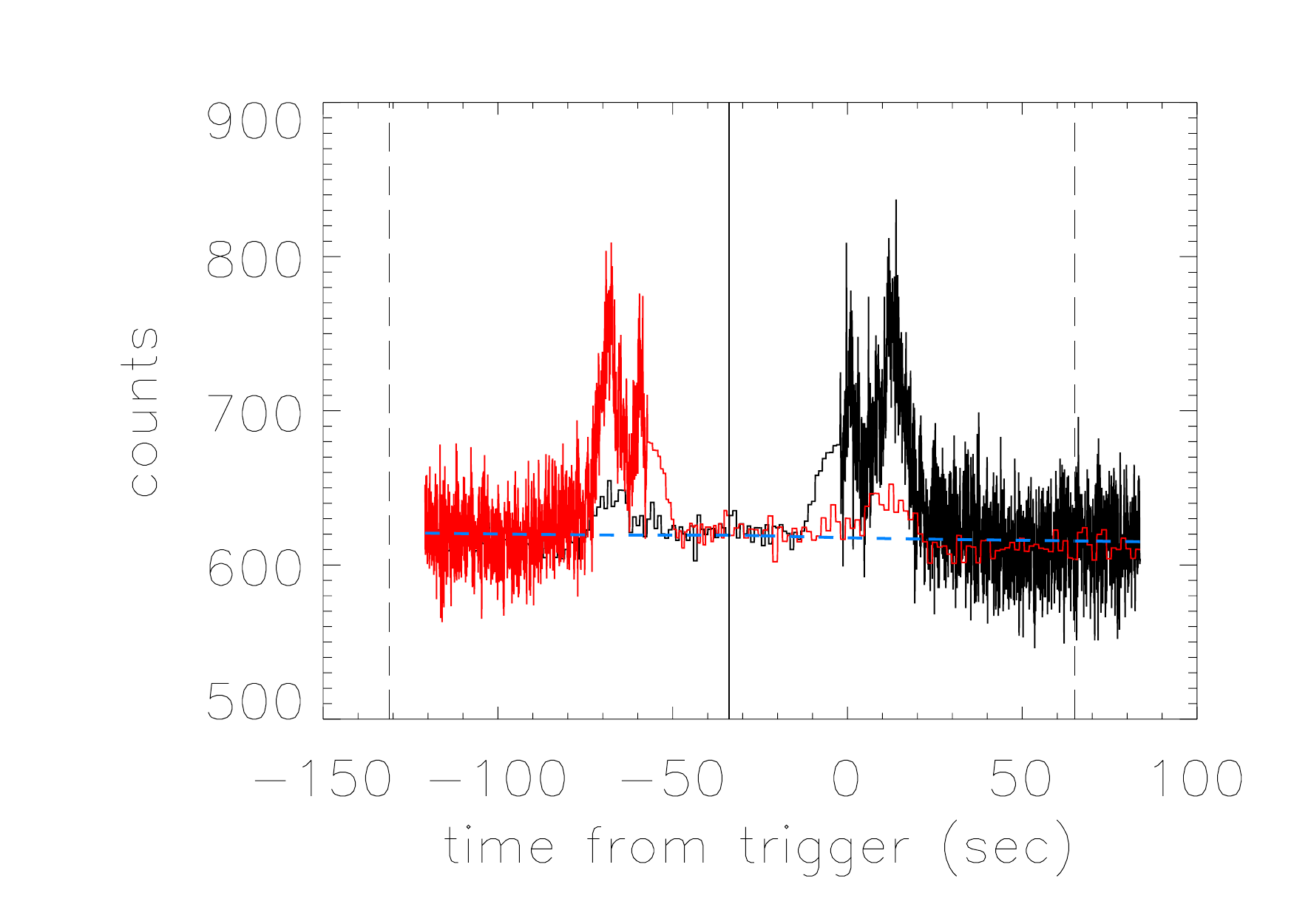} \\
    \includegraphics[width=.5\textwidth]{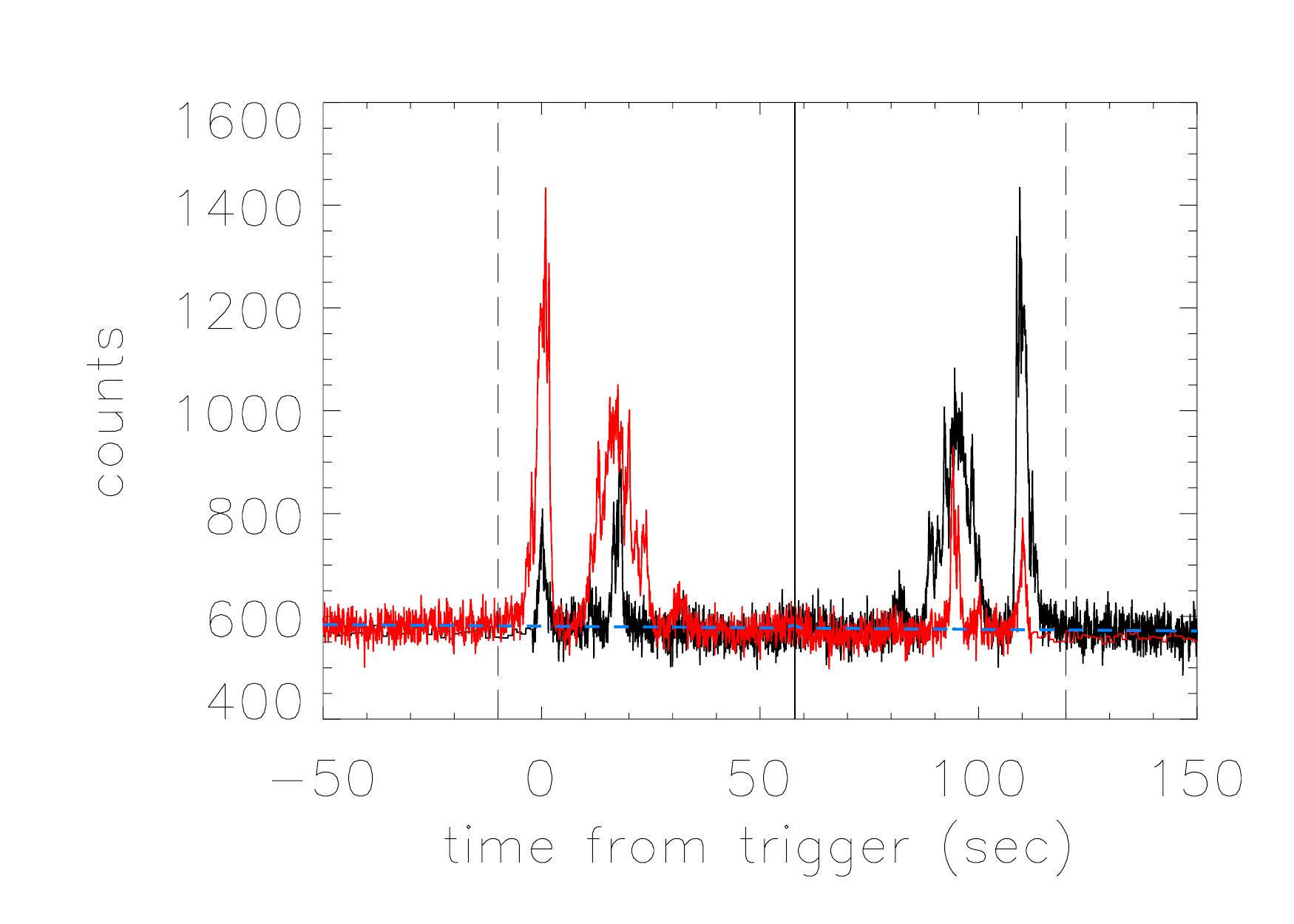}  & 
    \includegraphics[width=.5\textwidth]{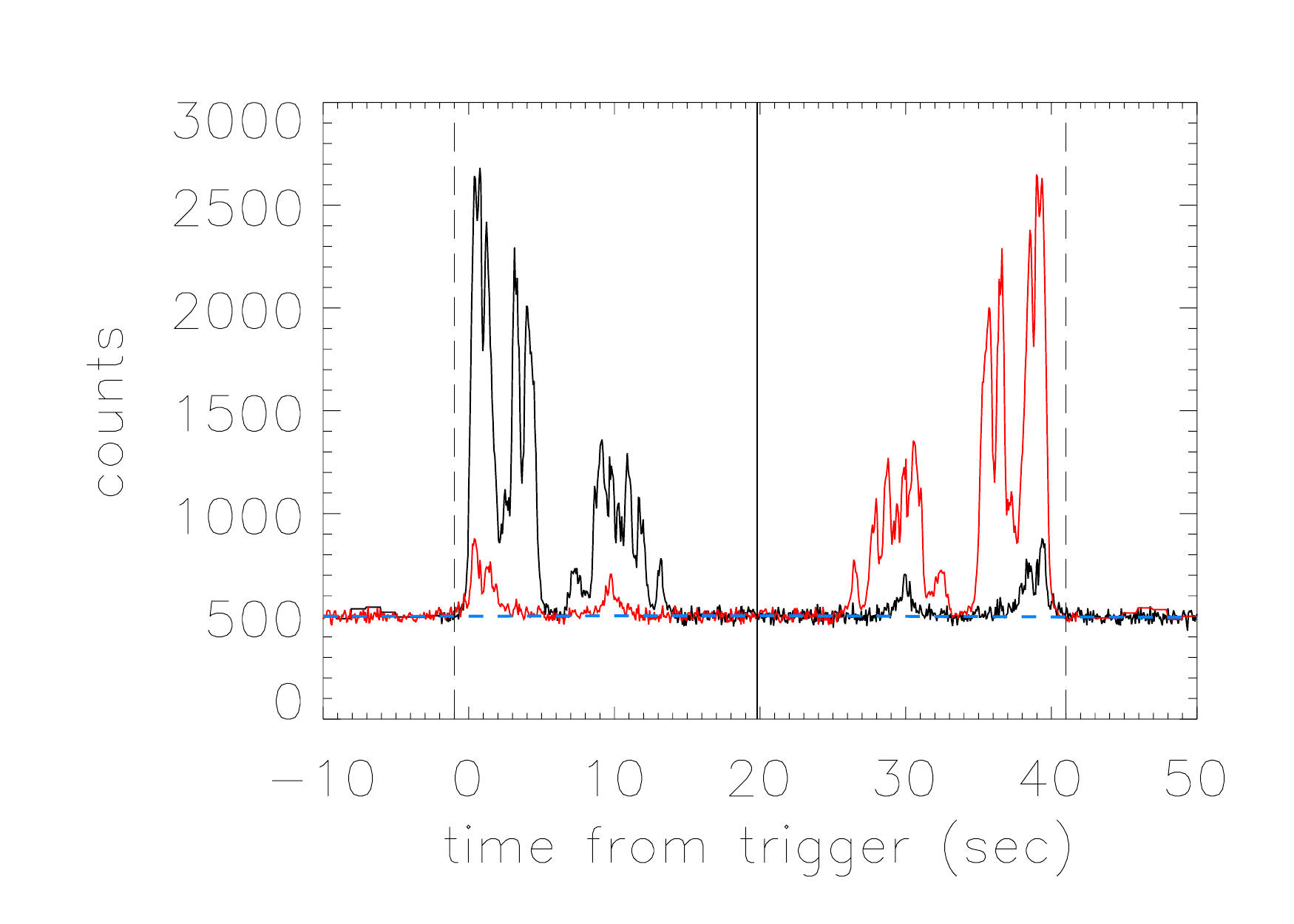}   \\
  \end{tabular}
  \caption{Temporally-symmetric models of no-pulse u-pulses BATSE 222 (upper left), BATSE 816 (upper right), BATSE 869 (lower left) and BATSE 2090 (lower right). Shown are the counts data (black), the fit to the background (blue dashed line), the time-reversed model (red), the duration window (vertical dashed lines), and the time of reflection (vertical solid line). \label{fig:fig16}}
\end{figure}

Since no-pulse u-pulses contain only structure and no measurable underlying monotonic pulse components, the asymmetry and offsets of these pulses cannot be ascertained. However, durations and stretching factors can be measured, and have these pulses have thus been included (as purple stars) in Figure \ref{fig:fig12}. Their $s_{\rm mirror}$ values and durations are indeed consistent with those of normal u-pulses. Examples of time-reversed models of no-pulse u-pulses are shown in Figure \ref{fig:fig16}.

\subsubsection{Unclassified pulses}
Only about 5\% of fitted pulses remain unclassified after placing pulses in the aforementioned categories. These are not problematic to the morphology scheme; they simply do not fit into the {\em ad hoc} morphologies defined in this manuscript. Some of these unclassified pulses have overlapping bumps on one side of the light curve but not on the others, suggesting either faint additional pulses or non-temporally symmetric structure. Others have time-symmetric structures or monotonic pulse components that are difficult to characterize given their low signal-to-noise ratios. It is possible that all unclassified pulses lack recognizable morphologies simply because of the small number of GRBs studied: with a larger pulse sample, additional pulse morphologies might be identified.


\subsubsection{Unfitted pulses}
Roughly 14\% of pulses have monotonic components and/or structure that cannot be fitted by the models. There are two main reasons why these GRBs might not have been fitted: 1) they have such complex structures that the number of pulses cannot be easily ascertained ({\em complex pulses}), or 2) they have slow-rise, rapid-decay asymmetries that cannot be characterized by either the \cite{nor05} model or the Gaussian model ({\em Crescendo pulses}).

\begin{itemize}

\item Complex pulses. The characteristics of these events cannot be determined because pulses and pulse structures are strongly intertwined. In the BATSE 64 ms data, complex pulses tend to be associated with very long, structured GRBs. Examples of unfitted complex pulses are shown in Figure \ref{fig:fig17}.

\begin{figure}[htb]
\centering
  \begin{tabular}{@{}cc@{}}
    \includegraphics[width=.5\textwidth]{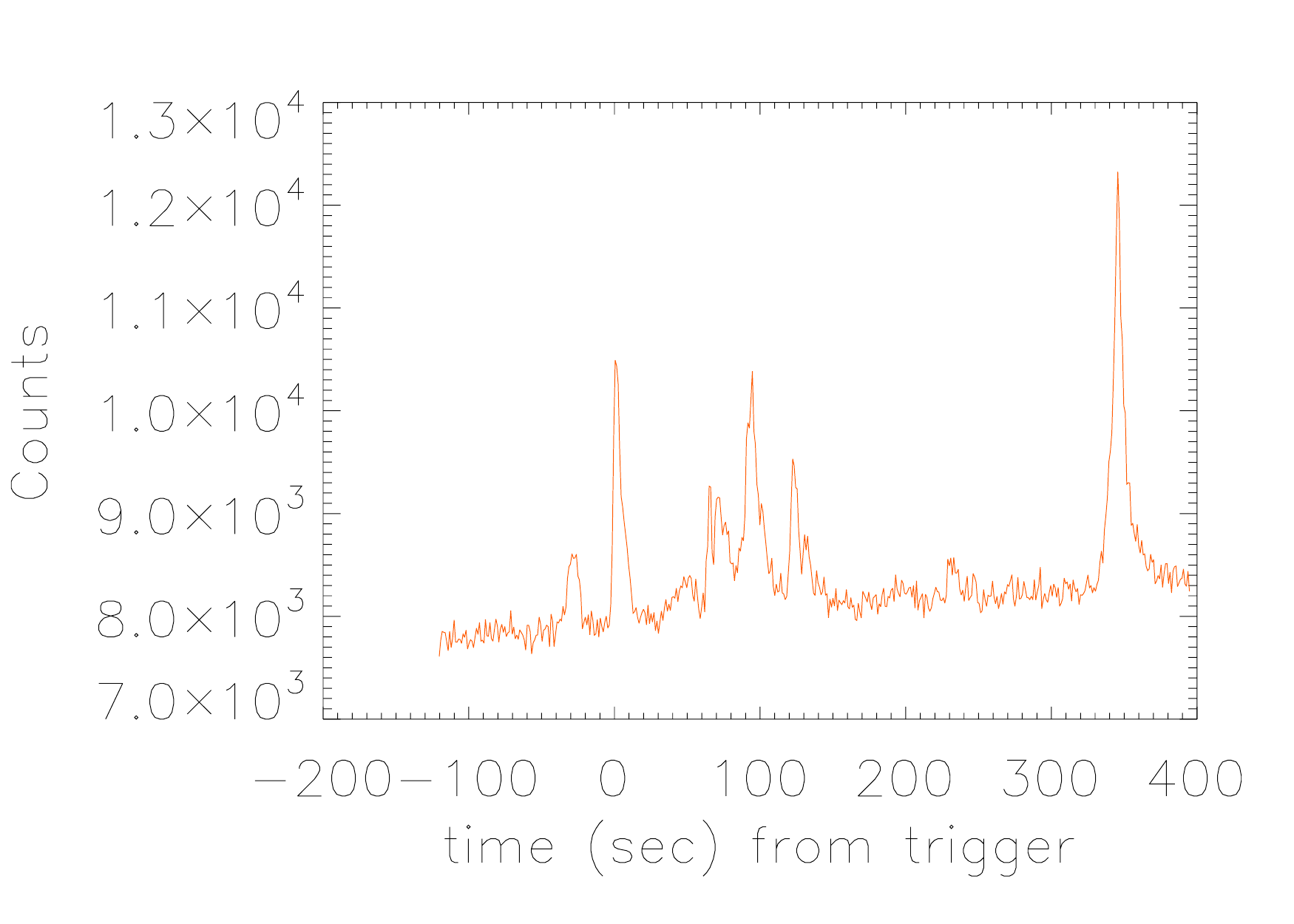}  & 
    \includegraphics[width=.5\textwidth]{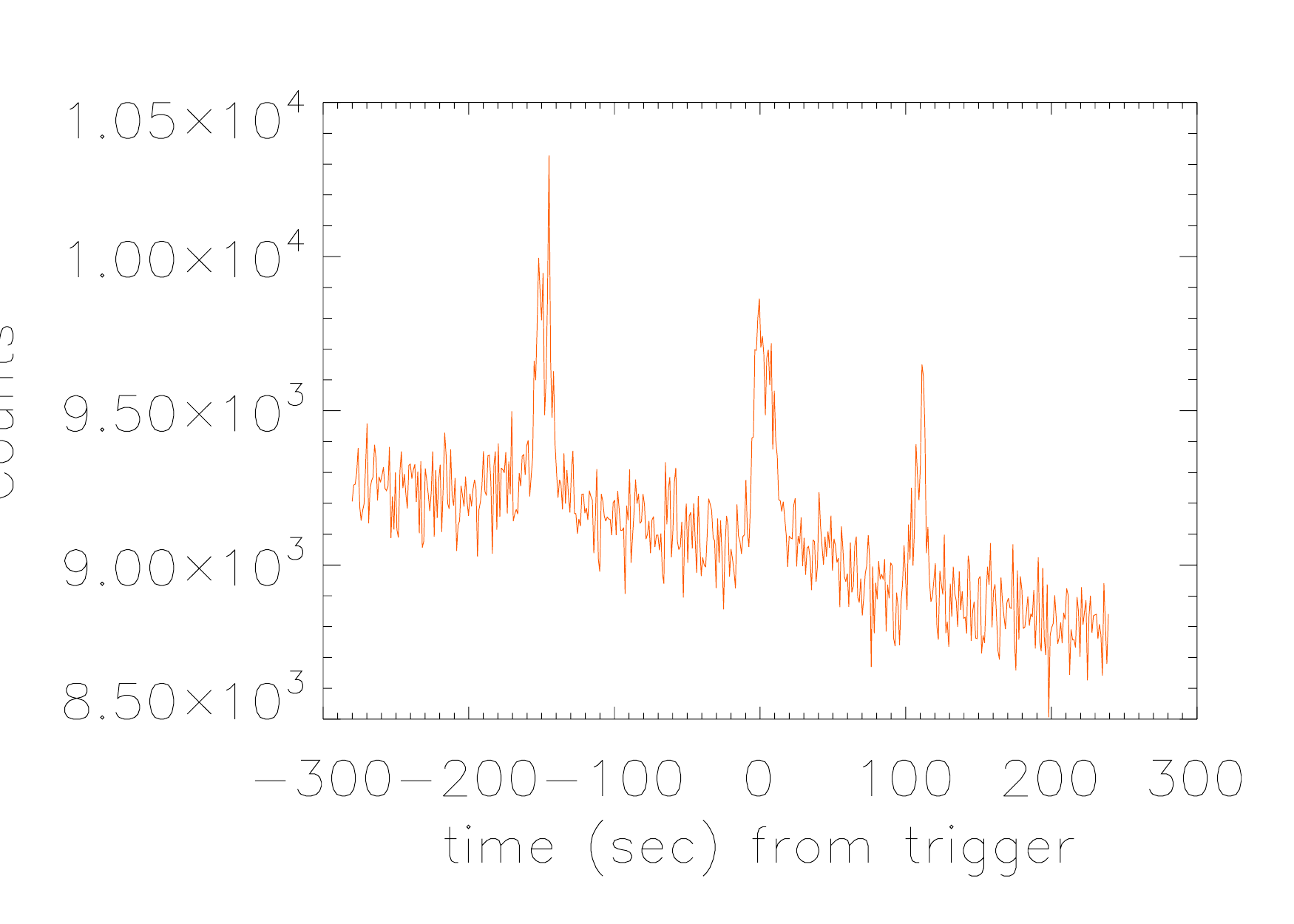} \\
    \includegraphics[width=.5\textwidth]{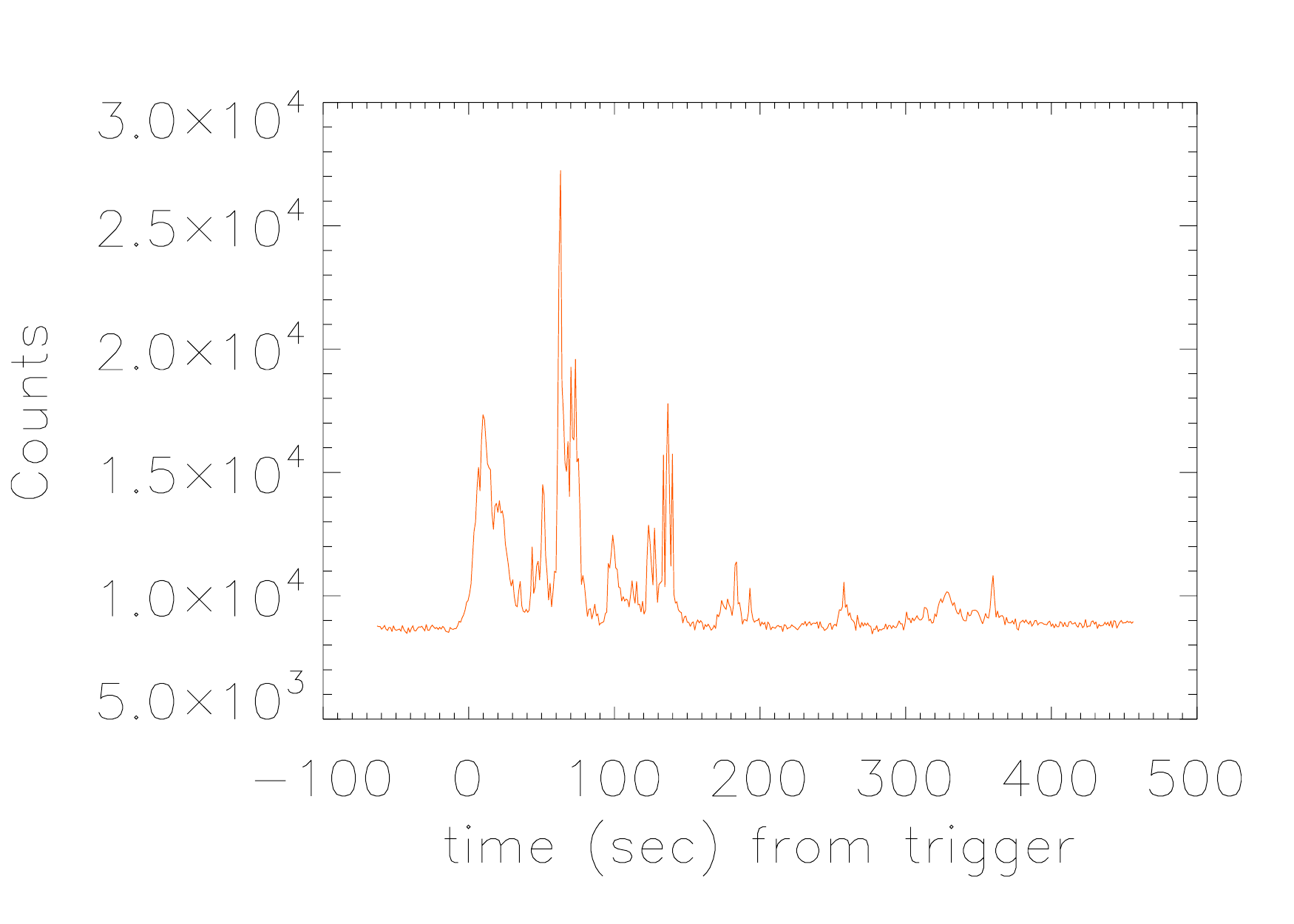}  & 
    \includegraphics[width=.5\textwidth]{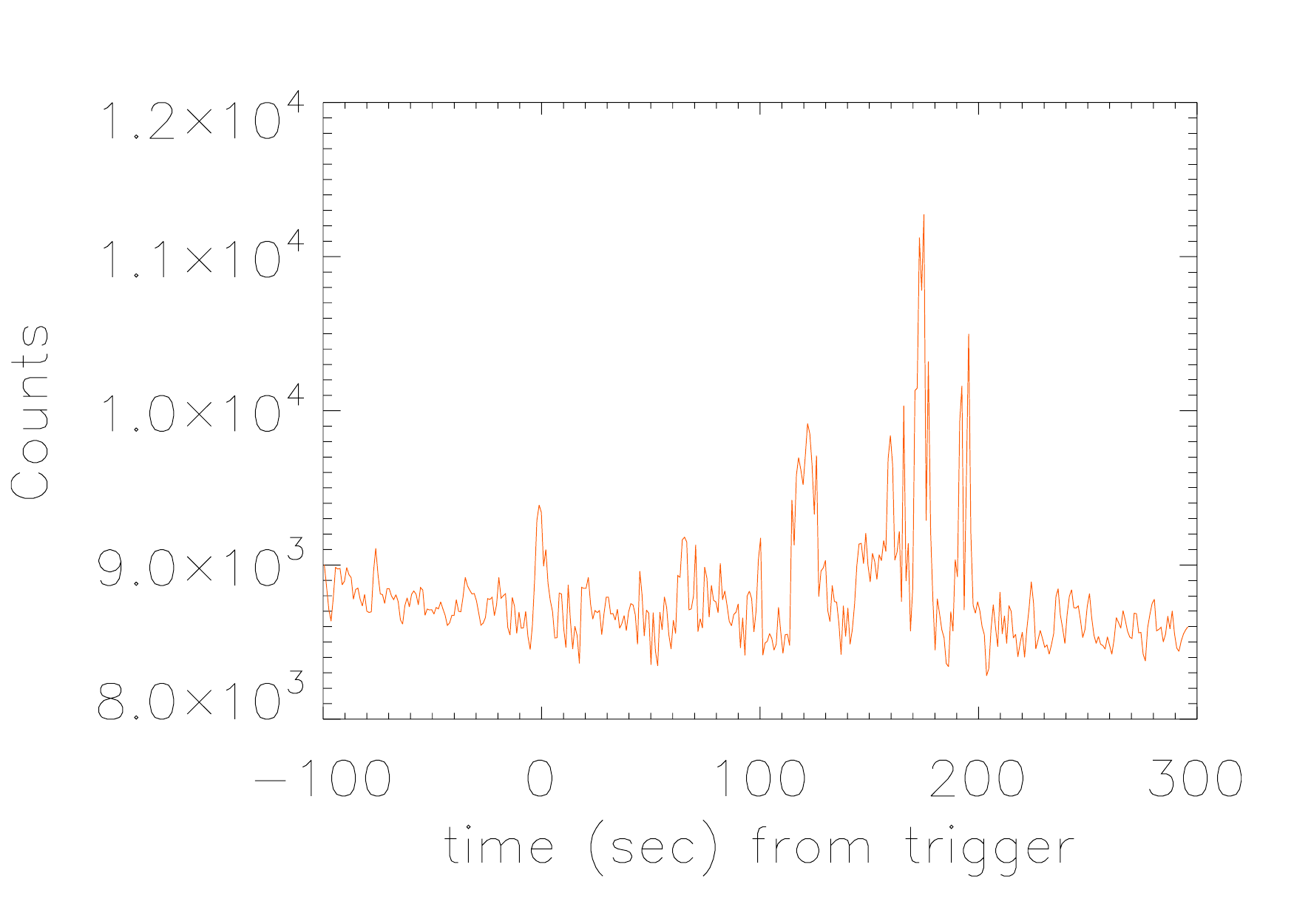}   \\
  \end{tabular}
  \caption{Light curves of unfitted complex pulses BATSE 1009 (upper left), BATSE 1152 (upper right), BATSE 1288 (lower left) and BATSE 1365 (lower right). \label{fig:fig17}}
\end{figure}

\item Crescendo pulses. Following the naming convention of \cite{hak18a}, crescendo pulses are characterized by a gradual emission buildup (often in the form of smaller, connected emission episodes) followed by a rapid decay; they have slow-rise, rapid-decay light curves that are asymmetric in a way that cannot be captured by the \cite{nor05} model. Examples of unfitted crescendo pulses are shown in Figure \ref{fig:fig18}.

\begin{figure}[htb]
\centering
  \begin{tabular}{@{}cc@{}}
    \includegraphics[width=.5\textwidth]{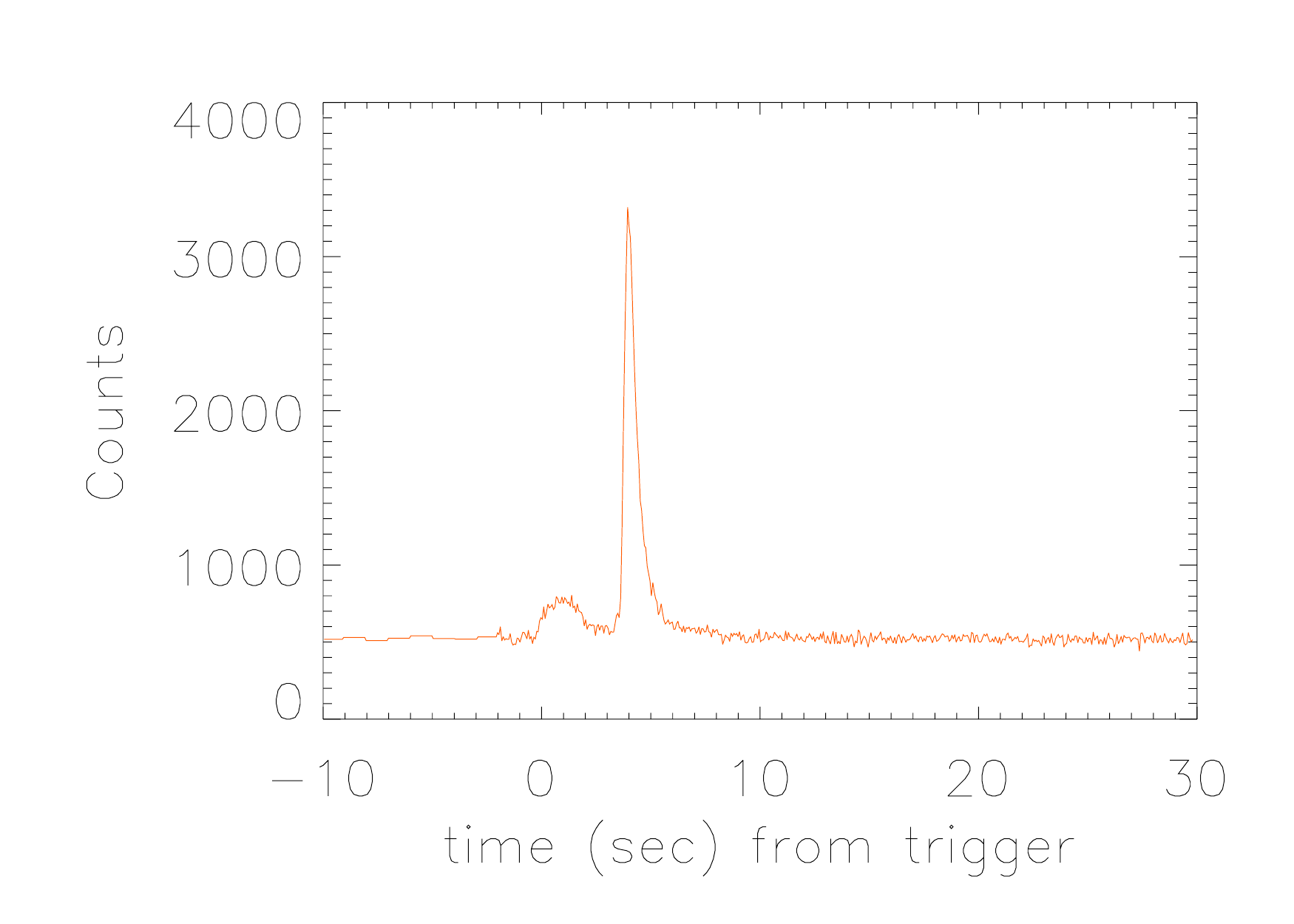}  & 
    \includegraphics[width=.5\textwidth]{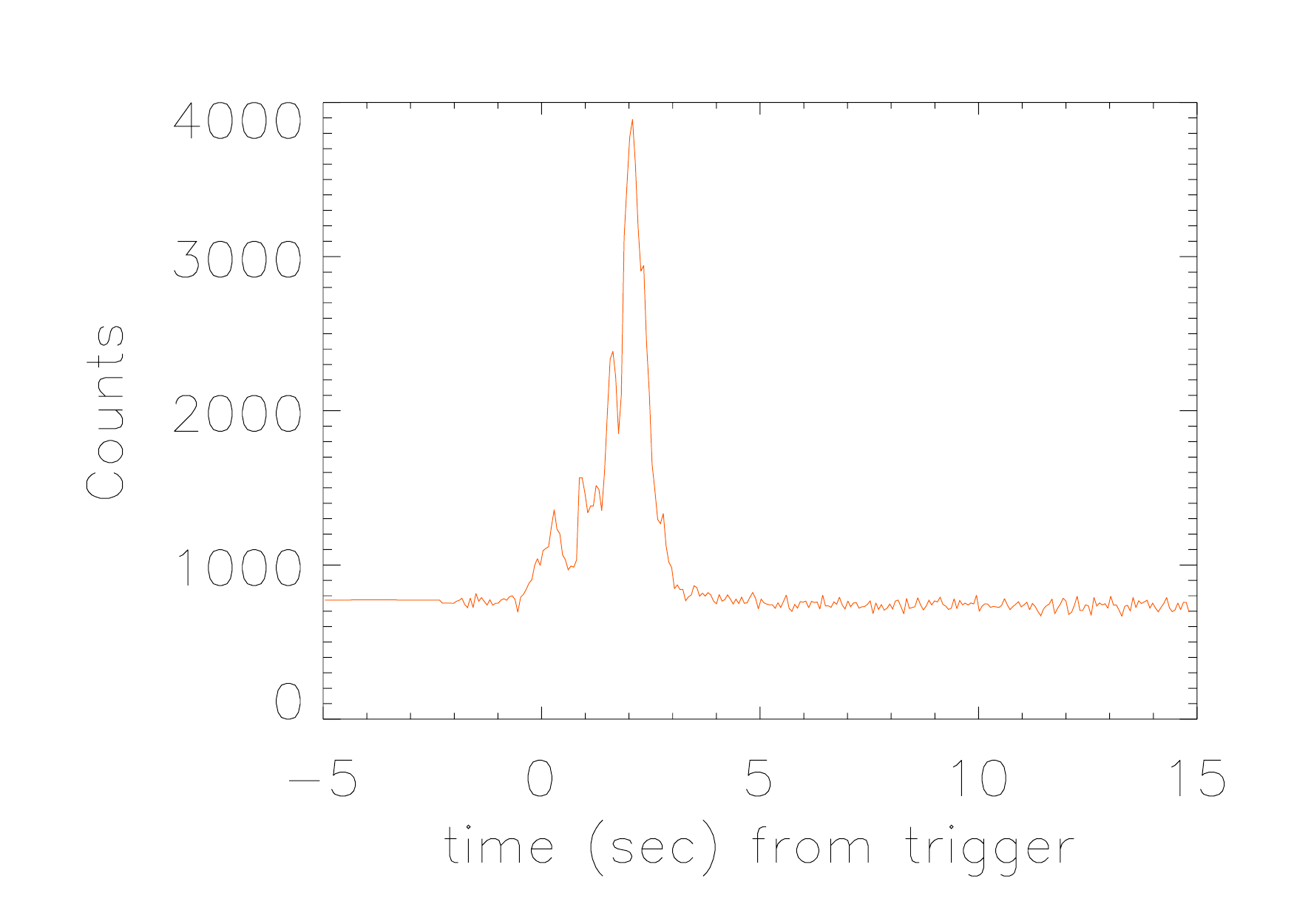} \\
    \includegraphics[width=.5\textwidth]{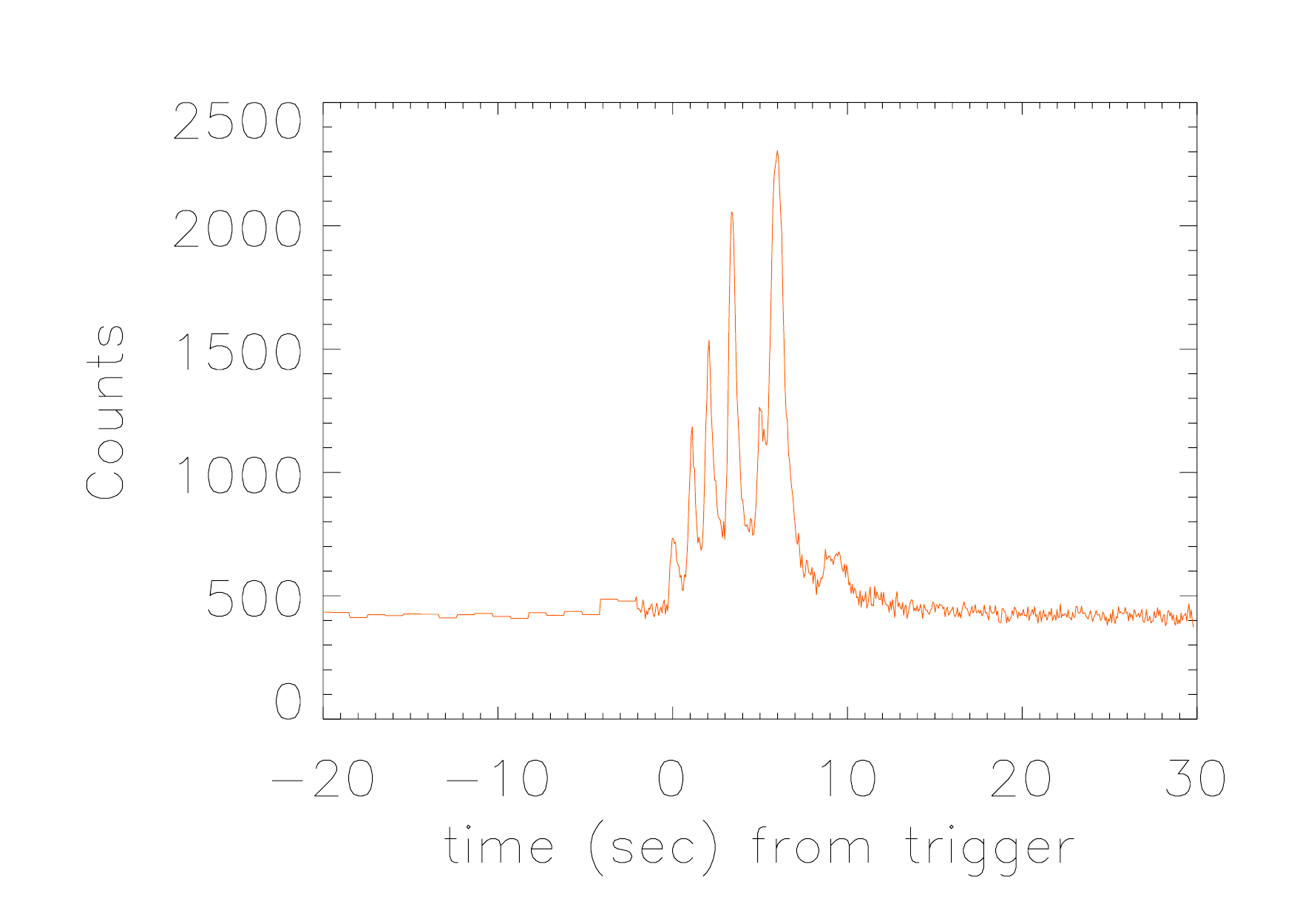}  & 
    \includegraphics[width=.5\textwidth]{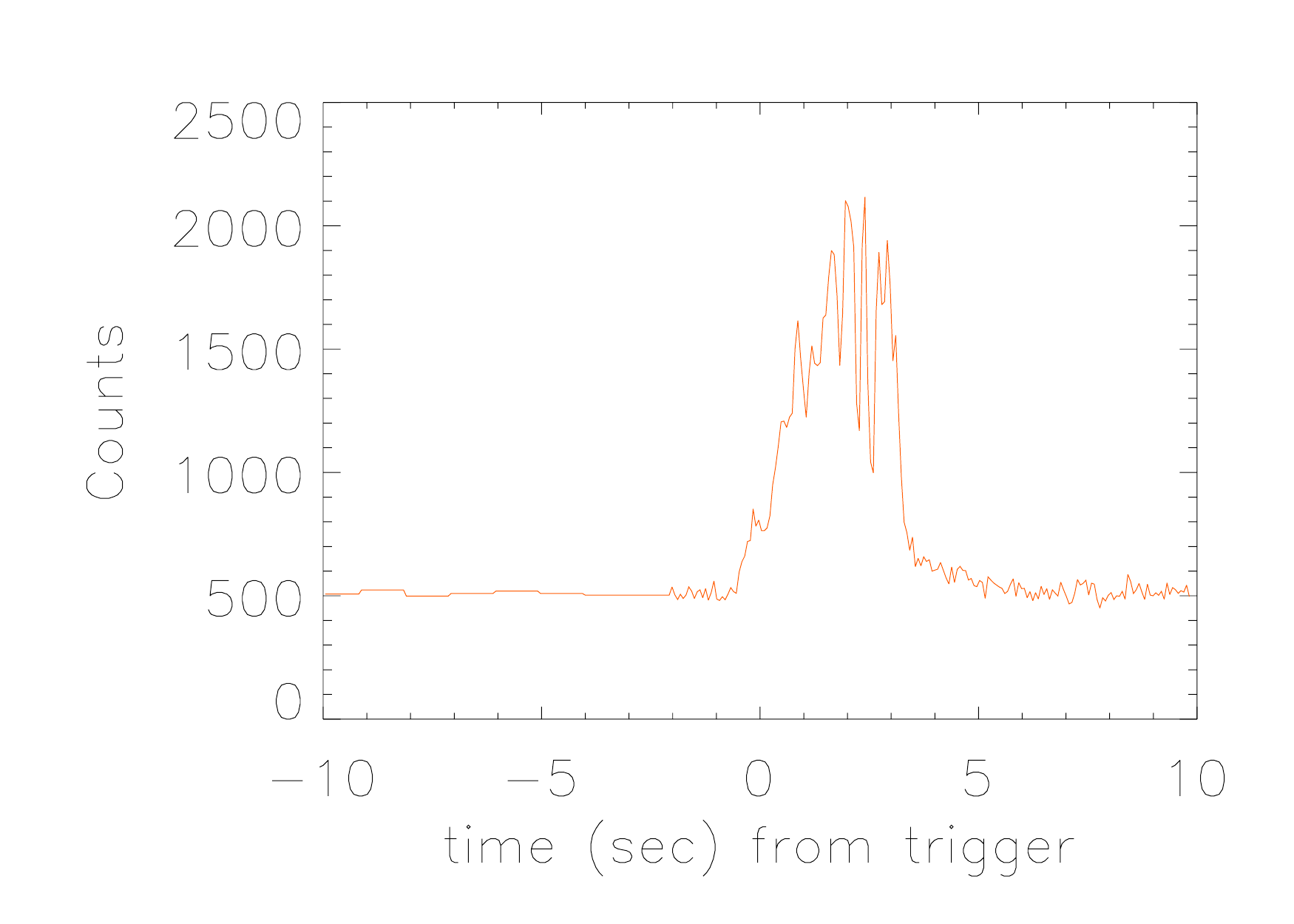}   \\
  \end{tabular}
  \caption{Light curves of unfitted crescendo pulses BATSE 999 (upper left), BATSE 1025 (upper right), BATSE 1425 (lower left) and BATSE 1683 (lower right). \label{fig:fig18}}
\end{figure}

\end{itemize}

\subsection{Delineating GRB pulses on the basis of their properties}

Even though these GRB pulse morphologies have been identified by the appearances of their light curves, the general bulk properties of the pulses in each group differ. This can be seen in Figures \ref{fig:fig8}, \ref{fig:fig11}, and \ref{fig:fig12}, and in the description of the centroids of these distributions in Table \ref{tab:tab7}. As described in the previous sections, the symmetric pulse types (crowns and u-pulses) have larger $s_{\rm mirror}$ values, smaller $\kappa$ values, and smaller offsets than the asymmetric pulse types (FREDs, rollercoaster pulses, and asymmetric u-pulses). Pulse durations typically increase from FREDs to crowns and rollercoaster pulses to u-pulses and asymmetric u-pulses.

\begin{deluxetable*}{lcccc}
\tablenum{7}
\tablecaption{Pulse Morphological Properties \label{tab:tab7}}
\tablewidth{0pt}
\tablehead{
\colhead{Morphology}  &  \colhead{$s_{\rm mirror}$}  &  \colhead{$\kappa$} &  \colhead{offset}  &  \colhead{log(duration~(s))}
}
\startdata
FRED  &  $0.37 \pm 0.22$  & $0.72 \pm 0.19$ & $0.08 \pm 0.22$  & $1.32 \pm 0.36$ \\
rollercoaster pulses & $0.74 \pm 0.33$ & $0.73 \pm 0.12$ & $0.15 \pm 0.10$  & $1.38 \pm 0.65$  \\
asymmetric u-pulses & $0.75 \pm 0.38$ & $0.69 \pm 0.20$ & $0.15 \pm 0.17$ & $1.68 \pm 0.42$  \\
u-pulses & $1.09 \pm 0.32$ & $0.01 \pm 0.02$ & $0.01 \pm 0.07$ & $1.68 \pm 0.42$ \\
crown pulses & $1.13 \pm 0.48$ & $0.05 \pm 0.09$ & $0.00 \pm 0.09$ & $1.28 \pm 0.49$ \\
\enddata
\tablecomments{Mean properties of successful and similar pulses described in Table \ref{tab:tab4}.} 
\end{deluxetable*}

Considerable overlap between morphological groups suggests that the visual pulse classification scheme explains most but not all of the scatter in Figure \ref{fig:fig8}. As a check on the efficacy of our morphological types, we apply a J48 decision tree \citep{qui93} as implemented in the WEKA 3.8 suite of data mining tools \citep{fra16} to 112 of the high-quality FREDs, rollercoaster pulses, u-pulses, crowns, and asymmetric u-pulses described in Table \ref{tab:tab5}. J48 applies its top-down, recursive, divide-and-conquer strategy to assign pulses to pre-defined classes, requiring that each IF-THEN-ELSE assignment statement contains at least two pulses. J48 is able to correctly classify 63\% of the pulses using only asymmetry, offset, duration, and $s_{\rm mirror}$ values. The success of this classification approach can be characterized by a Cohen's kappa statistic (which compares the J48 performance to that of random guesses that could be made according to the frequency of each class) value of 0.5, indicating that this is a ``moderate" reproduction of the visual classification scheme \citep{lan77}. J48's confusion matrix (shown in Table \ref{tab:tab8}) demonstrates that some morphological types overlap one another more than others. For example, J48 classifies 37 of 38 u-pulses correctly, while the other one is classified as a crown. On the other hand, only 5 crowns are classified correctly, with 22 being classified as u-pulses, and the other one being classified as an asymmetric u-pulse. Besides confusing crowns and u-pulses, J48 also confuses asymmetric u-pulses with rollercoaster pulses. These results suggest that close similarities exist between u-pulses and crowns, and between asymmetric u-pulses and rollercoaster pulses. On the other hand, symmetric pulse morphologies (crowns and u-pulses) are rarely confused with asymmetric pulse types (FREDs, rollercoasters, and asymmetric u-pulses), and {\em vice versa}. Overall, the J48 classification is supportive of the morphological class explanation, given measurement uncertainties coupled with the {\em ad hoc} visual classification scheme.

\begin{deluxetable*}{lccccc}
\tablenum{8}
\tablecaption{J48 Confusion Matrix \label{tab:tab8}}
\tablewidth{0pt}
\tablehead{
\colhead{classified as $->$}  &  \colhead{a}  &  \colhead{b} &  \colhead{c}  &  \colhead{d}  &  \colhead{e}
}
\startdata
a = asymmetric u-pulse   &  5 &  1  &  1 &  6 & 0  \\
b = crown  &  1 &  5  &  0 &  0 & 22  \\
c = FRED  &  1 &  0  &  13 &  4 & 0  \\
d = rollercoaster  &  3 &  0  &  2 &  9 & 0  \\
e = u-pulse  &  0 &  1  &  0 &  0 & 37  \\
\enddata
\end{deluxetable*}


\subsubsection{Properties of GRB pulses in short and long bursts}
All GRBs extracted from BATSE TTE data have durations shorter than 2 s (the duration of the TTE duration window), and thus all belong to the class of short hard GRBs. The extracted TTE pulses all have durations shorter than 1 s, whereas the shortest long GRB pulse has a duration of around 3 s. Thus, the $s_{\rm mirror}$ vs.~duration distribution of short GRB pulses might well be a parallel distribution to that of the long GRB pulses. By expanding our sample of short burst pulses to include measurements with larger uncertainties, we find evidence supporting this idea. The longest duration short pulses tend to be crown and u-pulses, similar to long GRB pulses. Likewise, short FRED and rollercoaster pulses tend to have the smallest $s_{\rm mirror}$, similar to long GRB pulses. These observations support the idea that short and long GRBs are produced by similar mechanisms occurring on different timescales.

\subsubsection{Symmetric vs.~asymmetric GRB pulse properties}

A gap exists between $s_{\rm mirror}$ vs.~$\kappa$ properties of symmetric and asymmetric pulses (Figure \ref{fig:fig10}). Crowns and u-pulses have symmetries less than $\kappa < 0.3$, whereas FREDs, rollercoaster pulses, and asymmetric u-pulses have $\kappa > 0.3$. This delineation is consistent with the results of the J48 classification, strongly suggesting that symmetric and asymmetric pulses are intrinsically different from one another. This delineation is less pronounced when fainter pulses are included in the sample; at low S/N ratios, everything looks like a monotonic bump that can be called a FRED.


\subsubsection{Spiky vs.~smooth GRB pulse properties}

The data in Table \ref{tab:tab7} suggest that spiky pulses ({\em e.g.,} those containing many structured episodes) have longer durations than smooth pulses. Spiky pulses include u-pulses, no pulse u-pulses, and asymmetric u-pulses. Smooth pulses are FREDs, rollercoaster pulses, and crowns.


\subsubsection{Shapes of time-reversed residual structures}

We cannot definitively state that ``temporal symmetry" includes not only the ordering of structures in a pulse, but also the shapes of those structures (for example, a FRED-shaped structure prior to the time of reflection might show up as a DERF-shaped structure following the time of reflection). We do occasionally observe features appear to exhibit these characteristics, but we cannot verify this due to the noisiness of GRB light curves, and due to the gradual replacement of high-energy photons in GRB pulses with lower-energy ones. Because high energy BATSE photons contain so much more average energy than low energy photons, this replacement over time gives many structural components low-energy ``tails," making it hard to unambiguously identify DERF-shaped structures. High temporal- and spectral-resolution GRB data may be needed to accurately address this issue.

\section{Discussion} \label{sec:theory}

Time series analysis of GRB photon counts data is a useful way to study the sequential ordering of GRB photons and the timescales on which GRB emission occurs: it provides direct and important information about GRB jet kinematics. In contrast, spectral analysis of photon energy distributions is the best tool for studying GRB emission mechanisms, but the time-ordering of photons is generally of secondary importance in these studies. Approaches have been used to reconcile time series and spectral analyses without the use of special spectroscopy detectors; \cite{cri99,pre16} deconvolve photon models through DRMs (detector response matrices) to account for the non-proportional relationship between the actual and recorded energies of detected photons. Excellent approaches such as this have their own advantages and limitations: they provide useful information about spectral evolution at the cost of losing some spectral and temporal resolution.

The time series approach used in this study has allowed temporally symmetric GRB pulse features to be extracted. We see no evidence that these features are instrumental in nature, even though our photon counts analysis has not directly accounted for photon downscattering in the DRMs.  First, the smooth wavy features seen in FRED residuals have been found in GRB pulses observed by a variety of instruments ({\em e.g.,} BATSE, Swift, Suzaku, and GBM). Second, an effect capable of causing small-amplitude deviations like the ones in FREDs is unlikely to explain the hard, late, bright, spiky features seen in u-pulses and asymmetric u-pulses or the multi-peaked features observed in crowns and rollercoaster pulses. Finally, the residual structures found in all pulse morphologies are consistent with temporal symmetry: there is no known instrumental bias that can cause structure to repeat in a time-reversed fashion.

GRB prompt emission is thought to result from the relativistic ejection of material accompanying stellar black hole formation \citep{ree94}. General mechanisms proposed for prompt emission include synchrotron shocks \citep{ree94,kob97,dai98}, photospheric emission ({\em e.g.}, \cite{pac86a,goo86,pee08,bel11}), and/or magnetic reconnection \citep{spr01,dre02,gia05,zha11}. Synchrotron shocks occur in optically thin regions and models predict smoothly-varying light curves for which pulse duration reflects variations in the intensity of the central engine \citep{kob97,dai98,pan99}. Photospheric emission is generally expected to produce smoothly-varying light curves due to the quasi-thermal nature of the spectrum produced when the ejecta becomes transparent to its own radiation. However, these light curves can have more rapidly-varying, non-thermal components by adding low-energy synchrotron tails and/or high-energy comptonized tails ({\em e.g.}, \cite{tho94,ree05,pee06,bel10,ryd11}). Magnetic reconnection can produce rapidly-varying light curves characterized by nonthermal or quasi-thermal spectra \citep{spr01,dre02,gia05,zha11}; this process is favored in optically-thin regions where shocks cannot develop and where magnetization is large.

The temporal pulse properties identified in this study place important temporal constraints on GRB emission mechanisms. First, GRB prompt emission is generally confined to a small number of emission episodes, and each of these episodes exhibits some form of temporally-symmetric structure. It is surprising that the light curves of gamma-ray bursts exhibit temporal symmetry since the mechanisms responsible for producing these high-energy astrophysical sources are believed to be associated with relativistic, non-reversible, non-equilibrium mass outflow processes. The implication is that events in the jet cause some part of the pulsed emission to be repeated in reverse order: either the emission mechanism has to account for this through some sort of kinematic behavior of the jet \citep{hak18b}, or a separate emission mechanism is needed to explain the time-reversed part of the light curve \citep{hak19}. 

The variety of GRB pulse morphologies identified in this study adds a new twist to the problem. Time-symmetric GRB pulses appear to exhibit different types of structures than time-asymmetric GRB pulses, and pulse morphology appears related to both symmetry and duration.

\section{Conclusions and future work} \label{sec:results}

Temporal symmetry is a ubiquitous and perhaps defining characteristic of GRB pulse light curves.  Accepting that pulses can exhibit similar forward and time-reversed light curve components, without requiring that these components are necessarily identical to each other, has allowed us to more closely explore temporal symmetry characteristics. At least 86\% of the GRB pulses in our BATSE sample having bright structure exhibit temporally-symmetric characteristics; this percentage exceeds 90\% if we ask how many GRB pulses (including faint ones) have residuals that are simply consistent with temporally-symmetric structure.

We have demonstrated that temporally-symmetric GRB pulses can be fitted by a monotonic pulse component coupled with a time-symmetric structural component. Recognizing the existence of temporal symmetry allows us to redefine what GRB pulses are and to conclude that they are rare events: a burst typically contains only a few structured pulses rather than many small independent fluctuations. There is a strong correlation between the asymmetries of the structure and the monotonic pulse component, but individual pulses can have a large scatter relative to this relationship. Most of this scatter is accounted for by the pulse's light curve morphology. GRB pulse morphological types can be delineated into general groups based on their $\kappa$, $s_{\rm mirror}$, duration, and offset values. We have defined some of the more prevalent morphological groups as FREDs, rollercoaster pulses, crowns, u-pulses (with their relatives, no-pulse u-pulses), and asymmetric u-pulses. These groups can explain the light curves of roughly 86\% of all GRB pulses. Even unfitted and unclassified pulses exhibit similarities among their light curves, suggesting that a few less common pulse morphological types still await discovery. 

The development of a reliable GRB pulse classification system opens up many new potential avenues in GRB research. For example, how does luminosity vary across GRB pulse morphologies? What physical and/or environmental conditions produce different GRB morphologies? How closely related are spectral evolution and pulse morphology? What role does spectral evolution have in determining how similar pre-$s_{\rm mirror}$ structure is to post-$s_{\rm mirror}$ structure? What morphological relationships exist between pulses in multi-pulsed GRBs? Do x-ray flares exhibit morphologies similar to prompt GRB pulses? What physical mechanisms and/or physical conditions lead to temporal morphology in GRB pulses? What constraints are placed on GRB models by the existence of pulse temporal symmetry?

In the 1990's, BATSE's W. S. Paciesas notably summarized the inability of astrophysicists to find recognizable, repeatable patterns in gamma-ray burst (GRB) prompt emission by stating, ``Show me one gamma-ray burst, and I'll show you one gamma-ray burst." The identification of GRB pulse morphological classes may mean that it is time to update this as, ``Show me one gamma-ray burst pulse, and I’ll show you how it is related to other gamma-ray burst pulses.”

\acknowledgments

We thank Fahn Hakkila for her infinite patience while this project was slowly carried out at home during the Covid-19 pandemic. We acknowledge Thomas Cannon for coining the term ``crowns." We are grateful to Tim Giblin, Amy Lien, Rob Preece, and Bob Nemiroff for reviewing early drafts of this manuscript. Finally, we are indebted to the anonymous referee for suggestions that greatly improved the quality and clarity of the manuscript.

%



\appendix

\section{Effects of Signal-to-Noise Ratio on Pulse Structure}

GRB pulse structure becomes indistinguishable from instrumental noise at low signal-to-noise ratios \citep{hak18b}.  Our understanding of how noise affects our ability to characterize GRB pulse structure can be improved via the use of signal-to-noise ratio measures. The measures we use are the residual statistic ($R$), the signal-to-noise ratio ($S/N$), and the cross correlation function ratio ($CCF~ratio$). We define $S/N$ \citep{hak18b} as
\begin{equation}\label{eqn:SN}
S/N = (P_{t}-B)/\sqrt P_{t}
\end{equation}
where $P_{t}$ is the peak counts measured on the t-millisecond timescale (t=64 for BATSE 64 ms data and t=4 for binned BATSE TTE data) and $B$ is the mean background count measured on the same timescale. 

As described in Section \ref{sec:procedure}, the temporal symmetry parameters $t_{\rm 0; mirror}$ and $s_{\rm mirror}$ are found through application of the CCF to the pulse residuals. The CCF ($\rm CCF_{resids}$) assesses the strength of the correlation between the time-forward component of the residuals prior to $t_{\rm 0; mirror}$ and the stretched and time-reversed component following $t_{\rm 0; mirror}$.  It does not provide information about how the time-reversed and stretched model matches the complete GRB pulse light curve; this information can be better summarized by using the CCF to compare the time-reversed light curve to the light curve ($\rm CCF_{model}$). Both $\rm CCF_{resids}$ and $\rm CCF_{model}$ can be used to characterize temporal symmetry in GRB pulses, but both have limitations.  Faint pulses can produce large $\rm CCF_{resids}$ values due to chance alignments of random background fluctuations. Bright pulses with no residual structure can produce large $\rm CCF_{model}$ values because the monotonic pulse itself is time symmetric (its decay can be considered to be a time-reversed and stretched version of its rise). We introduce the ``CCF ratio" to overcome these limitations.

\begin{figure}[htb]
\centering
  \begin{tabular}{@{}cccc@{}}
    \includegraphics[width=.23\textwidth]{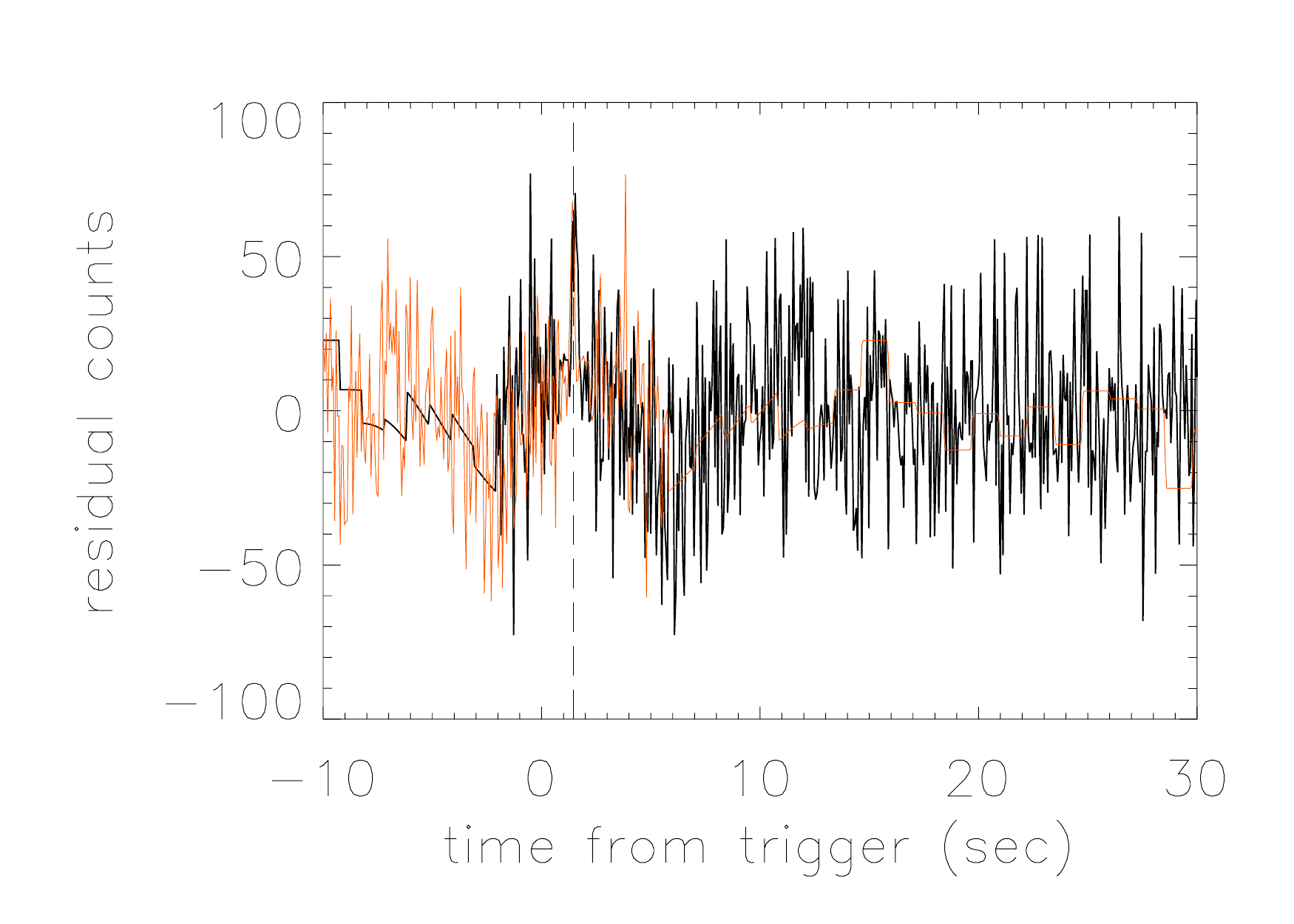} & 
    \includegraphics[width=.23\textwidth]{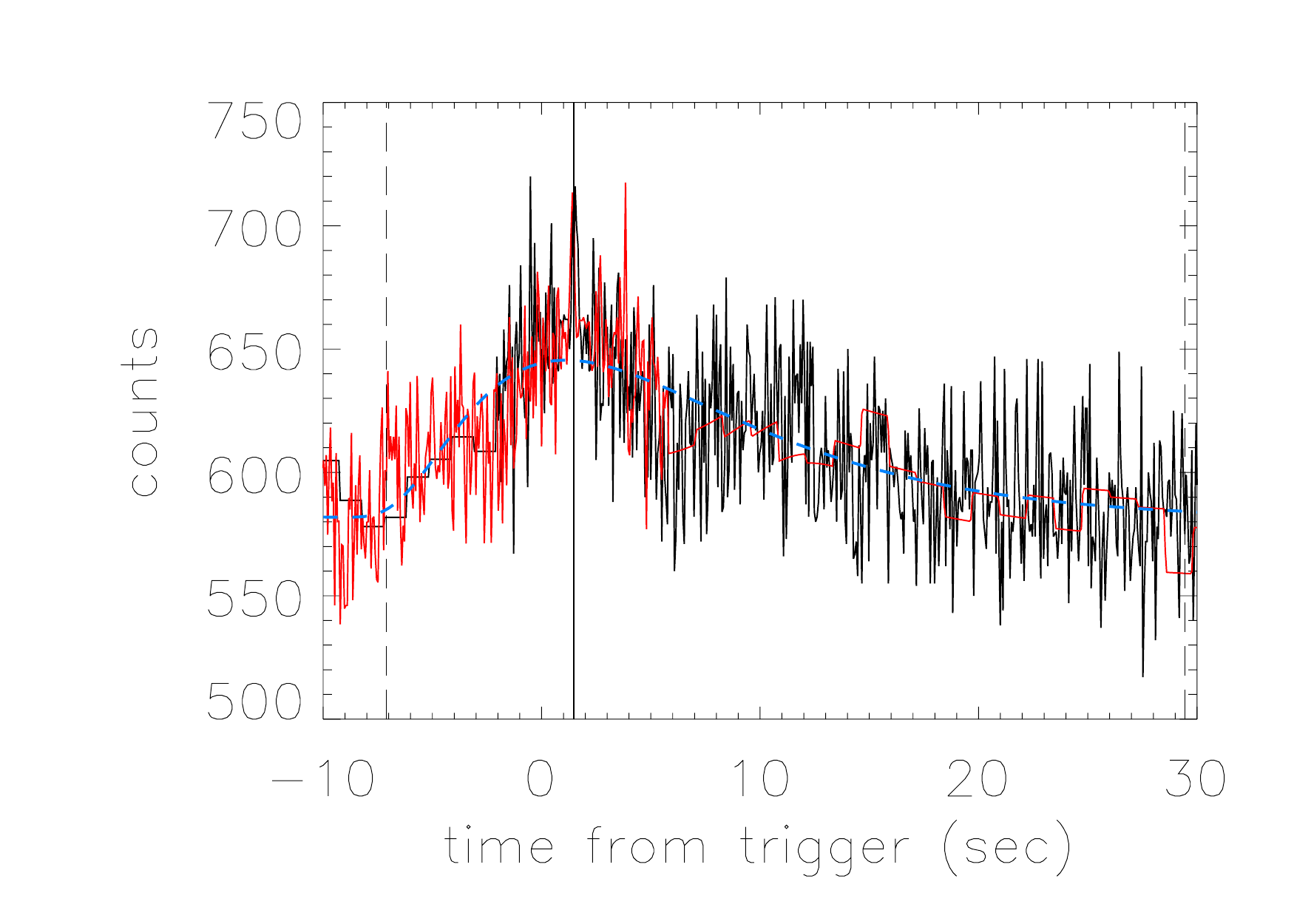} \\
    \includegraphics[width=.23\textwidth]{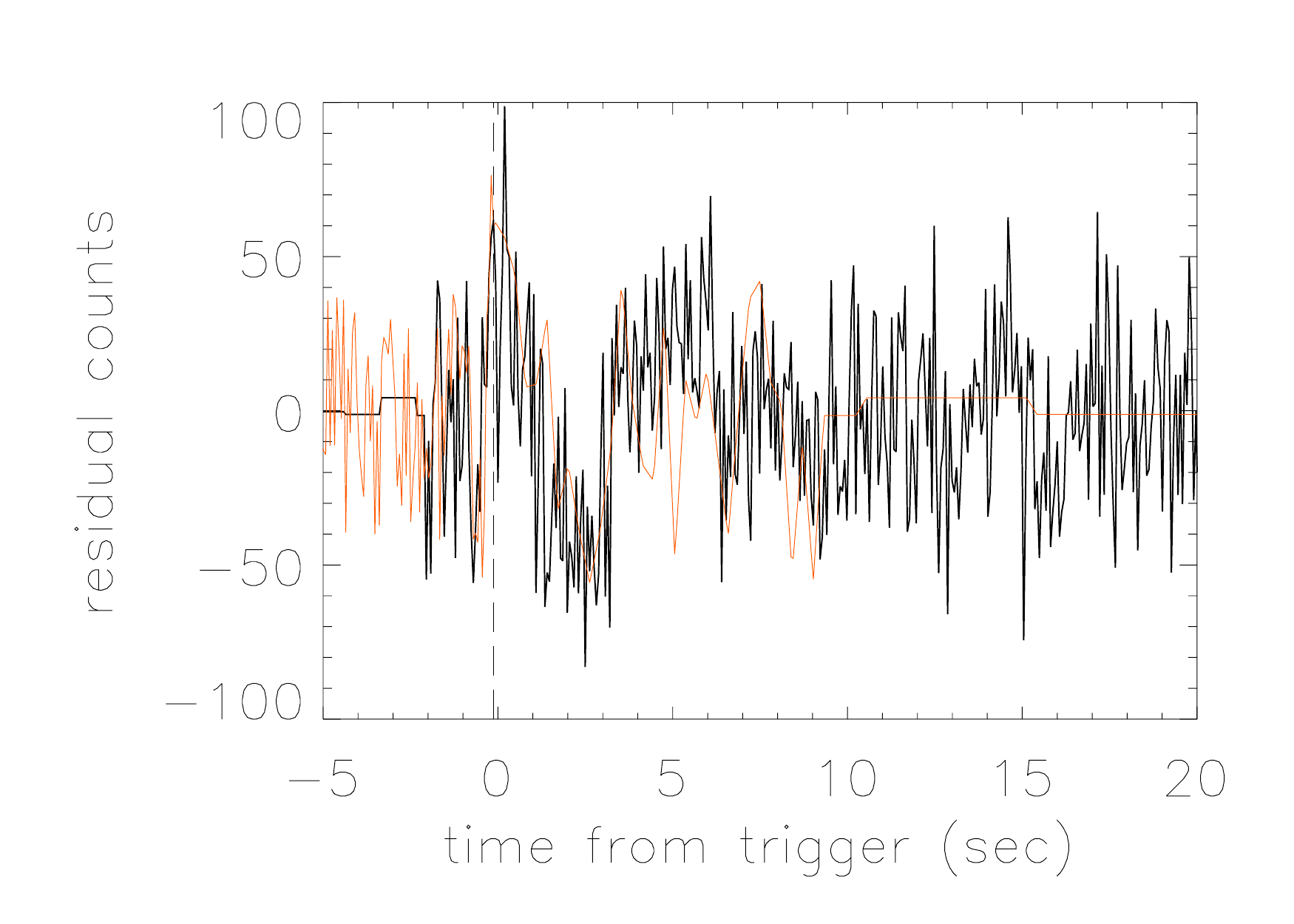} &
    \includegraphics[width=.23\textwidth]{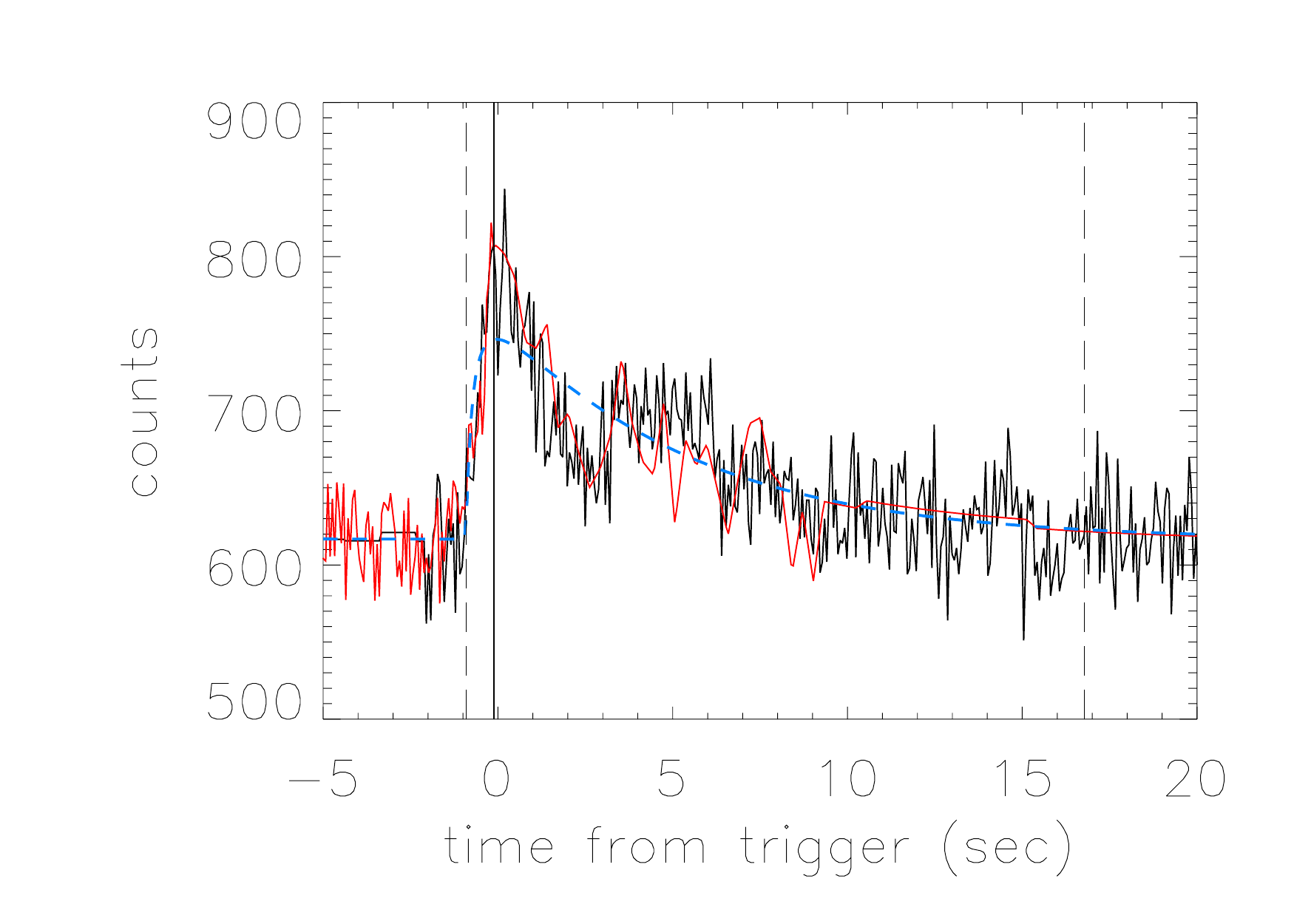} \\ 
    \includegraphics[width=.23\textwidth]{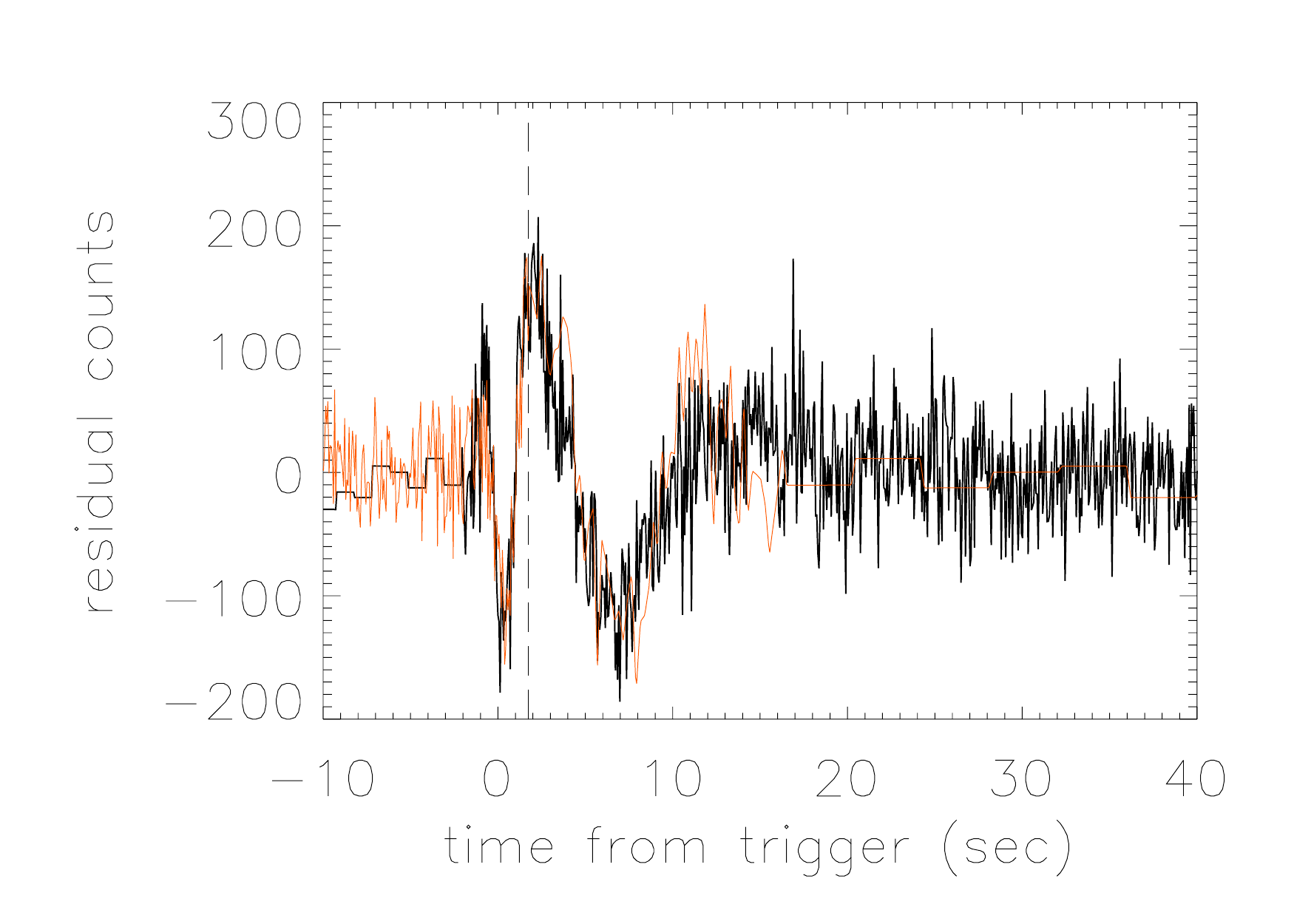} &
    \includegraphics[width=.23\textwidth]{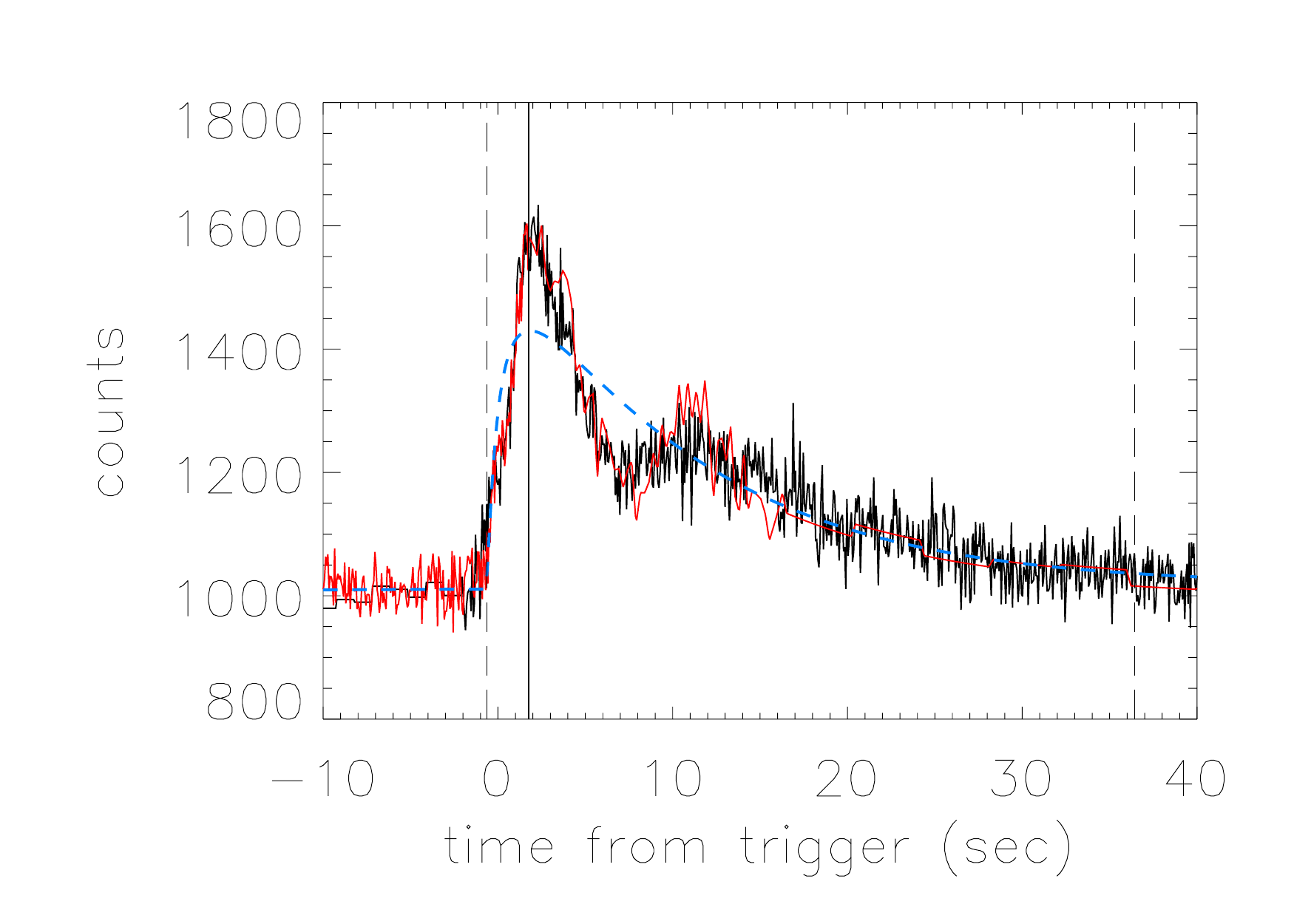} \\
    \includegraphics[width=.23\textwidth]{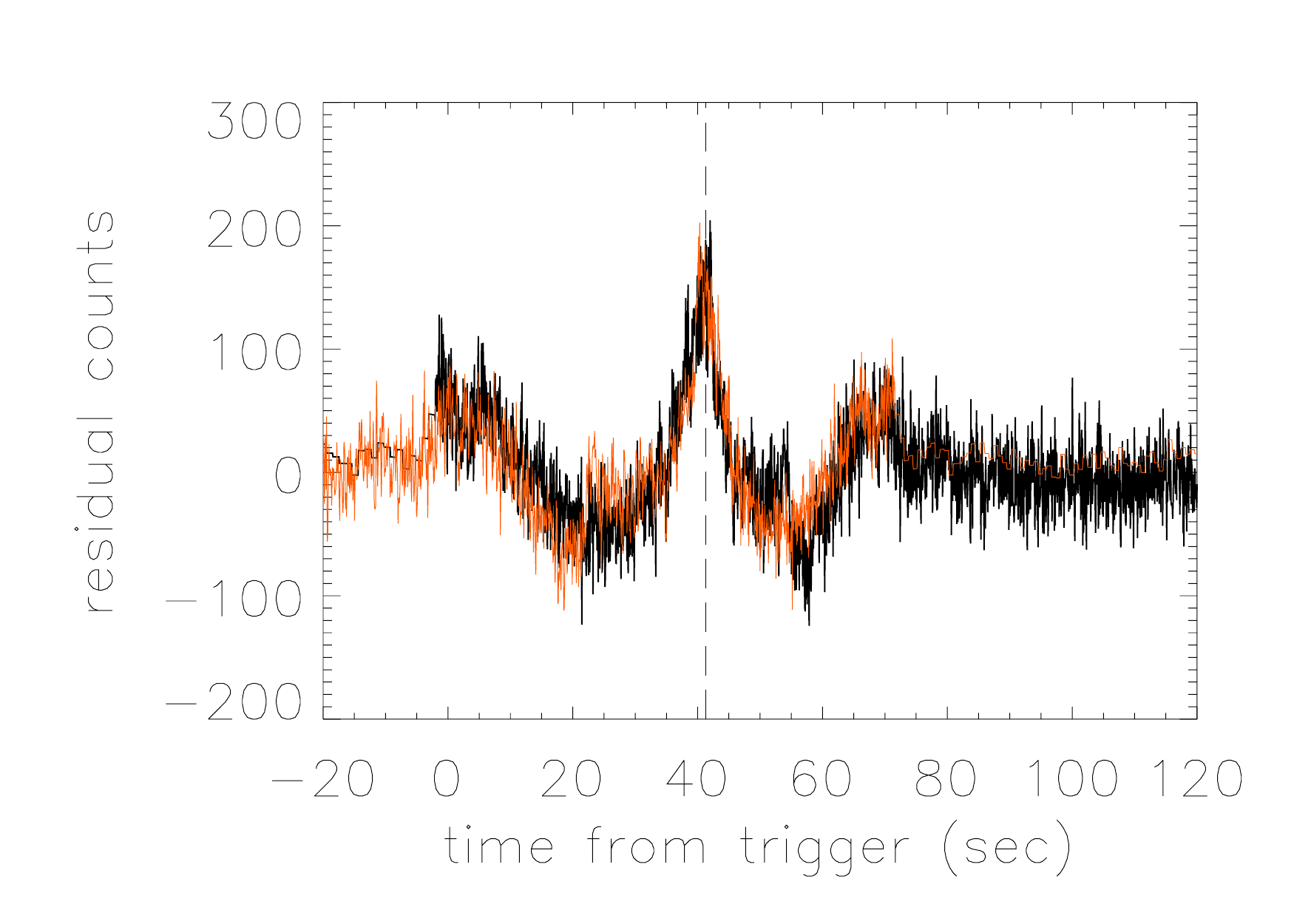} &
    \includegraphics[width=.23\textwidth]{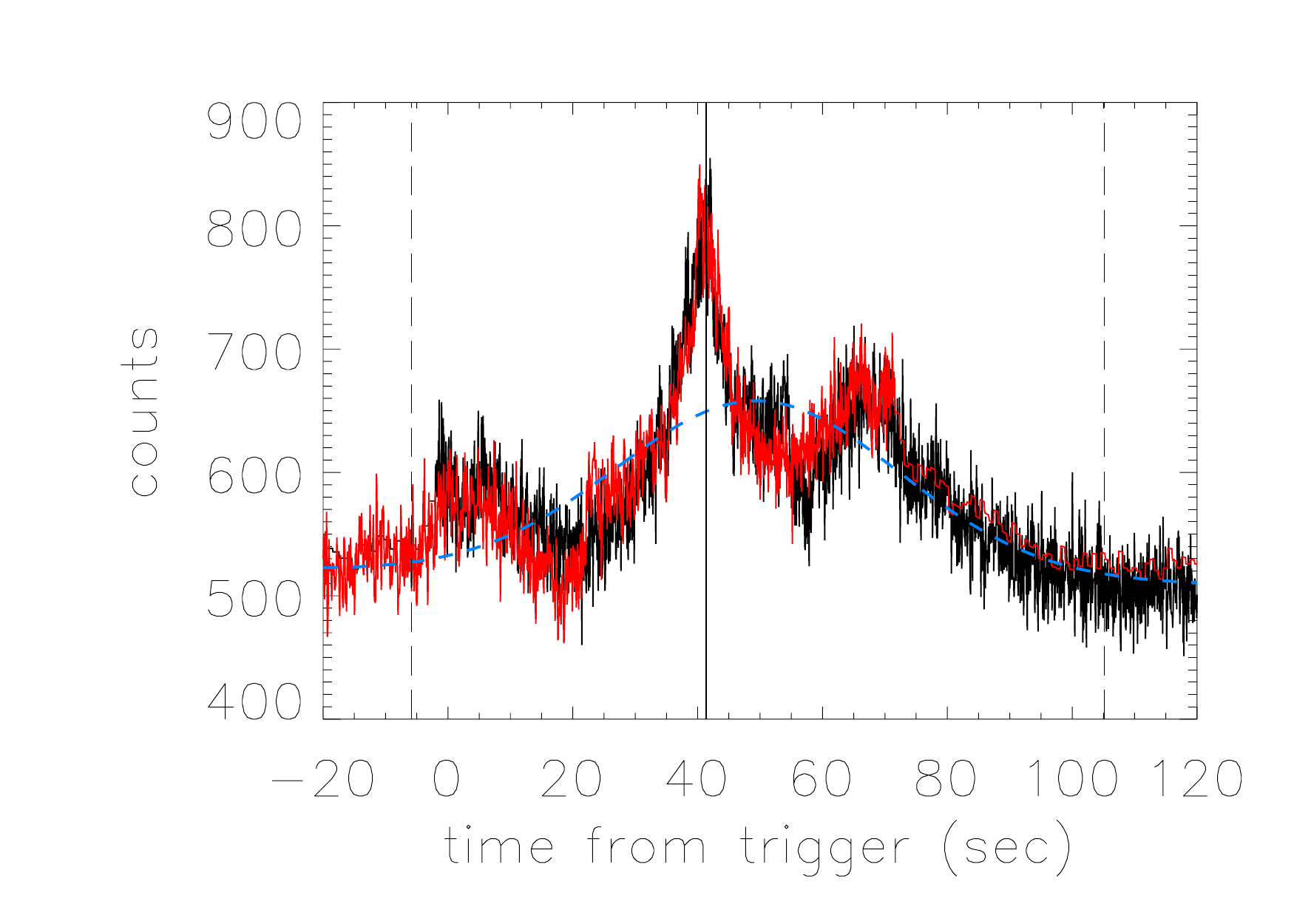} \\
  \end{tabular}
  \caption{Examples of CCF rations in temporally-symmetric GRB pulses. Residual fits (orange) are shown in the left column while corresponding model fits (red) are shown in the right column. In the residual plots, the time of reflection is identified by a vertical dashed line, while in the model plots, the  vertical dashed lines indicate the duration window, and the vertical solid line indicates the time of reflection. A CCF ratio of 0.21 is obtained by comparing BATSE 235p2's residual model having $\rm CCF_{\rm resids}=0.10$ (upper left) to its combined pulse model having $\rm CCF_{\rm model}=0.49$ (upper right).  A CCF ratio of 0.41 is obtained by comparing BATSE 686's residual model having $\rm CCF_{resids}=0.32$ (upper center left) to its combined pulse model having $\rm CCF_{model}=0.77$(upper center right). A CCF ratio of 0.65 is obtained by comparing BATSE 1447's residual model having $\rm CCF_{resids}=0.65$ (lower center left) to its combined pulse model having $\rm CCF_{model}=0.61$(lower center right). A CCF ratio of 0.83 is obtained by comparing BATSE 351's residual model $\rm CCF_{resids}=0.64$ (lower left) to its combined pulse model $\rm CCF_{model}=0.83$ (lower right). \label{fig:figA1}}
\end{figure}

The CCF ratio is defined as the CCF of the time-reversed residuals ($\rm CCF_{resids}$) divided by the CCF of the time-reversed model ($\rm CCF_{model}$), or
\begin{equation}\label{eqn:CCF}
    {\rm CCF~ratio}=\frac{\rm CCF_{resids}}{\rm CCF_{model}},
\end{equation}
where both of these values are defined over the pulse duration window.
CCF ratios range from values of 0 to 1, as CCF ratios of 0 characterize pulses in which temporally symmetric structure is not measurable whereas those near unity characterize pulses in which it is. CCF ratios indicate how pronounced the temporally-symmetric structure is relative to the pulse as a whole. Examples of a range of CCF ratios are shown in Figure \ref{fig:figA1}.

\begin{figure}
\plottwo{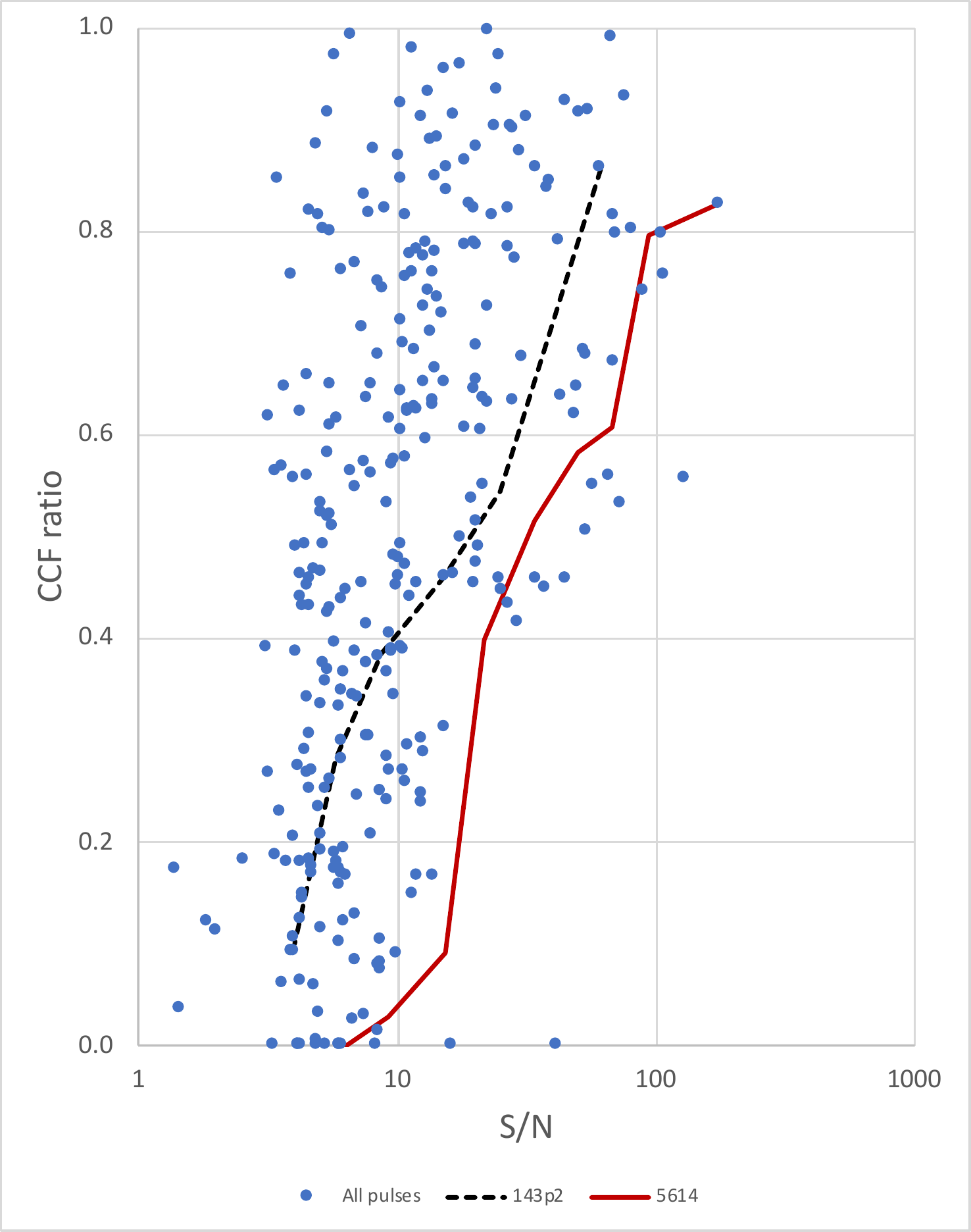}{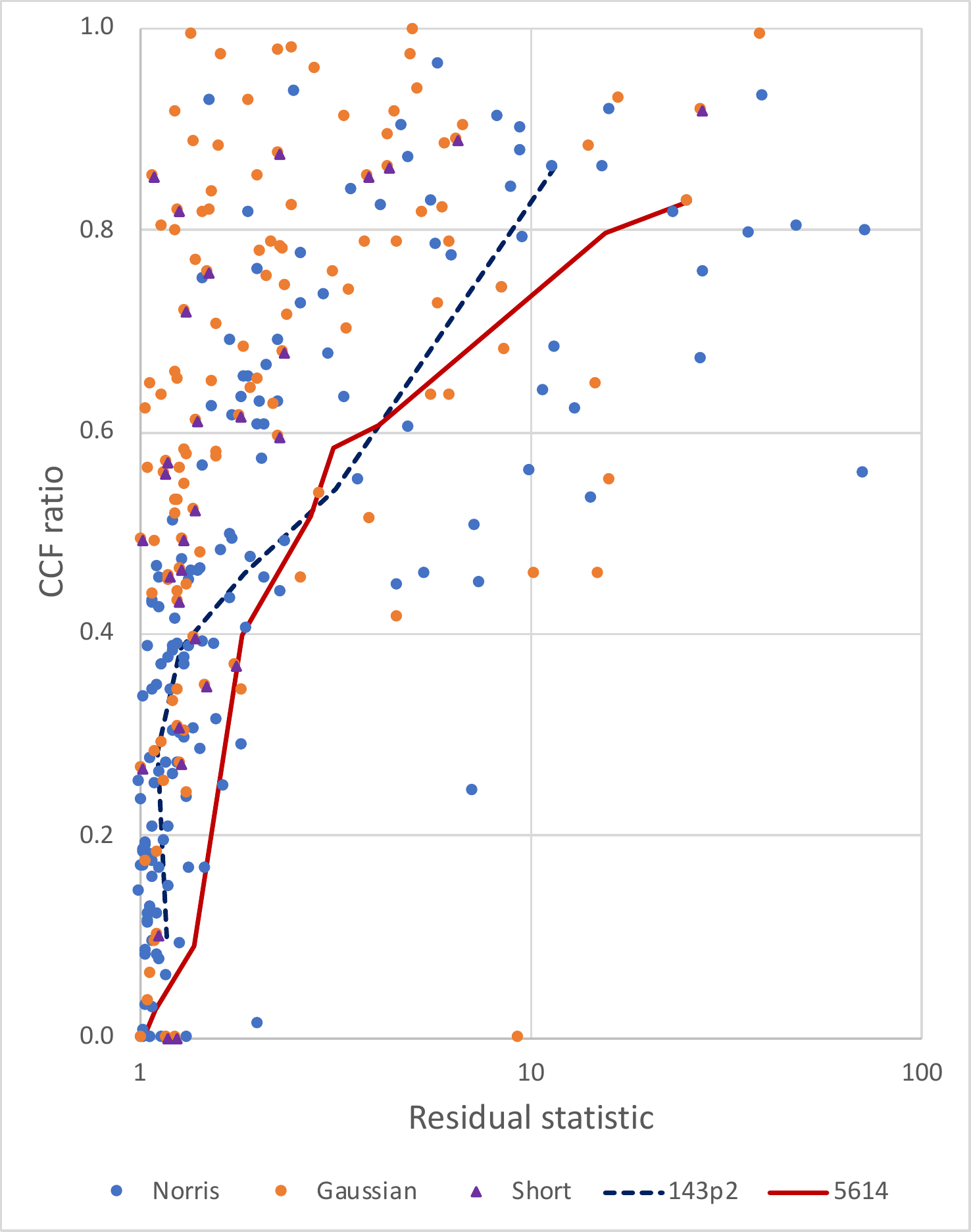}
\caption{The CCF ratios of successful and similar BATSE GRB pulses relative to the burst S/N ratio (left panel) and to the residual statistic (right panel). The intensities of two pulses described in \cite{hak18b} and shown in Figure \ref{fig:figA3} (black dotted line indicating BATSE 143p2 and red solid line indicating BATSE 5614) have been successively reduced, with their properties remeasured to demonstrate how the CCF ratio and residual statistic decrease with S/N ratio. These plots demonstrate the difficulty there is in separating structure from noise in pulses having CCF ratios less than 0.4. \label{fig:figA2}}\end{figure}

The CCF ratio allows us to measure the contribution that the temporally-symmetric residuals make to the overall temporal symmetry of the pulse. Presumably this value will be small for faint pulses near the detection threshold. The CCF ratio can be large for bright pulses exhibiting significant intrinsic structure, but it can also be small for bright pulses exhibiting little intrinsic structure. Examples of various GRB pulse CCF ratio can be seen relative to the pulse's S/N ratio in the left panel of Figure \ref{fig:figA2}, while the CCF ratio is plotted relative to each pulse's residual statistic $R$ in the right panel of Figure \ref{fig:figA2}. Indeed, pulses with pronounced temporally-symmetric residuals (characterized by large CCF ratios) are most likely to be found in bright pulses (those having large S/N ratios) and/or in pulses exhibiting pronounced structure (those having large R values).

\begin{figure}
\plottwo{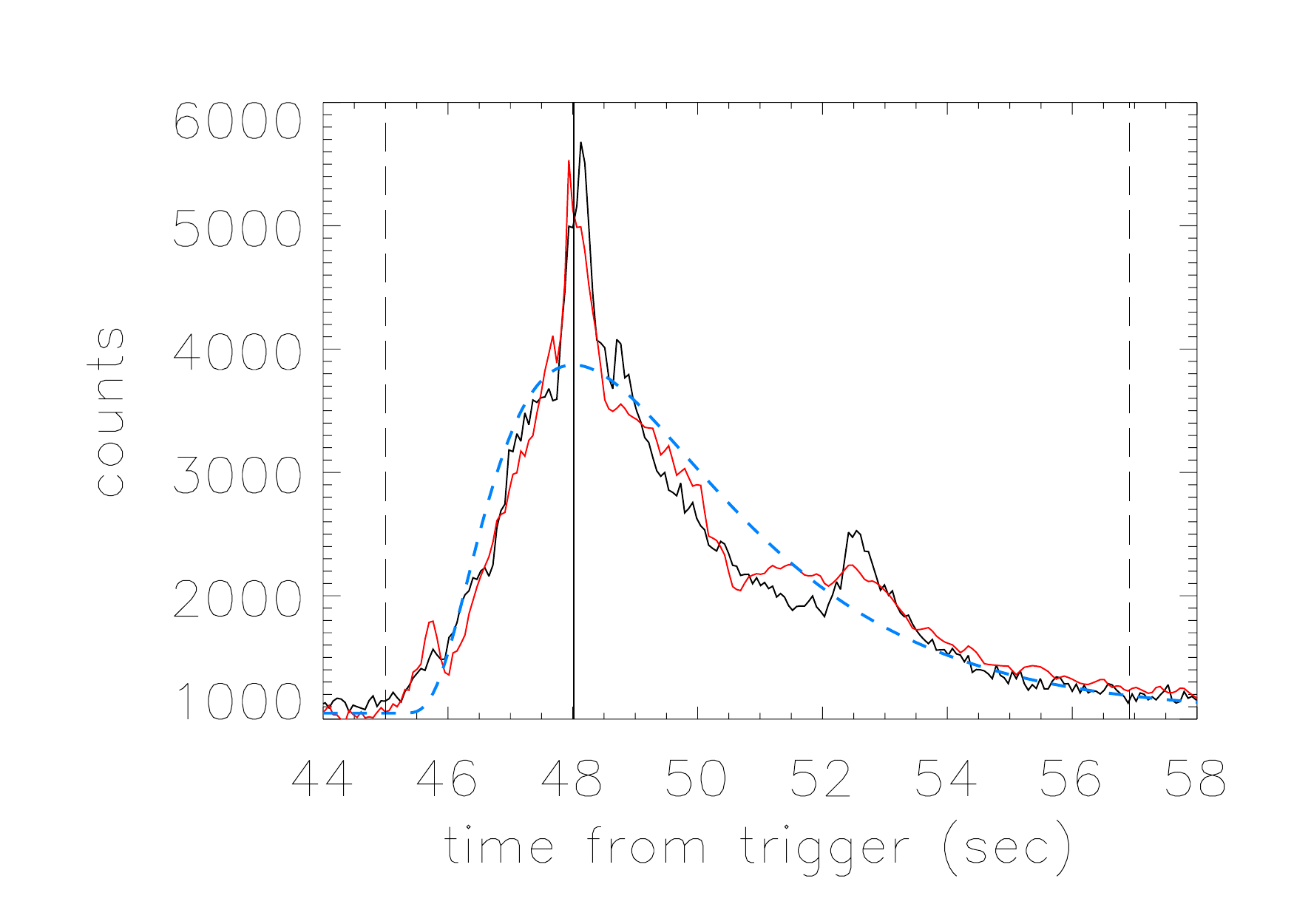}{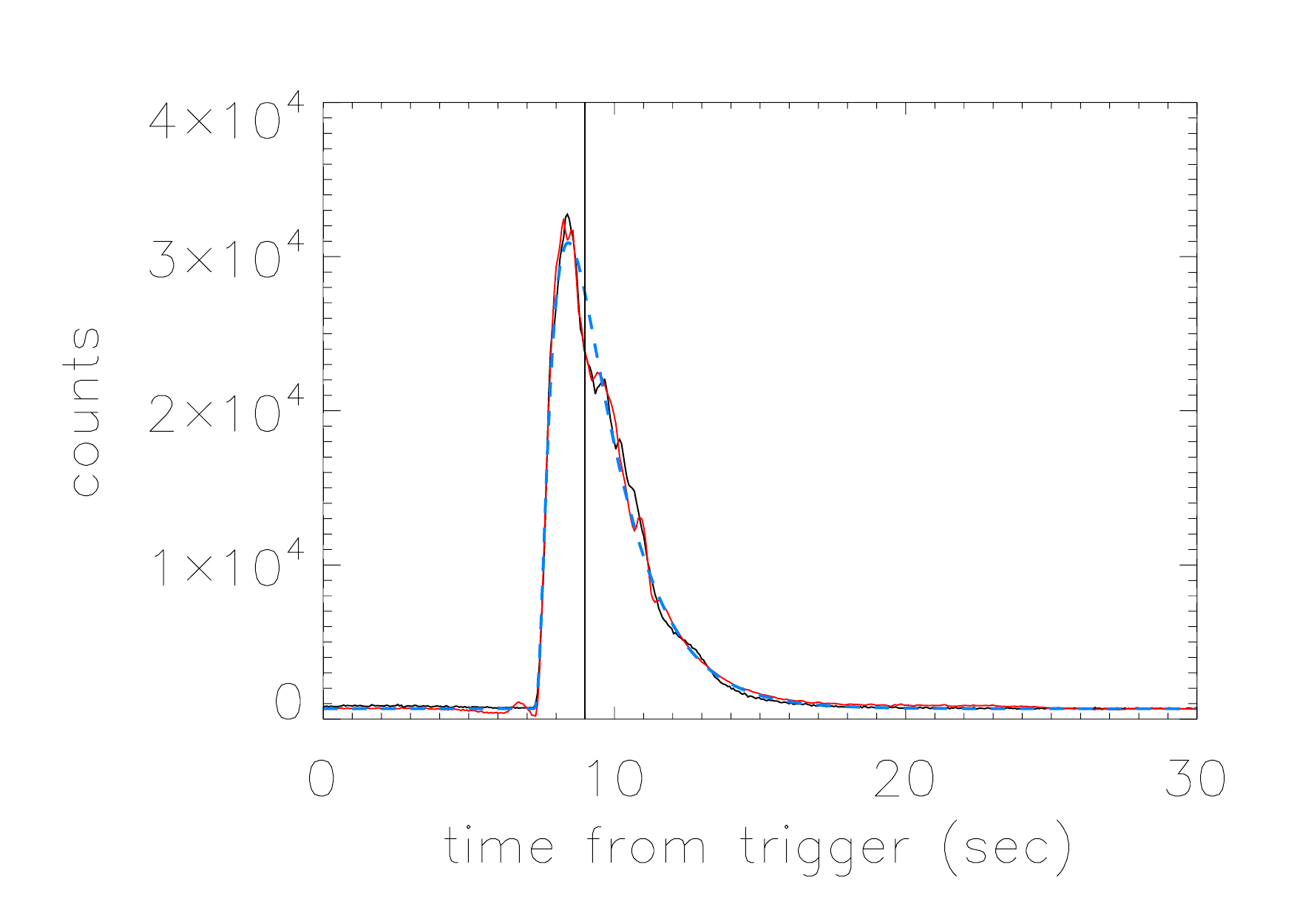}
\caption{Time-reversed models for BATSE 143p2 (left panel) and BATSE 5614 (right panel). Shown are the counts data (black), the fit to the Norris/Gaussian model (blue dashed line), the time-reversed model (red), the duration window (vertical dashed lines), and the time of reflection (vertical solid line).The structural properties of these pulses are shown as their light curves are reduced in intensity in Figure \ref{fig:figA2}. \label{fig:figA3}}\end{figure}

We demonstrate how the loss of structure can occur by examining two bright BATSE pulses from \cite{hak18b}) whose S/N ratios have been systematically reduced and their properties re-measured. The light curves and time-reversed models for these pulses (BATSE 143p2 and BATSE 5614) are shown in Figure \ref{fig:figA3}. As the S/N ratio of each pulse is reduced, so are their residual statistics and CCF ratios ({\em see} Figure \ref{fig:figA2}). The CCF ratios of both 143p2 (black dotted line) and 5614 (red solid line) become indistinguishable from other faint pulses once their S/N ratio has been sufficiently reduced. BATSE Pulses with CCF ratios less than 0.4 appear to have inadequate amounts of temporally-symmetric structure. 



\end{document}